\pdfoutput=1
\documentclass[preprint]{aastex}

\shorttitle{{\it Herschel} Observations of NGC~4631}
\shortauthors{Mel\'endez  et al.}

\begin{document}

%% LaTeX will automatically break titles if they run longer than
%% one line. However, you may use \\ to force a line break if
%% you desire.

\title{Exploring the Dust Content of Galactic Winds with
  {\it Herschel}. \\ I. NGC~4631\thanks{{\it Herschel} is an ESA space
    observatory with science instruments provided by European-led
    Principal Investigator consortia and with important participation
    from NASA.} }

%% Use \author, \affil, and the \and command to format
%% author and affiliation information.
%% Note that \email has replaced the old \authoremail command
%% from AASTeX v4.0. You can use \email to mark an email address
%% anywhere in the paper, not just in the front matter.
%% As in the title, use \\ to force line breaks.

\author{M. Mel\'endez\altaffilmark{1}, S. Veilleux\altaffilmark{1,2},
  C. Martin\altaffilmark{3}, C. Engelbracht\altaffilmark{4},
  J. Bland-Hawthorn\altaffilmark{5}, G. Cecil\altaffilmark{6},
  F. Heitsch\altaffilmark{6}, A. McCormick\altaffilmark{1},
  T. M{\"u}ller\altaffilmark{7}, D. Rupke\altaffilmark{8} \& S. H. Teng\altaffilmark{9} }

\altaffiltext{1}{Department of Astronomy, University of Maryland,
  College Park, MD 20742, USA; marcio@astro.umd.edu, 
  veilleux@astro.umd.edu}

\altaffiltext{2}{Joint Space-Science Institute, University of Maryland,
  College Park, MD 20742, USA}

\altaffiltext{3}{Department of Physics, University of California,
  Santa Barbara, CA 93106}

\altaffiltext{4}{Department of Astronomy, University of Arizona,
  Tucson, Arizona 85721 (deceased)}

\altaffiltext{5}{Department of Physics, University of Sydney, Sydney,
  NSW 2006, Australia}

\altaffiltext{6}{Department of Physics, University of North Carolina,
  Chapell Hill, NC 27599}

\altaffiltext{7}{Max-Planck-Institute for Extraterrestrial Physics
   (MPE), Giessenbachstra{\ss}e 1, 85748 Garching, Germany}

\altaffiltext{8}{Department of Physics, Rhodes College, Memphis, TN 38112}

\altaffiltext{9}{Science and Tech Division, Institute for Defense Analyses, Alexandria, VA 22311}

\begin{abstract}
We present a detailed analysis of deep far-infrared observations of
the nearby edge-on star-forming galaxy NGC~4631 obtained with the {\it
  Herschel Space Observatory}. Our PACS images at 70 and 160 $\micron$
show a rich complex of filaments and chimney-like features that
extends up to a projected distance of 6~kpc above the plane of the
galaxy. The PACS features often match extraplanar H$\alpha$,
radio-continuum, and soft X-ray features observed in this galaxy,
pointing to a tight disk-halo connection regulated by star formation.
On the other hand, the morphology of the colder dust component
detected on larger scale in the SPIRE 250, 350, and 500 $\micron$ data
matches the extraplanar H~I streams previously reported in NGC~4631
and suggests a tidal origin. The PACS 70/160 $\micron$ ratios are
elevated in the central $\sim$3.0 kpc region above the nucleus of this
galaxy (the ``superbubble''). A pixel-by-pixel analysis shows that
dust in this region has a higher temperature and/or an emissivity with
a steeper spectral index ($\beta > 2$) than the dust in the disk, 
  possibly the result of the harsher environment in the superbubble.
  Star formation in the disk seems energetically insufficient to lift
  the material out of the disk, unless it was more active in the past
  or the dust-to-gas ratio in the superbubble region is higher than
  the Galactic value. Some of the dust in the halo may also have been
  tidally stripped from nearby companions or lifted from the disk by
  galaxy interactions.

\end{abstract}

\section{Introduction}
The interstellar medium (ISM) of galaxies keeps a record of one of
their most important properties: their star formation history.  As
galaxies age, their dying stars enrich the ISM by injecting gas and
dust that eventually make up the next generation of stars. As a
consequence of this recycling process, interstellar dust plays a
fundamental role in the chemistry and appearance of a galaxy. Dust
grains are unambiguously associated with galaxy evolution because they
can work as a catalyst for chemical reactions resulting in the
creation of molecules, they can absorb and re-emit starlight thus
helping with the heating and cooling of the ISM, and they can remove
heavy elements from the gas phase.  For all these reasons, it is
crucial to understand the location and dust content of galaxies in
order to understand galaxy evolution. In this regard, galactic winds
are the primary mechanism in which gas and dust can be recycled
through the entire galaxy. Thus, it is important to study the
geometry, mass, and energy of dusty galactic winds \citep[see ][for a
  review]{2005ARA&A..43..769V}. Overall, dusty galactic winds can have
a direct impact on the morphology of the galaxy and could be a leading
factor in regulating the star formation rate (SFR) by removing the
material from which stars are formed
\citep{2010A&A...518L..66R,2013Natur.499..450B,2011ApJ...733L..16S,2013ApJ...776...27V,2014A&A...562A..21C}.
 
This first paper in a series focuses on the nearby \citep[7.4 Mpc,
][]{2011ApJS..195...18R}, edge-on ($i \sim 86\arcdeg$) starburst
galaxy, NGC~4631. This galaxy shows a prominent multi-phase halo that
comprises a relativistic radio component
\citep{1977A&A....54..973E,1988A&A...197L..29H,1990A&A...236...33H},
diffuse ionized gas
\citep{1992ApJ...396...97R,1995ApJ...450L..45D,1996A&A...313..439G,1999ApJ...522..669H},
dust \citep{2001ApJ...555..258M}, molecular gas
\citep{2000ApJ...535..663R,2011MNRAS.410.1423I}, and hot X-ray
emitting plasma
\citep{1987ApJ...315...46F,1995ApJ...439..176W,1996A&A...311...35V,2001ApJ...555L..99W,2004ApJS..151..193S,2006A&A...448...43T,2009PASJ...61S.291Y}.
      The dimension (vertical height) of the halo ranges from
        740~pc \citep[molecular gas;][]{2000ApJ...535..663R} up to 18
        kpc \citep[ionized gas;][]{1995ApJ...450L..45D}.  This object
      thus offers an unique opportunity to study the connection between the
      disk and the extended emission in galaxies. Galactic-scale winds
      or outflows may be responsible for ejecting the dusty gas from
      the disk into the halo \citep[e.g.,][]{1994A&A...284..777G} and,
      ultimately, heating and enriching the circumgalactic
      medium. There has been mounting evidence for this outflowing gas
      in NGC~4631
      \citep{1990A&A...236...33H,1994A&A...284..777G,1999A&A...343...51A,1999ApJ...522..669H,2000ApJ...535..663R,2009A&A...506..689I,2011MNRAS.410.1423I},
      thus the question: what powers this outflow? Given the fact that
      there is no active galactic nucleus in the center of NGC~4631, a
      starburst-driven wind seems appropriate. Contrary to M82
      \citep{1963ApJ...137.1005L,1999A&A...343...51A}, NGC~4631 does
      not possess a nuclear starburst, but rather a central molecular
      ring \citep{2001A&A...373..853D,2011MNRAS.410.1423I} with
      widespread H$\alpha$ and UV emission distributed throughout the
      entire stellar disk
      \citep{1969A&A.....2....1C,1992ApJ...396...97R,2001ApJ...546..829S}.
      Galaxy interactions may also have played a role in shaping the
      gaseous halo.  NGC~4631 lives in a group which has been
      extensively studied in H~I 21-cm, where tidal spurs and two
      highly energetic supershells have been detected
      \citep{1968ApJ...151..117R,1978A&A....65...37W,1978A&A....65...47C,1993AJ....105.2098R,1994A&A...285..833R}.
      NGC 4627, a dwarf elliptical galaxy ($\sim$3\arcmin~ NW of the
      nucleus), and NGC~4656, an edge-on spiral ($\sim$30\arcmin~ SE
      of the nucleus), are the most prominent companion galaxies of
      NGC~4631. The NGC~4631 group of galaxies also contains three
      dwarfs observed in H~I
      \citep{1993AJ....105.2098R,1994A&A...285..833R}, a young optical
      dwarf galaxy candidate \citep{2005AJ....129.1331S}, and an
      UV-bright tidal dwarf galaxy \citep{2012ApJ...750..171S}.

There is also some evidence for dust outside the main body of the
galaxy.  The first detection of a cold dust component in NGC~4631
was presented by \cite{1995A&A...295L..55B} at 1.3~mm where they found
a good correspondence between the CO \citep{1994A&A...286..733G} and
the 1.3 mm continuum in the inner 2~kpc of the galaxy. A deeper map
presented by \cite{1999PNAS...96.5360N} proved the existence of a cold
dust component at 1.2~mm outside the disk of the galaxy that follows
(in part) the more extended H~I extraplanar emission. The first maps
at 450 and 850~$\micron$ were presented by \cite{1999A&A...343...51A},
focusing on the inner 3\arcmin\, where they tentatively connected the
expulsion of dust grains (dust outflow) with various small ``chimneys"
or extraplanar dusty filaments associated with the enhanced and
widespread star formation activity in NGC~4631. Far-infrared (FIR) and
submillimeter maps and fluxes for NGC~4631 can be found in various
studies including, \cite{2003AJ....125.2361B,2006ApJ...652..283B},
\cite{2004A&A...414..475D},
\cite{2005ApJ...633..857D,2007ApJ...655..863D,2009ApJ...703..517D} and
\cite{2005MNRAS.357..361S}. However, despite the great effort to
understand the warm dust properties and structure in NGC~4631, all the
previous studies were limited by relatively poor angular resolution. The
superior angular resolution and sensitivity of {\it Herschel Space
  Observatory} \citep{2010A&A...518L...1P} provides an unique
opportunity to study the dust content and structure of nearby
galaxies.

In this paper we present a spatially resolved analysis of the
warm-to-cold dust components in NGC~4631 using our own {\it Herschel}
observations at 70, 100, 160, 250, 350, and 500 $\micron$. This paper
is structured as follows. In Section~\ref{OD}, we describe in detail
the different {\it Herschel} observations used in our analysis and the
data processing. In Section~3, we present a morphological comparison
between the {\it Herschel} observations and published data spanning
the entire radio to X-ray energy range in order to investigate the
connection between the various components in the halo. Section~4
presents the results from a more detailed analysis of our data,
including the dust temperature morphology of the warm and cold dust
components in the halo (Section~4.1 and 4.2), dust mass estimates
(Section~4.3), star formation rates (Section~4.4), and a spatially
resolved dust analysis of NGC~4631 (Section~4.5). In Section~5 we
develop a picture for the origin and properties of the dusty halo in
NGC~4631. Finally, the main conclusions are summarized in Section~6.
In addition, Appendix~\ref{apend} presents a detailed comparison
between the FIR flux and color maps and the brightest X-ray point
sources detected in this galaxy with {\em Chandra}. Throughout this
paper, we adopt a distance of 7.4 Mpc for NGC 4631
\citep{2011ApJS..195...18R}, corresponding to a linear scale of $\sim$36
pc per arcsecond.

\section{Observations and  Data Processing\label{OD}}

NGC~4631 was observed with the Photodetector Array Camera and
Spectrometer \citep[PACS, ][]{2010A&A...518L...2P} on board {\it
  Herschel}. The main observations presented in this work are from our
cycle 1 open-time program (OT1\_sveilleu\_2, PI: S. Veilleux).  For
this program we obtained simultaneous PACS imaging for the PACS blue
70 $\micron$ (60-85 $\micron$) and red 160 $\micron$ (130-210
$\micron$) channels in scan mode along seven position angles at
50\arcdeg, 60\arcdeg, 70\arcdeg, 110\arcdeg, 120\arcdeg, 130\arcdeg,
and 140\arcdeg~ (Obs IDs: 1342235130, 1342235131, 1342235132, 1342235133, 1342235134, 1342235135 and 1342235136). At each orientation angle, we requested 76 scan legs
of 6.0\arcmin\ length, 4.0\arcsec\ scan leg separation, a repetition
factor of two, and a scan speed of 20\arcsec\ s$^{-1}$. The
total time per position angle, including telescope overhead, was
$\sim$ 1.6 hours, for a total request of 11.1 hours. We complement our
program with archival observations at 70, 100 and 160
\micron\ obtained from the ``Key Insight of Nearby Galaxies: A
Far-Infrared Survey with Herschel" (KINGFISH) open-time key program
(KPOT\_rkennicu\_1). These observations were taken in scan mode along
two position angles at 45\arcdeg\ and 135\arcdeg~ (Obs IDs: 1342209672, 1342209673, 1342209674 and 1342209675). At each orientation
angle, the observations consisted of one scan leg of
21.0\arcmin\ length, 2.0\arcsec\ scan leg separation, a repetition
factor of three, obtained at a  scan speed of
20\arcsec\ s$^{-1}$. The total observation time per scan angle,
including telescope overhead, was $\sim$ 0.7 hours, for a total of
$\sim$ 1.4 hours. Observations from the Spectral and Photometric
Imaging Receiver \cite[SPIRE,][]{2010A&A...518L...3G} were taken
simultaneously at 250, 350, and 500 $\micron$ as part of the KINGFISH
open-time key program. NGC~4631 was observed in large map mode
covering an area of 23\arcmin\ $\times$ 23\arcmin\ with two
repetitions at nominal speed (30\arcsec\ s$^{-1}$).

For the PACS data reduction, we used the Herschel Interactive
Processing Environment \citep[HIPE,][]{2010ASPC..434..139O} version
8.0. The ``Level 0" observations (raw data) were processed through the
standard pipeline procedure to convert from Level 0 to Level 1
data. This procedure includes the extraction of the calibration tree
needed for the data processing, correction for electronic crosstalk,
application of the flat-field correction, and finally deglitching and
conversion from Volts to janskys per array pixel. In order to correct
for bolometer drift (low frequency noise), both thermal and
non-thermal (uncorrelated noise), and to create the final maps from
the Level 1 data, we used the algorithm implemented in {\it
  Scanamorphos} \citep[v21.0,][]{2013PASP..125.1126R}, which makes use
of the redundancy built in the observations to derive the brightness
drifts. Because of this, {\it Scanamorphos} is independent of any
pre-defined noise model because it relies on the fact that each
portion of the sky is scanned by multiple bolometers at different
times. All final PACS maps have a pixel size of $\sim$1/4 of the PSF
FWHM, i.e., 1.4\arcsec\ at 70 \micron\ and 2.85\arcsec\ at 160
$\micron$.  {\it Scanamorphos} also produces error and weight
maps. The error map is defined as the error on the mean brightness in
each pixel. It is built using the weighted variance because weights
are used for the projection of the final map. The error map does not
include any error propagation associated with the different steps
performed on the pipeline.  On the other hand, the weighted map is
built by co-adding the weights and is normalized by the average of the
weights. Given the relatively small field of view of our observations,
relative to the size of the galaxy, and the observing strategy, we
created the {\it Scanamorphos} maps with the ``minimap" and ``flat"
options.  On the other hand, the SPIRE data was processed from ``level
0" up to level 1 with the HIPE scripts included in the {\it
  Scanamorphos} distribution. The preprocessing by the pipeline
includes the same steps as in the PACS pipeline except that
the conversion to brightness is in janskys per beam and the thermal
drifts are subtracted by using the smoothed series of thermistors
located on the detector array as the input of the drift model
\citep[see ][for details]{2010ASPC..434..139O}. The final SPIRE maps
have a pixel size of $\sim$1/4 of the PSF FWHM, i.e., 4.5\arcsec,
6.25\arcsec, and 9.0\arcsec\ at 250, 350, and 500 $\micron$,
respectively.

To extract the global fluxes, we chose an elliptical aperture that
visually encloses all of the observed emission at 500 $\micron$. The
parameters of this ellipse are: major and minor semi-axes of 514 and
150\arcsec, and position angle of 84\arcdeg, measured anti-clockwise
from north to east (i.e. aligned along the major axis of the
disk). In addition, background subtraction is performed locally with
  a circular or elliptical annulus around the source where the
  background annulus was set to encompass a clean, uncontaminated sky
  region close to the source. For the large field of view observations
  (KPOT\_rkennicu\_1), the background contribution is measured in an
  elliptical annulus with inner major and minor semi-axis equal to the
  global flux aperture and outer major and minor axis equal to 1.2
  times the inner values. For the smaller field of view observations
  (70 and 160~$\micron$, OT1\_sveilleu\_2), where the global
  elliptical aperture is larger than the field of view, the sky
  background (and standard deviation) was measured from 10 circular
  apertures placed above and below the galaxy plane, not too close to
  the galaxy (to avoid galaxy contamination) and far from the edges of
  the map (to avoid unreliable fluxes due to elevated noise). The
  total uncertainty on the integrated photometric measurements is a
  combination of the error on the mean brightness in each pixel added
  in quadrature within the source aperture (the error map produced by
  {\it Scanamorphos}), the standard deviation of all the pixels in the
  background aperture, and the PACS photometer flux calibration
  accuracy. For the calibration uncertainties (extended sources), we
  adopted a large (conservative) value of 10\% for both PACS and
  SPIRE. This value comes from adding the systematic (4-5\%),
  statistical (1-2\%), and PSF/beam size uncertainties (4\%).  To
  derive aperture corrections, we used the higher resolution image
  from {\it Spitzer}/IRAC 8.0 $\micron$ \citep{2013ApJ...774..126M}
  and measured the total flux with the same aperture employed for our
  analysis. Then we convolved the same image with the appropriate
  kernel to bring it to the PACS
  resolution\footnote{http://www.astro.princeton.edu/$\sim$ganiano/Kernels.html}
  \citep{2011PASP..123.1218A} and remeasured the flux in the same
  aperture. The ratio of the unconvolved to the convolved (the same
  PSF as the {\it Herschel} PACS) flux is used as an estimate of the
  aperture correction. For the global flux, the big aperture resulted
  in small corrections with values of 1.01, 1.01, and 1.02 for PACS
  70, 100, and 160 \micron, respectively, and 1.00 for all of the
  SPIRE images. We color-corrected PACS fluxes assuming a modified
  blackbody with $\beta = 2$ and a blackbody temperature of $T = 30$~K
  (0.977/0.974/1.037 at 70/100/160~$\micron$)\footnote{PACS Photometer
    Passbands and Colour Correction Factors for Various Source
    SEDs. Document PICC-ME-TN-038, version 1.0}.

Our PACS flux densities for the entire galaxy are 141.9 $\pm$ 14.2,
232.9 $\pm$ 23.3, and 247.0 $\pm$ 24.7 Jy at 70, 100, and 160
$\micron$, respectively (see Table~\ref{fluxes}).  These fluxes are
consistent, within the uncertainties, with the (color uncorrected)
{\it Herschel} fluxes from \cite[][137 $\pm$ 7, 223 $\pm$ 11, and 246
  $\pm$ 12 at 70, 100, and 160 $\micron$,
  respectively]{2012ApJ...745...95D}. Our values are also in agreement
with the {\it Spitzer} MIPS photometry of NGC~4631 presented by
\cite[][138 $\pm$ 17 and 269 $\pm$ 42 Jy at 70 and 160 $\micron$,
  respectively]{2009ApJ...703..517D}.  In addition, there are reported
values for the (color-corrected) {\it IRAS} total flux densities at 60
and 100 $\micron$ of 89.3$\pm$ 13.4~Jy and 216.4 $\pm$ 32.5~Jy from
\cite{1988ApJS...68...91R}, and 90.7$\pm$18.2 and 170.4$\pm$34.1 at 60
and 100 $\micron$, respectively \citep{1989ApJS...70..699Y}. To
convert the SPIRE data from Jy beam$^{-1}$ to Jy pixel$^{-1}$, we used
the pipeline beam area as provided in the SPIRE Data Reduction Guide,
Table 6.7, i.e., 465.39, 822.58, and 1768.66 arcsec$^2$ at 250, 350,
and 500 \micron, respectively. For the SPIRE fluxes, we applied color
corrections for extended sources assuming a gray body with emissivity
index $\beta = 2.0$ and modified blackbody temperature of 30 K
(0.9790/0.9687/0.9783 at 250/350/500~$\micron$)\footnote{Table 5.7 in
  SPIRE Handbook, version 2.5, March 17, 2014}.  The parameters
  used for the PACS and SPIRE color corrections are based on our full
  analysis of the spectral energy distribution of NGC~4631 (see
  Section~\ref{DM}).  Our SPIRE global flux densities are 111.4 $\pm$
11.1, 50.0 $\pm$ 5.0, and 19.9 $\pm$ 2.0 Jy at 250, 350, and 500
$\micron$, respectively. Our SPIRE fluxes are consistent, within the
uncertainties, with the (color-uncorrected) {\it Herschel} fluxes from
\cite[][with 124 $\pm$ 9, 54.5 $\pm$ 3.9, and 24.0 $\pm$ 1.7 Jy at
  250, 350, and 500 $\micron$,
  respectively]{2012ApJ...745...95D}.\footnote{Note that the SPIRE
  fluxes presented in \cite{2012ApJ...745...95D} were estimated with a
  different set of beam areas than the ones used in the present work,
  namely, 423, 751, and 1587 arcsec$^2$ at 250, 350, and 500~\micron.}

% 1.38 ± 0.17E+2	2.69 ± 0.42E+2

\section{Morphological Comparisons with Ancillary Data Sets\label{adata}}

In addition to the PACS and SPIRE observations, we assembled a set of
multi-wavelength ancillary observations for the purpose of multiphase
comparisons. In the mid-infrared, we used images from {\it Spitzer}
IRAC 4.5 and 8 \micron\ and MIPS 24
\micron\ \citep{2009ApJ...703..517D,2013ApJ...774..126M}.  In the
optical, we compared the {\it Herschel} data with observations of the
warm ionized gas in H$\alpha$ \citep{1992ApJ...396...97R}, and in the
radio, we used the atomic hydrogen H~I 21 cm maps of
\citep{1994A&A...285..833R} and the {\it L}-band (1.5~GHz) images from
\cite{2012AJ....144...44I}. Finally we also compared the {\it
  Herschel} data with soft X-ray images (0.3 -- 2.0 keV) from our own
analysis of archival {\it Chandra} observations of NGC~4631, where the
final image was created following the procedure described in
\cite{2004ApJS..151..193S}. For these comparisons, we enhanced the
fainter features in our PACS and SPIRE images by using the adaptive
median smoothing code ADAPTSMOOTH with a minimum signal-to-noise ratio
(S/N) equal to 10 per pixel with a direct local noise estimate
\citep[version V.2.7.1.15.11.2010,
][]{2009arXiv0911.4956Z,2009MNRAS.400.1181Z}. This adaptive smoothing
scheme retains the image effective resolution and does not alter the
photometric fluxes.

The results of the comparisons with the {\it Spitzer} data are shown
in Figure~\ref{mosaic_ngc4631}.  Note that the closest companion
galaxy, NGC~4627, is not detected in any of the {\it Herschel} maps,
not even in our very deep PACS images whereas it is clearly detected
in the IRAC maps at 4.5 and 8 \micron. This result indicates that
NGC~4627 is highly depleted in (warm and cold) dust, consistent with
its classification as a dwarf elliptical galaxy.

In Figure~\ref{NGC4631_PACS_Halpha} we present the comparison between
the PACS images at 70 and 160~$\micron$ and the diffuse ionized gas in
H$\alpha$ \citep[see][for details of the
  observations]{1992ApJ...396...97R}. The H$\alpha$ image shows a rich
complex of filaments and worm-like structures extending vertically
from the disk of the galaxy.  These structures coincide in position
with similar structures in our deep exposure PACS images and high
angular resolution C-band (6 GHz) images \citep{2012AJ....144...44I},
suggesting the existence of a close connection, {\it via} star
formation, between the dust, the relativistic plasma, and the warm
ionized gas \citep[e.g.,][]{2008ApJ...674..157C}.  This match is
particular good above the center of the galaxy:
Figure~\ref{NGC4631_PACS_Halpha_zoom} shows a close-up comparison of
the central region of NGC~4631 between our PACS observations at
70~\micron\ and H$\alpha$. Near the center of the galaxy, two
worm-like structures can be seen extending above the northern edge of
the H$\alpha$ \citep[below label ``A", ][]{1992ApJ...396...97R}. This
double-worm structure can also be seen in the PACS image at
70~\micron.  This structure has been associated in the past with the
breakout of a superbubble produced by a  nuclear star formation
  event, but not a starburst \citep{1992ApJ...396...97R}. In
  addition, this split H$\alpha$ filament appears to be associated
  with strong diffuse X-ray emission and two radio plumes of similar
  size. \cite{1995ApJ...439..176W} have suggested that magnetic
  pressure plays a role in confining the hot gas bubble above the
  plane of the galaxy.

To look further into this issue, we compare the smoothed PACS images
with the 1.5 GHz and soft X-ray data in
Figure~\ref{comp_PACS_x-ray_radio_smooth}. We find a good match
between the FIR, radio and soft X-ray morphological features. In this
comparison, the X-ray contours are from the soft X-ray (0.3-2.0 keV)
{\it Chandra} observations whereas the radio contours show the 1.5 GHz
image as obtained from \cite{2012AJ....144...44I}. While the agreement
is more evident in the bright disk of the galaxy, it is also seen in
the halo at fainter flux levels. Note the loop-like feature $\sim$5 kpc
north and east of the galaxy's major axis in the PACS images. This
extraplanar structure is enhanced in our smoothed maps and it is
co-located with a large radio loop at 1.5~GHz
\citep{2012AJ....144...44I} and a similar soft X-ray loop, interior to
the bigger radio continuum loop. Given the fact that the X-ray loop is
interior to the radio loop, \cite{2012AJ....144...44I} suggested that
the hot X-rat emitting plasma is confined by magnetic pressure. In a
similar way, the FIR loops is slightly smaller than the radio loop,
but a better match with the soft X-ray loop. Magnetic pressure
confinement may therefore also influence the warm dust component in
this region. Finally, note in
Figure~\ref{comp_PACS_x-ray_radio_smooth} the FIR emission extending
towards the south at the eastern end of the disk, in direction of the
companion, NGC~4656. 

Next, in order to test the connection between the cold dust and gas,
we compare our SPIRE observations with those from atomic hydrogen, H~I
21-cm emission \citep[see][for details of the observations and data
  processing]{1994A&A...285..833R}.
Figure~\ref{NGC4631_SPIRE_HI_LR_smooth_units} shows the H~I gas as a
contour plot with 45\arcsec\ $\times$ 87\arcsec\ resolution overlaid
on the SPIRE 250, 350, and 500 $\micron$ images. For this comparison,
we labeled four H~I tidal spurs, following the nomenclature of
\cite{1978A&A....65...37W}. Figure~\ref{NGC4631_SPIRE_HI_LR_smooth_units}
shows that Spur 4 is also detected in the SPIRE bands, especially at
250 $\micron$. This result suggests that Spur 4 can be traced back to
emission from the disk, as first proposed by
\cite{1994A&A...285..833R}. A similar correspondence was found between
the extraplanar cold dust distribution at 1.2~mm and H~I 21-cm
\citep{1999PNAS...96.5360N}. Note that the magnetic field orientation
in this region (north-east of the galaxy center) also seems to be
aligned with Spur 4 and the cold dust component
\citep{1994A&A...284..777G,2013A&A...560A..42M}. On the other hand,
there is no match between the cold gas and dust in Spur 3, suggesting
that there is no gas and dust associated with NGC~4627 (at the
sensitivity of the H~I and SPIRE observations). Note that numerical
simulations of the tidal interactions within the NGC~4631 group of
galaxies require NGC~4627 as a source of gas in the past to create
Spurs 2 and 3 \citep{1978A&A....65...47C}.

% Spur~1 and 2 have a complicated structure joins the disk over a large
% area in agreement with the observed SPIRE emission which extends

\section{Results}
\subsection{PACS Colors\label{SB}}

Typically, the FIR emission from galaxies peaks between $\sim$70 and
$\sim$160 $\micron$.  Given our deep PACS 70 and 160 \micron\ images,
the $S_{70}$/$S_{160}$ ratio can be used as a stand-alone color
diagnostic to provide useful information on the peak of the FIR
spectral energy distribution (SED), a proxy for the dust temperature
and the dust
properties. Figure~\ref{ratio_radio_and_x-ray_smooth_flat_s3}~
compares the $S_{70}/S_{160}$ ratio map with the X-ray and radio
emission (shown as contours) in NGC~4631.  For this comparison, we
convolved the 70 $\micron$ map to the PACS kernel at 160 $\micron$
\citep{2011PASP..123.1218A} and then applied the adaptive median
smoothing code ADAPTSMOOTH with a S/N = 10 per pixel with a direct
local noise estimate. Thus, the resulting $S_{70}$/$S_{160}$ ratio
image has a 2.85\arcsec\ pixel$^{-1}$ resolution, corresponding to the
 resolution of the PACS 160 \micron\ {\it Scanamorphos}
map. Note that at this scale each pixel in the ratio map is not
independent of its neighbors. However, this map can be used to look
for relative changes in temperature and dust properties independent of
any modeling.

The upper panel of Figure~\ref{ratio_radio_and_x-ray_smooth_flat_s3}~
nicely shows the extent of the dusty halo in NGC~4631 and the
existence of a warm central structure extending up to 6 kpc north of the
major axis of the galaxy. This structure (hereafter called
``superbubble'' to distinguish it from smaller bubble-like structures
seen in the disk) appears to be partially enclosed by the dusty, giant
arch observed by \cite{1999PNAS...96.5360N} in the 1.2~mm cold-dust
continuum emission. However, this cold arch is not a single structure
but a projection between the tidal H~I Spur 4 and an H~I plume
extending up from the disk \citep{2003AJ....125.1204T}. This result
emphasizes that caution must be taken with any visual interpretation
of the data as projection effects may confuse the picture. The
southern edge of the superbubble coincides with the location of the
nuclear double-worm structure observed in the H$\alpha$ and PACS~70
\micron\ maps (see Figure~\ref{NGC4631_PACS_Halpha_zoom}), suggesting
a chimney mode that ejects hot dust out of the disk
into the halo along the magnetic field lines
\citep{1991A&A...248...23H,2012AJ....144...44I,2013A&A...560A..42M}. In
addition, the superbubble seems to coincide with a large shell of
emission or loop observed in the H$\alpha$ + [N~II] image of NGC~4631,
extending 3.7~kpc into the halo \citep[we adjust the size of the loop
  by our adopted distance, ][]{1999ApJ...522..669H}.  The lower panels
of Figure~\ref{ratio_radio_and_x-ray_smooth_flat_s3}~ show a good
correspondence between the far-infrared ratios and the soft X-ray (0.3
-- 2.0~keV) and radio (1.5 GHz) emission, reinforcing the idea of a
physical link between the hot and relativistic gas phases and the warm
dust above the disk of the galaxy.

%
% In particular, there is a stream of warm dust that extends directly
% below NGC~4627 at the southwest end of the radio, FIR and soft X-ray
% loop that we mentioned above.

% Similar to the PACS monochromatic images, the $S_{70}/S_{160}$ ratio
% shows the same prominent loop north and east from the galaxy center.
% The overall morphology of the warm dust halo is in agreement with all
% the other ISM component. The high PACS ratios observed in this region
% suggest a hot enough dusty component collocated with the soft X-ray
% emission but perhaps too warm and dusty to produce an intense
% H$\alpha$ halo
% \citep{1992ApJ...396...97R,1995ApJ...450L..45D,1999ApJ...522..669H}.

The upper panel of Figure~\ref{ratio_radio_and_x-ray_smooth_flat_s3}~
also shows the position and size of the two expanding H~I supershells
found by \cite{1993AJ....105.2098R} in the disk of NGC~4631. It can be
seen that the positions of both supershells coincide with luminous
X-ray sources and prominent regions of star formation (discussed in
Appendix~A). Higher dust temperatures (higher $S_{70}$/$S_{160}$
values) are also seen near these positions, especially in the unbroken
southern half of the larger ($\sim$3~kpc in diameter) and more
energetic eastern supershell \citep[see Figure 3
  in][]{1993AJ....105.2098R}. The 1.5~GHz emission shown in the
  bottom panel is predominantly non-thermal emission,  except in two
  specific star-forming regions, one of which coincides with this
  larger H~I supershell \citep{2012AJ....144...44I}.
% The extended FIR component inside the supershell seems to be
% ``leaking" towards the broken part at the top of the shell. 
Interestingly, the dust temperatures are lower and the X-ray emission
is weaker outside of the expanding shock from the H~I
supershells. This X-ray ``gap" between the expanding shell and the
superbubble region is more noticeable around the larger shell. 

Before we end this section, we wish to better quantify the size and
morphology of the superbubble region. For this, we generated a radial
profile of the $S_{70}/S_{160}$ ratio perpendicular to the mid-plane
passing through the galaxy center (slice width of one pixel,
2.85\arcsec). Figure~\ref{radial_profile} shows that the highest
  ratios, $S_{70}/S_{160} \sim 0.9$, are reached within the disk of
  the galaxy about $\sim$10\arcsec\ from the disk mid-plane.  A value
  of $S_{70}/S_{160} \sim 0.5$ is a good lower limit ($-1 \sigma$) for
  the typical distribution of values observed in starburst galaxies
  \citep{2014ApJ...794..152M}.  In the rest of the paper, we define
  the superbubble as the region where $S_{70}/S_{160} \ge 0.5$,
  excluding any FIR emission from within one scale height of the
  galaxy stellar continuum observed at 4.5 \micron\ \citep[see][for
    details]{2013ApJ...774..126M}. The superbubble region defined in
  this way is shown in Figure~\ref{ratio_smooth_bubble}.  This figure
  emphasizes the fact that the superbubble is not symmetrically
  located with respect to the nucleus of NGC~4631, extending nearly
  $\sim$5 kpc north and west from the galaxy center.

\subsection{SPIRE Colors}

Figure~\ref{NGC4631_SPIRE_ratios_250_500_250_350_smooth} shows a
comparison between the $S_{250}/S_{350}$ and $S_{250}/S_{500}$ ratios
map and the neutral hydrogen gas. The superbubble is barely visible in
the $S_{250}/S_{350}$ ratio map (left panels). On the other hand, we
see a good correspondence between the cold dust component and the
large-scale H~I spurs. This corroborates the results from our previous
morphological comparisons of the SPIRE images and H~I 21-cm maps and
the cold dust distribution observed at 1.2~mm
\citep{1999PNAS...96.5360N}.  However, the SPIRE color map makes the
connection between the cold extraplanar dust and the H~I emission from
Spurs 1, 2 and 4 more obvious, emphasizing a disk origin for these
features. Again, Spur 3 appears to be disconnected from the galaxy
disk, in agreement with previous results \citep{1994A&A...285..833R}.

The superbubble is more evident in the $S_{250}/S_{500}$ ratio map
(right panels in
Figure~\ref{NGC4631_SPIRE_ratios_250_500_250_350_smooth}), where
slightly higher $S_{250}/S_{500}$ values are observed in the northern
region above the disk, suggesting a change in the spectral slope
between 250 and 500 $\micron$. We will return to this result in
Section 4.5 using a more quantitative analysis. There is also an
excellent spatial correspondence between the larger H~I supershell
(east of the central region) and flatter spectral indices in
Figure~\ref{NGC4631_SPIRE_ratios_250_500_250_350_smooth}, suggesting
that the expanding H~I shell is affecting the associated cold dust
component by increasing its temperature (see discussion in
Section~\ref{TB}). For these comparisons, we convolved the SPIRE
  images at 250 and 350 $\micron$ to the SPIRE kernel at 500 $\micron$
  with a common pixel size of 18.\arcsec4 pixel$^{-1}$ (FWHM/2 of the
  SPIRE PSF at 500 $\micron$) and then applied an adaptive median
  smoothing with a S/N = 10 per pixel with a direct local noise
  estimate.

\subsection{Dust Masses\label{DM}}

We determined the dust masses in NGC~4631 by fitting a
single-temperature modified black-body (MBB) to the PACS and SPIRE
data. For this, we used
\begin{equation}
F_\nu = \frac{{\rm M_d}\kappa_\nu B_\nu({\rm T_d})}{D^2},
\label	{MBB}
\end{equation}

\noindent
where ${\rm M_d}$ is the dust mass, $B_\nu$ is the Planck function,
${\rm T_d}$ is the dust temperature, $D$ is the distance to the
galaxy, and $\kappa_\nu$ is the dust emissivity, $\kappa_\nu=
\kappa_o(\nu/\nu_o)^\beta$, where $\kappa_o$ is the dust opacity at
350 \micron, 0.192 m$^2$ kg$^{-1}$ \citep[][hereafter,
  D03]{2003ARA&A..41..241D}. This value of the dust opacity is based
on the best fit of the average far-infrared dust emissivity for the
Milky Way model presented in D03, which yields a best-fit spectral
index value of $\beta = 2.0$. Caution must be taken as the normalized
dust model opacity cross-section, $\kappa_o$, is associated with a
dust model with $\beta = 2.0$ and thus, discrepancies may arise
between the results from a single-temperature MBB fit with a fixed
emissivity and with an emissivity as a free parameter \citep[see][for
  a review]{2013A&A...552A..89B}. In Figure~\ref{figure_1}, we present
the best fit to the {\it Herschel} observations of the global
emission from the galaxy with $\beta = 2.0$ and the dust mass and
temperature as free parameters. At wavelengths longer than $\sim$20
$\micron$, the infrared continuum is dominated by thermal radiation
from dust grains. Time-dependent radiation from small dust grains
(grains with radii, $a$ $\leq$ 50 \AA) dominates the continuum at
wavelengths shorter than $\sim$60 $\micron$ while, at longer
wavelengths, the FIR continuum is dominated by larger size dust grains
emitting at a more steady temperature \citep[e.g.,
][]{2003ARA&A..41..241D,2012ARA&A..50..531K}. Given the uncertainty on
an ``average temperature" for the small grain population and to avoid
contamination to the larger dust grain emission, we use the
70~\micron~ flux as an upper limit in our fits.

The best single-temperature MBB fit with $\beta = 2$ (fixed) gives a
global dust mass of log($M_d / M_\odot$) = $7.61^{+0.04}_{-0.05}$ and
$T_d = 22.26\pm 0.55$ K with $\chi_r^2 = 0.59$. If $\beta$ is a free
parameter then we obtain a dust mass of log($M_d / M_\odot$ =
$7.51^{+0.05}_{-0.06}$, $\beta = 1.65 \pm 0.18$, and $T_d = 24.98\pm
44$~K with $\chi_r^2 = 0.32$.  This latter fit thus results in a
slightly hotter and lower dust mass than the single-temperature MBB
fit with fixed $\beta$. In order to estimate the uncertainties in the
fitted parameters we generated 200 SED created by randomly choosing a
photometric data point within the observed errorbars.  A best fit
model is found for each SED and the uncertainties are taken as the
1-$\sigma$ distribution of the values for each parameter. 
% , confirming that changes in the dust opacity and emissivity may
% have an effect on the predicted dust masses.

We also estimated the physical parameters associated with the dust in
the disk and superbubble using the regions shown in
Figure~\ref{ratio_smooth_bubble} (see Section 4.1). For the disk, the best
single-temperature MBB fit with a fixed $\beta$ gives a dust mass of
log($M_d / M_\odot$) = $7.56^{+0.04}_{-0.05}$ and $T_d = 22.17\pm
0.53$ K with $\chi_r^2 = 0.63$. Allowing $\beta$ to vary as a free
parameter we obtained a disk mass of log($M_d / M_\odot$) =
$7.44^{+0.04}_{-0.05}$ and $T_d = 25.17\pm 1.45$ K with $\chi_r^2 =
0.28$. These values imply that the disk mass represents $\sim 85-89\%$
of the total dust mass of the galaxy. A single-temperature MBB fit to
  the PACS and SPIRE values in the superbubble region gives: log($M_d
  / M_\odot$) = $6.16^{+0.05}_{-0.05}$ and $T_d = 24.46 \pm 0.7$~K for
  $\beta = 2.0$ (fixed) and, for $\beta$ as a free parameter, log($M_d
  / M_\odot$) = $6.11^{+0.08}_{-0.10}$, $\beta = 1.85 \pm 0.23$ and
  $T_d = 25.82 \pm 2.21$~K (see Table~\ref{MBB}). Thus, the dust mass
  of the superbubble represents $\sim$4$\%$ of the total dust mass of
  the galaxy. Figure~\ref{figure_1} shows a comparison between the
  best-fit models for the global, disk and the superbubble
  observations. See Tables~\ref{fluxes} and \ref{MBB} for details on
  the fluxes in each region and the resulting fit parameters.

Our values for the dust masses in NGC~4631 can be compared with those
in the literature (adjusted to our adopted distance of 7.4 Mpc).  By
fitting a single-temperature MBB to the observed global 70 -- 450
$\micron$ fluxes with an emissivity law with $\beta= 2$,
\cite{2006ApJ...652..283B} estimated the total dust mass for this
galaxy to be log($M_d / M_\odot$) = 7.52, in good agreement with our
results. Using a similar approach, \cite{2011ApJ...738...89S}
performed a fit to the 70 -- 500 $\micron$ fluxes densities in
NGC~4631 and found a dust mass of log($M_d / M_\odot$) = 7.23 $\pm$
0.05 and $T_d = 27.7 \pm 0.8$~K. In their models they assumed a
smaller spectral slope, $\beta=1.5$, which yielded a higher
temperatures than our two models ($\beta = 2$ and $\beta={\rm 1.65 \pm
  0.18}$ derived from our best fit with $\beta$ as a free
parameter). The discrepancies in the temperature and dust mass may be
explained by the different $\beta$ and adopted dust opacity,
$\kappa_{o} (500 \micron)$ = 0.095 m$^2$ kg$^{-1}$
\citep{2013A&A...552A..89B}. Alternatively, the dust mass content in
galaxies can be estimated by fitting the observed flux densities with
a set of physical models of interstellar dust heated by a stellar
radiation field \citep[see][for details]{2007ApJ...657..810D}. From
this, \cite{2006ApJ...652..283B} found a dust mass of log($M_d /
M_\odot$) = 7.82 by fitting the 1.0 - 1230 \micron\ fluxes of
NGC~4631. This value agrees with the results from the best-fit models
presented by \cite{2007ApJ...663..866D} with a dust mass log($M_d /
M_\odot$) = 7.94, for a range of the maximum starlight intensity scale
factor, and log($M_d / M_\odot$) = 7.78, for a fixed scale factor
\cite[see ][for details]{2006ApJ...652..283B}. Overall, the dust
masses derived from the best-fit physical models are a factor $\sim 2$
higher than our estimates. As we mentioned before, the discrepancy
between the two mass determinations has been studied in detail by
\cite{2013A&A...552A..89B} where it was found to be the result of a
different (not consistent) absorption cross section between the two
approaches. In our case, the absorption cross section normalization,
$\kappa_o$, used in our MBB fit is taken from the grain model in D03
(with an implicit spectral slope of $\beta=2$, see discussion above)
whereas results from \cite{2007ApJ...663..866D} used a different grain
size and composition.

 The total gas mass in NGC~4631, adjusted for heavy elements, is
  log($M_{HI+H_2} / M_\odot$) = 10.05 \citep{2011MNRAS.410.1423I},
thus implying a gas-to-dust ratio of $275^{+34}_{-24}$ if we use the
dust mass derived from our best fit with a fixed $\beta = 2.0$. Using
the best-fit dust mass with $\beta$ as a free parameter, the
gas-to-dust ratio is $345^{+51}_{-38}$. Overall, these values are a
factor of $\sim$2 higher than the ratio for the Milky Way
\citep[$\sim$ 140, see Table~2 in ][]{2007ApJ...663..866D}. Our simple
calculation does not include the contribution from the warmer dust
component, which may bring the gas-to-dust ratio closer to the Milky
Way value, which is regarded as typical in spiral galaxies.

In an attempt to quantify the effect of the warm dust component, we
attempted to fit the FIR fluxes of NGC~4631 as a combination of two
MBB of different temperatures. However, given the number of free
parameters (six in total) and the number of observed points, this
two-temperature MBB fit is somewhat less constrained. For this reason,
we also include in our analysis the fluxes at 450 and 850~$\micron$
from the Submillimeter Common-User Bolometric Array
\citep{2005ApJ...633..857D}, see Table~\ref{fluxes}. We perform the
SED fit in the 70 -- 850 $\micron$ range with a fixed spectral index
of $\beta = 2$ for both components.  The best-fit model is shown in
Figure~\ref{two_comp_three_comp} (see Table~\ref{MBB1} for details).  We obtain a dust
mass of log($M_d / M_\odot$) = $7.43^{+0.12}_{-0.16}$ and $T_d = 24.45
\pm 3.75$~K in the warm component and a dust mass of log($M_d /
M_\odot$) = $8.66^{+0.26}_{-0.60}$ and $T_d = 6.68 \pm 3.81$~K in the
colder component. Our results agree with those presented by
\cite{2006ApJ...652..283B} and their two thermal component
approximation within our (large) uncertainties. The temperature of our
cold dust component is higher, implying a gas-to-dust ratio for this
component of $25^{+73}_{-11}$, which is closer to the Milky Way value
within its uncertainty. Comparing our values for the warm and cold
dust components, we find that our temperatures are lower than the
values derived by \cite{2005MNRAS.357..361S} using a similar
two-temperature greybody approximation with $\beta = 2.0$ (38~K and
17~K for the warm and cold dust components, respectively). However,
for this analysis, they used fluxes from {\it IRAS} (60 and 100
$\micron$) that underpredict the observed global fluxes by a factor of
$\sim 1.5$ when compared with values from {\it Spitzer} and {\it
  Herschel}.

\subsection{Star Formation Rates}

Finally, in order to estimate the SFR for this galaxy via the total
infrared (IR; integrated over the 8 -- 1000 $\micron$ range)
luminosity, we add a third component to fit the observed SED. Here, we
also include in the fit the {\it Spitzer} observations at 5.8 and 8.0
$\micron$ from IRAC and at 24 $\micron$ from MIPS
\citep{2009ApJ...703..517D}. Figure~\ref{two_comp_three_comp} shows
the best-fit model with log($M_d / M_\odot$) = 8.66 and $T_d = 6.66$~K
for the cold dust component, log($M_d / M_\odot$) = 7.43 and $T_d =
24.42$~K for the warm dust component, and log($M_d / M_\odot$) = 1.35
and $T_d = 221.8$~K for the hot dust component (see Table~\ref{MBB1}
for details). From a direct integration of the SED, the total 8 --
1000 \micron\ luminosity is log($L_{IR} / L_\odot$) = 10.30, which
implies a SFR of 3.47 $M_\odot$ yr$^{-1}$ when using the SFR -- IR
calibration of \cite{1998ARA&A..36..189K}. Our value is slightly
higher than those estimated from the total infrared luminosity based
on the observed {\it IRAS} fluxes at 12, 25, 60, and 100 $\micron$,
$\sim 2.9$ $M_\odot yr^{-1}$ \citep[corrected by our adopted distance
  to the source, ][]{2004ApJS..151..193S,2006A&A...457..779T}.  The
SFR derived from a combination of the $H\alpha$ and 24 $\micron$
luminosities yields a value of 1.6 $M_\odot$ yr$^{-1}$
\citep[corrected by our adopted distance to the source,
][]{2009ApJ...703.1672K,2010ApJ...714.1256C}. Note that dust
extinction, especially in edge-on galaxies, can hamper the use of some
of the star formation indicators that rely on short-wavelength
emission such as H$\alpha$. Thus, we adopt our IR-based SFR value for
the remainder of the paper.  For comparison, in M82, the SFR derived
  from the total IR based on the {\it IRAS} fluxes is 9.2 $M_\odot$
  yr$^{-1}$ \citep[][]{2004ApJS..151..193S}, largely in the nuclear
  starburst whereas in NGC~4631 the star formation is widely
  distributed \citep[e.g.,][]{2011MNRAS.410.1423I}.

\subsection{Pixel-by-Pixel SED fit}

We also performed a pixel-by-pixel SED fit, where we only considered
pixels with detection greater than 3-$\sigma$ at all wavelengths.  For
this exercise, we converted the higher resolution PACS images to that
of the SPIRE 500~\micron~ PSF with a pixel size of
18\arcsec\ pixel$^{-1}$ (FWHM/2 of the SPIRE PSF at 500 \micron). Note
that at 18\arcsec, each pixel is not independent from its
neighbors. However, a pixel size comparable to that of the 500
\micron\ PSF (36\arcsec) would have resulted in too few pixels to have
a meaningful comparison. Therefore, results from our pixel-by-pixel
approximation are intended for a quantitative analysis of the dust
properties and distribution across the galaxy.
Figure~\ref{mosaic_ngc4631_NO70_SED_fix_beta} shows the distribution
of values for $M_d$ and $T_d$ by fitting all the observations with a
single-temperature MBB and $\beta = 2$. Similarly, we performed a
pixel-by-pixel SED fit with $\beta$ as a free parameter, see
Figure~\ref{mosaic_ngc4631_NO70_SED}. As before, we performed the
pixel-by-pixel SED fit assuming the observed flux at 70 $\micron$ is
an upper limit to the actual flux at that wavelength.
Figures~\ref{mosaic_ngc4631_NO70_SED_fix_beta}-\ref{mosaic_ngc4631_NO70_SED}
show that the dust mass is symmetrically distributed along the
galaxy's major axis. However, the temperature maps depend
significantly on our assumption on fixing $\beta$ or not.

The temperature map with $\beta = 2$ follows roughly the morphology of
the superbubble observed in the PACS $S_{70}$/$S_{160}$ ratio map,
with higher temperatures north of the nucleus, but it fails to show
details in the active star formation regions across the galaxy
disk. In contrast, the temperature map with floating spectral index
shows more detail in the disk, with higher temperatures derived near
the active star-forming regions in the disk observed in the PACS 70
$\micron$ and H$\alpha$ images.  In this case, however, it is the map
of the spectral index, $\beta$, rather than the temperature map, that
shows a closer spatial correspondence with the superbubble region
(indicating $\beta > 2$). Clearly, there is a degeneracy between $T_d$
and $\beta$ in these fits, so we cannot distinguish between the
following two plausible scenarios:  (1) the dusty material in the
  superbubble region is heated by star formation in the disk, the hot
  X-ray emitting material, or shocks, or (2) the dust in the
  superbubble region has a different (steeper) emissivity slope than
  that in the disk, perhaps indicating different intrinsic dust
  properties (e.g., grain size distribution and composition).

To further look into this issue, we produce pixel-by-pixel color-color
diagrams to investigate the dust properties across the galaxy, see
Figure~\ref{ratio1_ratio2}. For this comparison, we adopt the
superbubble region defined in Section~\ref{SB}. We also show the
expected color-color relationship for a single-temperature MBB with
two different emissivities, $\beta=2.0$ and 1.5, and a range of
temperatures (15K -- 40K). The dusty superbubble follows a color
distribution that is different from that of the disk in the
$S_{250}$/$S_{500}$ versus $S_{100}$/$S_{250}$ diagram (see Left panel
in Figure~\ref{ratio1_ratio2}), with higher $S_{250}$/$S_{500}$
values, thus a steeper spectral slope, in the superbubble, suggesting
a higher dust opacity. The right panel in Figure~\ref{ratio1_ratio2}
shows the SPIRE colors, $S_{250}$/$S_{350}$ versus
$S_{350}$/$S_{500}$, where it can be seen that values from the
superbubble have a steeper slope than those from the disk emission,
therefore favoring scenario \#2 over scenario \#1.

% Figure~\ref{comp_DALE_all_fluxes} shows a pixel-by-pixel comparison
% between the {\it Spitzer} MIPS 24~\micron~ flux and the PACS and
% SPIRE fluxes for the disk and the dusty, hot `bubble" region. Again
% for this comparison we converted the higher resolution MIPS and PACS
% images to that of the SPIRE 500~\micron~ PSF with a pixel size of
% 18” pixel$^{-1}$. Figure~\ref{comp_DALE_all_fluxes} shows that the best
% agreement between the disk and the hot, dusty ``bubble" is at
% shorter wavelengths suggesting that the origin of the hot bubble can
% be associated with a star formation
% event. Figure~\ref{comp_DALE_all_fluxes} also shows the very tight
% correlation between MIPS 24 and PACS~70~\micron, suggesting that the
% 70~\micron~ emission is dominated by hot dust, thus, validating our
% use of this emission as an upper limit in order to estimate the
% physical parameters of the colder dust component via a single MBB
% for NGC~4631.

\subsection{The Temperature and Spectral Index Relation\label{TB}}

In our single-temperature MBB approximation, the fitted spectral
index, $\beta$, carries important information about the dust opacity
and composition. However, this simple approximation, used to describe
the optically thin thermal emission from large grains radiating in
thermal equilibrium, comes at a price: a degeneracy between $\beta$
and temperature. Part of this degeneracy comes from the SED fitting
\citep[fitting technique, limited number of observations and flux
  uncertainties, e.g.,][]{2009ApJ...696..676S,2012ApJ...752...55K},
the single-temperature simplification employed to describe the dust
temperature distribution along the line of sight \citep[e.g.,
][]{2009ApJ...696.2234S}, and real variations of the physical
properties of the dust grain population
\citep[e.g.,][]{2007A&A...468..171M,2011A&A...535A.124C,2012A&A...542A..21Y}.
An anti-correlation between the dust temperature and the spectral
index, $\beta$, has been observed in a number of sources based on a
variety of experiments and observatories
\citep[e.g.,][]{2003A&A...404L..11D,2008A&A...481..411D,2010A&A...520L...8P,2010ApJ...713..959V,2011MNRAS.412.1151B,2012ApJ...756...40S,2014MNRAS.438.1922D,2014A&A...565A...4H}. For
these reasons, one has to be cautious when interpreting the T-$\beta$
relation.

Our pixel-by-pixel analysis allows us to investigate the spatial
distribution of the T-$\beta$ relation throughout the
galaxy. Figure~\ref{beta-temp} shows the fitted temperatures as a
function of the dust spectral index across the disk and the
superbubble in NGC~4631. Our values show an inverse correlation
between the dust temperature and spectral index with two different and
well-separated relationships for the disk and the superbubble. This
result is in excellent agreement with the resolved analysis of the
dust and gas content in Andromeda (M31), where
\cite{2012ApJ...756...40S} found that the T-$\beta$ relation measured
within 3.1~kpc from the center of the galaxy is distinct from that
derived further out (green solid and dot-dashed lines,
Figure~\ref{beta-temp}). The T-$\beta$ relation from
\cite{2012ApJ...756...40S} for the outer region of M31 ($3.1 < R < 15
kpc$) is in good agreement with our values for the outer parts of the
disk in NGC~4631 and with values derived from the global properties of
the {\it Herschel} Virgo Cluster Survey
\citep[][]{2014MNRAS.438.1922D}, suggesting that the range of dust
properties and conditions in NGC~4631 and M31 are similar to those
found in the late-type galaxies that comprises the Virgo sample. On
the other hand, the $T-\beta$ relation for the inner region in M31 ($R
<$ 3.1 kpc) agree with our values for the superbubble in NGC~4631.

Interestingly, the radial break in the T-$\beta$ relation observed in
NGC~4631 and M31 is not present in the edge-on spiral NGC~891
\citep{2014A&A...565A...4H}. NGC~891 is almost perfectly edge-on ($i >
89\arcdeg$), thus making radial variations harder to detect.
Nevertheless, there are also no statistically significant variations
of the spectral index along the vertical direction
\citep{2014A&A...565A...4H}. The best-fit relation for NGC~891 lies
slightly below the distribution of points from the superbubble in
NGC~4631 (dashed blue line in Figure~\ref{beta-temp}). Submillimeter
point sources from the ARCHEOPS catalog \citep{2008A&A...481..411D},
mostly of Galactic origin, lie close to the T-$\beta$ relation of the
outer disk region of NGC~4631, whereas the best-fit T-$\beta$ relation
for the PRONAOS-based data on different regions of the Galactic ISM
\citep[][]{2003A&A...404L..11D} provides a good dividing line between
the data points in the outer disk of NGC~4631 and those in the
superbubble (see Figure~\ref{beta-temp}).

The excellent agreement between the T-$\beta$ relations of NGC~4631
and M31 and their radial dependences suggests a similar physical
origin. In both cases, the different relations between the inner and
outher disk regions are likely an indication that the dust in these
regions has physically distinct properties. This result is consistent
with the different color distributions observed between the
superbubble and the disk of NGC~4631 (Figure~\ref{ratio1_ratio2}) and
the observed spatial variations of the spectral index between 250 and
500$\micron$
(Figure~\ref{NGC4631_SPIRE_ratios_250_500_250_350_smooth}). We return
to this issue in the next section.

\section{Discussion}

The origin of the FIR emission in galaxies is unambiguously associated
with thermal emission from heated dust grains. However, the
identification of the heat source for the dust component requires
further considerations. In galaxies without AGN, like NGC~4631, the
amount of thermal emission from dust grains is often related to the
ultraviolet continuum emission provided by young, massive stars.  The
good agreement between the morphology of the ionized gas (H$\alpha$)
and the warm-to-cold dust component in the disk of NGC~4631 is
consistent with this picture. The origin and heat source of the
extraplanar dust in NGC~4631 is less clear, however. The filamentary,
chimney-like, and superbubble structures seen above the disk in the
FIR, warm and hot ionized gas, and relativistic plasma suggest that
some of the dust in the disk has been lifted out into the halo
\citep[e.g.,][]{2008ApJ...674..157C}. However, one must ask if
stellar activity in the disk is energetically capable of lifting this
material up to its observed position. In order to estimate the
potential energy involved in lifting this material up to height $z$
above the mid-plane of the galaxy, we assume an isothermal sheet model
for the vertical distribution of light and mass in galactic disks
\citep{1988A&A...192..117V}, as presented by
\cite{1997AJ....114.2463H} for the dusty clouds in NGC~891 and later,
by \cite{2000ApJ...535..663R} for the molecular clouds in NGC~4631:

\begin{equation}
\Omega = 10^{52}~{\rm ergs}\left ( \frac{M}{10^5 M_\odot} \right )
\left( \frac{z_o}{700~{\rm pc}} \right ) \left (\frac{\rho_o}{0.185~{\rm
    M_\odot~pc^{-3}}}\right )\ln \left [\cosh \left (
  \frac{z}{z_o}\right ) \right ],
\label{PE}
\end{equation} 

\noindent where $\rho_o$ is the mass density at the mid-plane, $M$ is
the mass of lifted material, and $z_o$ is the mass scale height of the
stellar disk. Here we use a mass scale height for the stellar disk of
1.5~kpc, as measured from the stellar disk observed at 4.5~\micron~
\citep{2013ApJ...774..126M}. This value is in excellent agreement with
the fit of the vertical distribution of the resolved stellar
population for NGC~4631 observed with the {\it Hubble Space Telescope
  Advanced Camera for Surveys} \citep[up to $z_o$ = 1.4~kpc,
][]{2005AJ....130.1574S}. 
% Note that, for NGC~4631, the fitted scale
% height could be subject to a greater uncertainty due to its irregular
% shape and close companion. 
Assuming an average mid-height for the superbubble region above the
plane of about 3~kpc, and a conservative starting point at $z$ =
$z_0$, then the gain in potential energy is:

\begin{equation}
\Delta\Omega = 4\times 10^{53}~{\rm ergs}\left (\frac{\rho_o}{0.185~{\rm
    M_\odot~pc^{-3}}}\right ).
\label{PE1}
\end{equation} 

Note that the main uncertainty on the prediction of the potential
energy comes from the mass density at mid-plane, where we adopted a
value of 0.185 ${\rm M_\odot~pc^{-3}}$, representative of the total
mass density at the solar position \citep[see ][for
  details]{1984ApJ...276..169B}. The potential energy required to lift
the dust in the superbubble is thus comparable to that found by
\cite{2011MNRAS.410.1423I} for the CO (J = 3-2) molecular clouds
observed in the central region of NGC~4631, and an order of magnitude
smaller than predictions from the CO (J = 1-0) molecular clouds
presented by \cite{2000ApJ...535..663R}. Note that we have
conservatively assumed that the material come from $z = z_0$ kpc. If
the material starts at a distance ten times closer to the mid-plane,
$z_0=150$~kpc, then the energy needed to lift the material would be
$\sim$1.5 times higher (note that eqn~\ref{PE} is undefined at the
mid-plane).

Assuming that dust and gas are mixed inside the superbubble with
typical Galactic ratios \citep[140, ][]{2007ApJ...663..866D}, then the
required energy to lift the material (gas and dust) into the halo is
two orders of magnitude higher, $\sim 6 \times 10^{55}$~ergs. Taken at
face value, this energy is equivalent to the kinetic energy of
$\sim$60,000 type II supernovae.  Given the current SFR of NGC~4631
($\sim$3.5 $M_\odot$ yr$^{-1}$, Section 4.4), and assuming SFR $\sim$
50-100 SNR, where SNR is the supernova rate, then this is equivalent
to the kinetic energy injected by all supernovae produced in NGC~4631
in the past $\sim$10$^6$ yrs. This last number should be increased to
take into account the fact that the superbubble structure is seen only in
the inner $\sim$6 kpc, whereas star formation activity in NGC~4631 is
distributed throughout the disk.  The current level of star formation
activity thus seems insufficient to lift the dust and gas out of the
disk unless 100\% of the kinetic energy from supernovae produced in the
past $\sim$10$^6$ is tapped to lift this material above the disk. This
scenario seems unlikely. The tension is reduced if the dust-to-gas
ratio inside the superbubble is higher than the Galactic value.  If
the dust-to-gas ratio is Galactic, then one has to invoke that (1) the
star formation rate was (much) higher in the past
\citep[e.g.,][]{2011MNRAS.410.1423I} or (2)  some of the dust
  inside the superbubble comes from material tidally stripped from nearby
  companions or disk material lifted during past interactions, a
  scenario that has also been evoked to explain the large-scale dusty halo
  of M82 \citep{2010A&A...518L..66R}.

% for the larger (shell 1, expanding at
% 45km~s$^{-1}$) and smaller (shell 2, expanding at 35km~s$^{-1}$)
% supershells, respectively. 

Our data do not allow us to distinguish between these two scenarios.
The presence of two highly energetic supershells in the disk with
kinetic energies of 2 -- 5 $\times$ 10$^{54}$ ergs (Shell 1 on the
east side, expanding at 45 km~s$^{-1}$) and 0.6 -- 1 $\times$
10$^{54}$ ergs (Shell 2 on the west side, expanding at 35 km~s$^{-1}$)
is difficult to reconcile with the current level of star formation
\citep{1993AJ....105.2098R,1995ApJ...439..176W} and thus favors
scenario (1) (unless these shells are produced instead in an oblique
impact of a high velocity cloud with the galactic disk
\citep{1996AJ....111..190R}). These supershells are no doubt releasing
large amounts of energy into the surrounding ISM via shocks from the
shells expanding into the surrounding ISM, changing the shape of the
dusty halo, and perhaps even causing the asymmetric geometry of the
superbubble (Section 4.1).

On the other hand, there is also considerable evidence that past
galaxy interactions are at least partly responsible for the existence
and peculiar morphology of the giant gaseous halo of NGC~4631
\citep[e.g.,][]{1988A&A...197L..29H,1990A&A...236...33H,2000ApJ...535..663R,2001ApJ...555L..99W}. Numerical
simulations for the H~I tidal streams in the three-galaxy system
NGC~4631, NGC~4656, and NGC~4627 have suggested that NGC~4627 has been
a source of cold gas in the past \cite{1978A&A....65...47C}. This
dwarf elliptical galaxy is the closest companion to NGC~4631 and is
undetected in our {\it Herschel} images and thus dust depleted.  This
picture is also supported by the morphology of the superbubble which
extends towards NGC~4627. In particular, there is a stream of warm
dust that extends directly below NGC~4627 at the southwest end of the
FIR, radio, and soft X-ray loop, as discussed in Section~\ref{adata}.

{Regardless of the exact origin of the dust in the halo of NGC~4631,
the FIR ratio maps presented in Section 4 indicate that dust in the
superbubble region has a different (steeper) emissivity spectral slope
than dust in the disk. Dust in the superbubble may be heated by UV
radiation from the on-going star formation activity in the disk, X-ray
bremsstrahlung radiation from the extraplanar hot plasma at the
position of the superbubble, or by shocks from a large-scale
star-formation driven wind \citep[e.g., ][]{2004ApJS..151..193S}. All
of these processes may alter the intrinsic dust properties inside the
superbubble by destroying the more fragile grains, e.g., via thermal
sputtering and grain shattering (grain-grain collision). These
mechanisms often favor the destruction of the smaller grains
\citep[e.g., sputtering, ][]{1998ApJ...503..247J,2013ApJ...778..161R},
and thus may change the overall dust grain size distribution in favor
of larger grains. This may help explain the different SPIRE colors and
T-$\beta$ relations observed in the disk and superbubble region.

\section{Conclusions}

We present the results from a detailed analysis of deep {\it Herschel}
PACS and SPIRE images in the nearby edge-on star-forming galaxy
NGC~4631.  Given the superior angular resolution and sensitivity of
{\it Herschel}, we were able to carry out a spatially resolved
analysis of the dust distribution in this object. We complement our
analysis of these data with published observations at other
wavelengths to build a more complete picture of the multi-phase ISM
in this galaxy. The main results can be summarized as follows:

\begin{itemize}

\item The PACS images at 70 $\micron$ (PSF FWHM $\sim$ 5.6\arcsec)
  and 160 $\micron$ (PSF FWHM $\sim$ 11.4\arcsec) show a rich complex
  of filaments and chimney-like features that extend up to $\sim$6
  kpc above the galaxy disk.  Many of the structures seen in the FIR
  coincide spatially with those seen in H$\alpha$, radio-continuum,
  and soft X-ray emission.  This spatial match suggests a tight
  disk-halo connection regulated by star formation.

\item The extraplanar emission detected on larger scale in the SPIRE
  images shows a good match with many of the extraplanar H~I 21-cm
  spurs of NGC~4631. Most of these features likely are left-over tidal
  debris from recent interactions between NGC~4631 and its galaxy
  companions.

\item A bubble-like structure that extends up to $\sim$6.0 kpc above
  the plane of the galaxy is detected in the PACS 70/160 $\micron$
  ratio map.  A pixel-by-pixel fit of the spectral energy distribution
  indicates that the dust in this region has a higher temperature
  and/or an emissivity with a steeper spectral index than the dust in
  the disk. The superbubble region also shows higher
  $S_{250}$/$S_{500}$ and $S_{350}$/$S_{500}$ flux ratios than those
  observed in the disk, consistent with a steeper spectral index in
  this region. Dust inside the superbubble is heated by UV
    radiation from young stars in the disk, X-ray emission from the
    hot plasma in the halo, or shocks associated with a large-scale
    outflow into the halo. All of these processes may alter the dust
    grain size distribution inside the superbubble, and explain the
    higher spectral index of the dust emissivity in this region.

\item 
  The dust mass of the superbubble represents about $\sim$4\% of the
  total dust mass in this galaxy.  The energy required to lift this
  dusty material in the superbubble region is at least $\sim 6\times
  10^{55}$ ergs, assuming a Galactic dust-to-gas ratio. The current
  rate of energy injection into the ISM by star formation activity in
  the disk appears to be energetically insufficient to lift this
  material into the halo, unless the dust-to-gas ratio is
  significanlty higher than the Galactic value.  One possible
  explanation is that the star formation rate was significantly higher
  in the past. The presence of two highly energetic H~I supershells in
  the disk of this galaxy is consistent with this scenario and may
  still be shaping the gaseous halo. Another possibility  is that
    some of the dust observed in the halo was deposited there from
    past galaxy interactions, either lifted directly from the disk of
    NGC~4631 or stripped from now dust-depleted companions (e.g.,
    dwarf elliptical NGC~4627). In this regard, NGC~4631 could be a
    twin of M82 where tidal interactions seem to be more efficient at
    depositing dust into the halo than the star-formation driven wind
    \citep{2010A&A...518L..66R}.

\end{itemize}

\clearpage

\acknowledgements

We are very grateful to H\'el\`ene Roussel for her help with the use
of {\it Scanamorphos}. We also thank Richard Rand and Judith Irwin for
generously making their published data available to us. Support for
this work was provided by NASA through Herschel contracts 1427277 and
1454738 (S.V. and M.M.). This work has made use of NASA’s Astrophysics
Data System Abstract Service and the NASA/IPAC Extragalactic Database
(NED), which is operated by the Jet Propulsion Laboratory, California
Institute of Technology, under contract with the National Aeronautics
and Space Administration. PACS has been developed by a consortium of
institutes led by MPE (Germany) and including UVIE (Austria); KU
Leuven, CSL, IMEC (Belgium); CEA, LAM (France); MPIA (Germany);
INAF-IFSI/OAA/OAP/OAT, LENS, SISSA (Italy); IAC (Spain). This
development has been supported by the funding agencies BMVIT
(Austria), ESA-PRODEX (Belgium), CEA/CNES (France), DLR (Germany),
ASI/INAF (Italy), and CICYT/MCYT (Spain). SPIRE has been developed by
a consortium of institutes led by Cardiff University (UK) and
including Univ. Lethbridge (Canada); NAOC (China); CEA, LAM (France);
IFSI, Univ. Padua (Italy); IAC (Spain); Stockholm Observatory
(Sweden); Imperial College London, RAL, UCL-MSSL, UKATC, Univ. Sussex
(UK); and Caltech, JPL, NHSC, Univ. Colorado (USA). This development
has been supported by national funding agencies: CSA (Canada); NAOC
(China); CEA, CNES, CNRS (France); ASI (Italy); MCINN (Spain); SNSB
(Sweden); STFC, UKSA (UK); and NASA (USA). HIPE is a joint development
(are joint developments) by the Herschel Science Ground Segment
Consortium, consisting of ESA, the NASA Herschel Science Center, and
the HIFI, PACS and SPIRE consortia.

\clearpage

\appendix

\section{Nature of the Brightest X-ray sources in NGC~4631\label{apend}}

Many bright X-ray point sources have been detected in the disk of
NGC~4631
\citep[e.g.,][]{1996A&A...311...35V,1995ApJ...439..176W,1997MNRAS.286..626R,2006ApJ...649..730W,2007ApJ...655..163W,2009ApJ...696..287S}.
These X-ray sources represent possible candidates for blowing out gas
and dust from the disk into the halo, and heating the
dust. Figure~\ref{ratio_radio_and_x-ray_smooth_flat_s10_apendix}~ shows
some of the X-ray sources overlaid on top of the PACS
$S_{70}$/$S_{160}$ ratio map.  South of the
dusty loop at the edge of the superbubble (at $\alpha$
12\fh42\fm04\fs2, $\delta$ +32\arcdeg28\arcmin50\arcsec, J2000.0 and
$\sim$ 4~kpc from the nucleus), there is a X-ray source, source
\#5. This source was identified by \cite{1996A&A...311...35V} (H10)
in the X-ray (0.1 - 2.4 keV) as a very bright source tentatively
located in the spiral arm of NGC~4631.  The location outside the halo
is consistent with the difference in X-ray spectral index of the
source relative to that of the surrounding diffuse emission
\citep{1997MNRAS.286..626R}. This source is thus unlikely to
  provide any contribution to the FIR bright halo.

Interestingly, two of the X-ray point sources, namely source \#1
\citep{1997MNRAS.286..626R} and source \#7 \citep[H7
  in][]{1996A&A...311...35V}, approximately coincide with the centers
of the two large supershells observed in H~I
\citep{1993AJ....105.2098R}. It
is clear from the PACS $S_{70}$/$S_{160}$ ratios that the regions
enclosed in the supershells show higher dust temperatures (ratios),
suggesting that the supershells may be filled with warm dust despite
being depleted of cold gas (to the extent of the sensitivity of H~I
observations).  Source \#7 is at the center of the smaller H~I
supershell (1.8~kpc in diameter, Shell 2) and is a fairly typical
power-law ultra luminous X-ray source (ULX), perhaps powered by
accretion onto a black hole (BH) with a mass $\leq$50$M_\sun$
\citep{2009ApJ...696..287S}. The source has been observed several
times and it shows little evidence for X-ray variability within the
{\it Chandra} and {\it XMM-Newton} exposures with an intrinsic X-ray
luminosity in the 0.3-10.0~keV band of $(3.8 \pm 0.1) \times 10^{39}$
ergs s$^{-1}$ and $(5.0 \pm 0.2) \times 10^{39}$ ergs
  s$^{-1}$ as estimated from {\it Chandra} and {\it XMM-Newton}
observations, respectively \citep{2009ApJ...696..287S}. In another
study, \cite{2006ApJ...649..730W} estimated a BH mass of 5.5 $M_\sun$
for this source and an accretion rate $L/L_{Edd}$ of 0.84, consistent
with a normal stellar black hole X-ray binary. As we mentioned in
Section 5, the required input energy to produce the kinematics of the
H~I shell 2 is about 0.6 -- 1.0 $\times$ 10$^{54}$ erg s$^{-1}$ and
cannot be uniquely explained by the cumulative action of stellar winds
and supernova explosions.
%        explosions from OB associations because the extremely high
%        number of OB stars required in the lifetime of the shell,
%        moreover, assuming that the shell stopped expanding the
%        calculated lower limit is 700-1700 OB stars
%       \citep{1993AJ....105.2098R}.

On the other hand, the larger supershell (Shell 1) surrounds a
prominent star formation complex approximately at the same position as
X-ray source \#1 \citep{1997MNRAS.286..626R}. This region is dominated
by the radio thermal emission with a 1.5 GHz spectral index of
$\alpha=-0.16\pm0.009$ \citep{2012AJ....144...44I}. It also shows an
extended FIR morphology characterized by high dust temperatures
indicative of UV heating from stellar activity. Of all the X-ray
sources in the disk, this source is the weakest, with an X-ray
luminosity in the 0.1-2.0~keV band of 3 $\times$ 10$^{38}$ ergs
s$^{-1}$ \citep{1997MNRAS.286..626R}. Because of this, the nature of
this X-ray source is still unknown.

Another interesting X-ray source is \#2, first detected as a luminous
supersoft source by {\it ROSAT}/PSPC observations
\citep{1996A&A...311...35V}. \cite{2009ApJ...696..287S} have suggested
that this highly variable supersoft ULX could be the result of
transient super-Eddington outbursts powered by non-steady nuclear
burning on the surface of a massive white dwarf, perhaps as an extreme
subclass of supersoft ULXs. However, the four-hour modulation observed
in this source remains unexplained
\cite{2007A&A...471L..55C,2009ApJ...696..287S}. Adding to the
interesting properties of this source is the fact that it is invisible
in our PACS ratio map. This source is in the lower PACS ratio region
between the larger H~I surpershell and the superbubble region.

The other two sources, \#3 and \#4, are close to the center of the
galaxy and appear as bright hot spots in our PACS ratio map, below the
superbubble. The X-ray analysis of these sources suggest stellar black
holes in different accretion states
\citep[e.g.,][]{2009ApJ...696..287S}.

\clearpage 

\begin{deluxetable}{lcc}
\tablecolumns{3}
\tabletypesize{\scriptsize}
\tablewidth{0pt}
\tablecaption{Flux Densities measured in NGC~4631}
\tablehead{
\colhead{Instrument}           & \colhead{$\lambda$}   & \colhead{Flux}       \\
                               &  \colhead{$\micron$}      & \colhead{Jy}          \\
\tableline
\colhead{(1)} & \colhead{(2)} & \colhead{(3)} \\
}
\startdata
\cutinhead{Global}
PACS & 70 & 141.9$\pm$ 14.2  \\
PACS & 100 & 232.9 $\pm$ 23.3   \\
PACS & 160 & 247.0 $\pm$ 24.7   \\
SPIRE & 250 & 111.4 $\pm$ 11.1 \\
SPIRE & 350 & 50.0 $\pm$ 5.0   \\
SPIRE & 500  & 19.9 $\pm$ 2.0   \\
SCUBA\tablenotemark{a} & 450  & 30.7 $\pm$ 10.0  \\
SCUBA\tablenotemark{a} & 850 & 5.7 $\pm$ 1.2   \\
\cutinhead{Disk}
PACS & 70 & 121.4 $\pm$  12.1 \\ 
PACS & 100& 202.8 $\pm$ 20.3 \\
PACS & 160 & 211.9  $\pm$ 21.2  \\
SPIRE & 250 & 98.3  $\pm$ 9.8 \\
SPIRE & 350 & 43.9  $\pm$ 4.4 \\
SPIRE & 500 & 17.7  $\pm$ 1.8 \\

\cutinhead{Superbubble}
PACS & 70 &  8.2 $\pm$ 0.8 \\
PACS & 100 &  15.4 $\pm$ 1.54  \\
PACS & 160 &   11.7 $\pm$ 1.2 \\
SPIRE & 250 &  5.8 $\pm$  0.6 \\
SPIRE & 350 &   2.2 $\pm$ 0.2\\
SPIRE & 500 &   0.8 $\pm$ 0.1 \\ 
\enddata
\tablecomments{Column 1: Instrument. Column 2: Central wavelength of the observed band. Column 3: Flux density}
\tablenotetext{a}{ Flux densities measured with the Submillimeter Common-User Bolometric Array \citep[SCUBA][]{2005ApJ...633..857D}}
\label{fluxes}
\end{deluxetable}

\begin{deluxetable}{lcccc}
\tablecolumns{5}
\tabletypesize{\scriptsize}
\tablewidth{0pt}
\tablecaption{Parameters derived from Single-temperature Modified Blackbody Fits}
\tablehead{
\colhead{Region}           & \colhead{$\log M_d$}      & \colhead{$T_d$}           & \colhead{$\beta$}      & \colhead{ $\chi_r^2$}    \\
                               & \colhead{$M_\odot$}       & \colhead{(K)}             &                        &    \\
\tableline
\colhead{(1)} & \colhead{(2)} & \colhead{(3)}& \colhead{(4)} & \colhead{(5)}\\
}
\startdata
\cutinhead{$\beta = 2.0$}
Total & ${\rm 7.61^{+0.04}_{-0.05}}$ & ${\rm 22.26\pm 0.55}$ & \nodata & 0.59\\
Disk & ${\rm 7.56^{+0.04}_{-0.05}}$ & ${\rm 22.17\pm 0.53}$ & \nodata & 0.63\\
Superbubble & ${\rm 6.16^{+0.05}_{-0.05}}$ & ${\rm 24.46\pm 0.77}$ & \nodata & 0.51\\
\cutinhead{$\beta$ Free}
Total & ${\rm 7.51^{+0.05}_{-0.06}}$ & ${\rm 24.98 \pm 1.44}$ & ${\rm 1.65 \pm 0.18}$ & 0.32\\
Disk & ${\rm 7.44^{+0.05}_{-0.06}}$ & ${\rm 25.17 \pm 1.45}$ & ${\rm 1.62 \pm 0.16}$ & 0.28\\
Superbubble & ${\rm 6.11^{+0.08}_{-0.10}}$ & ${\rm 25.82 \pm 2.21}$ & ${\rm 1.85 \pm 0.23}$ & 0.58\\
\enddata
\tablecomments{Column 1: Region where the fluxes are extracted. Column 2: Dust mass. Column 3: Dust temperature. Column 4: Spectral slope of the dust emissivity. Column 5: reduced chi-square of the fit.}
%\tablenotetext{a}{ The mean values and standard deviations are given by the Kaplan-Meier estimator. The two-sample test probability is given by  Gehan's Generalized Wilcoxon test. \label{table2}}
\label{MBB}
\end{deluxetable}

\begin{deluxetable}{lccc}
\tablecolumns{4}
\tabletypesize{\scriptsize}
\tablewidth{0pt}
\tablecaption{Global Parameters derived from Multiple Component Modified Blackbody Fits with $\beta = 2.0$}
\tablehead{
\colhead{Component}           & \colhead{$\log M_d$}      & \colhead{$T_d$}           & \colhead{ $\chi_r^2$}    \\
                               & \colhead{$M_\odot$}       & \colhead{(K)}                           &    \\
\tableline
\colhead{(1)} & \colhead{(2)} & \colhead{(3)}& \colhead{(4)} \\
}
\startdata
\cutinhead{Two-temperature MBB}
Component 1& ${\rm 7.43^{+0.12}_{-0.16}}$ & ${\rm 24.45 \pm 3.75}$ &1.04\\
Component 2& ${\rm 8.66^{+0.26}_{-0.60}}$ & ${\rm 6.68 \pm 3.81}$ &\nodata\\
\cutinhead{Three-temperature MBB}
Component 1& ${\rm 1.35^{+0.13}_{-0.17}}$ & ${\rm 221.81 \pm 5.15}$ &3.17\\
Component 2& ${\rm 7.43^{+0.13}_{-0.14}}$ & ${\rm 24.42 \pm 3.75}$  &\nodata\\
Component 3& ${\rm 8.66^{+0.24}_{-0.61}}$ & ${\rm 6.66 \pm 3.81}$ &\nodata\\
\enddata
\tablecomments{Column 1: Dust component. Column 2: Dust mass. Column 3: Dust temperature.  Column 4: reduced chi-square of the fit.}
%\tablenotetext{a}{ The mean values and standard deviations are given by the Kaplan-Meier estimator. The two-sample test probability is given by  Gehan's Generalized Wilcoxon test. \label{table2}}
\label{MBB1}
\end{deluxetable}

\begin{figure*}
\epsscale{1.0}
\plotone{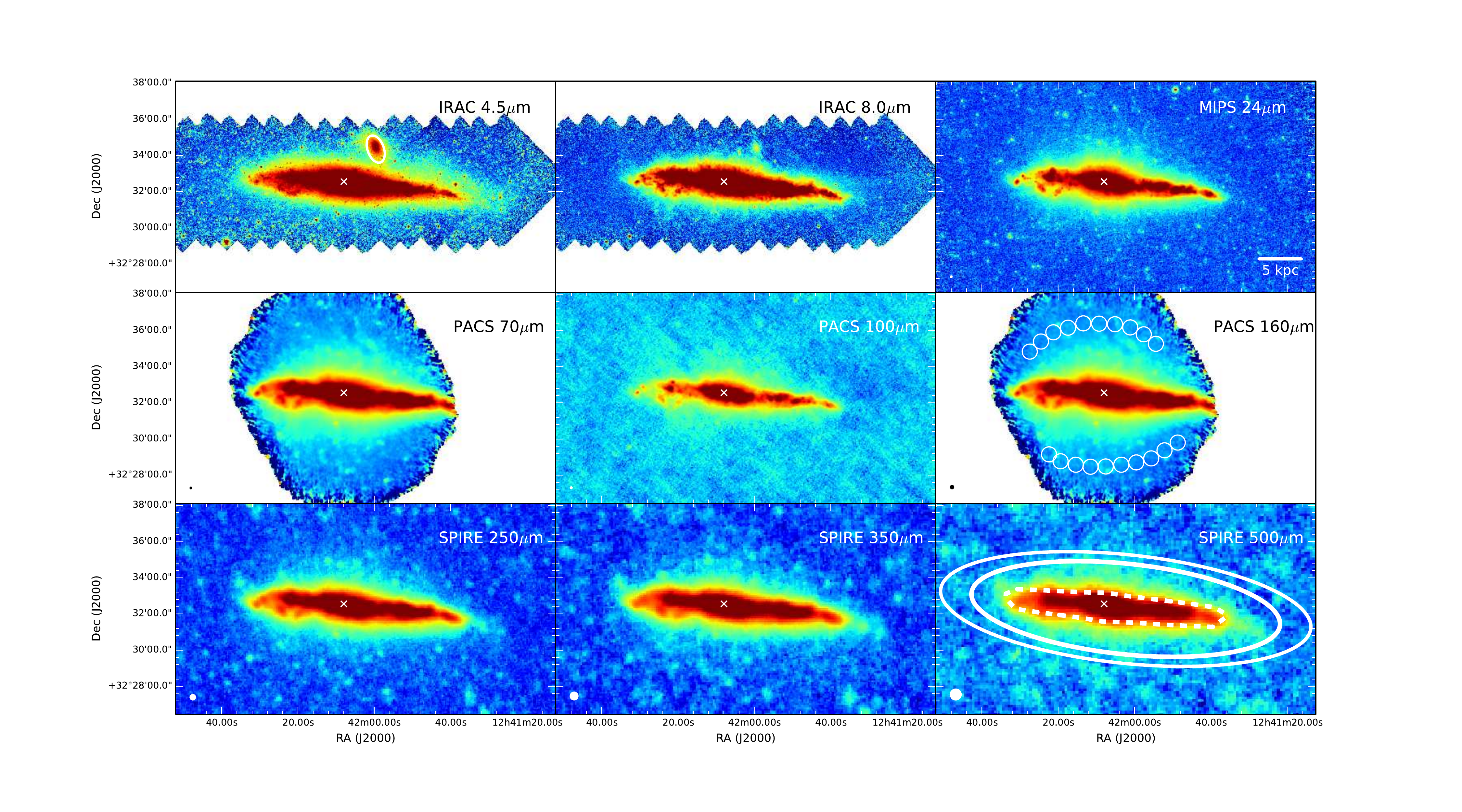} 
\caption{Comparison between Spitzer IRAC 4.5 and 8
  \micron\ \citep{2013ApJ...774..126M}, MIPS 24
  \micron\ \citep{2009ApJ...703..517D}, PACS 70, 100, and 160 \micron,
  and SPIRE 250, 350, and 500 \micron\ for NGC~4631.  The images are
  displayed with an inverse hyperbolic sine scaling. All the images
  are presented in their native resolution and pixel size:
  1.22\arcsec\ pixel$^{-1}$ for the IRAC data, 1.4 and
  2.85\arcsec\ pixel$^{-1}$ (FWHM/4) for the PACS data at 70 and 160
  \micron, respectively, 4.5, 6.26, and 9\arcsec\ pixel$^{-1}$
  (FWHM/4) for the SPIRE data at 250, 350, and 500 \micron,
  respectively.  North is up and east is to the left in all of the
  images. The horizontal line in the lower right corner of the upper
  right panel represents 5 kpc. The panel of the 160 $\micron$
    emission shows the sky apertures used for the smaller field of
    view observations (OT1\_sveilleu\_2).  The panel of the 500
    $\micron$ emission shows the photometric apertures used for the
    global flux and sky background (solid line) and the disk region
    within one vertical scale height of the galaxy mid-plane, derived
    from the IRAC 4.5 \micron\ image \citep[dashed line,
    ][]{2013ApJ...774..126M}. For the sake of comparison throughout
  the paper, the solid white ellipse in the 4.5 $\micron$ panel
  encompasses the companion galaxy NGC~4627. The beam size at each
  wavelength is indicated by a white/black filled circle in the bottom
  left corner of each panel.  In all of the panels, the white
    ``X'' indicates the position of the infrared nucleus
    \citep{1978PASP...90...28A,1981PASP...93..535A}.
\label{mosaic_ngc4631}}
\end{figure*}

\begin{figure*}
\epsscale{1.0}
\plotone{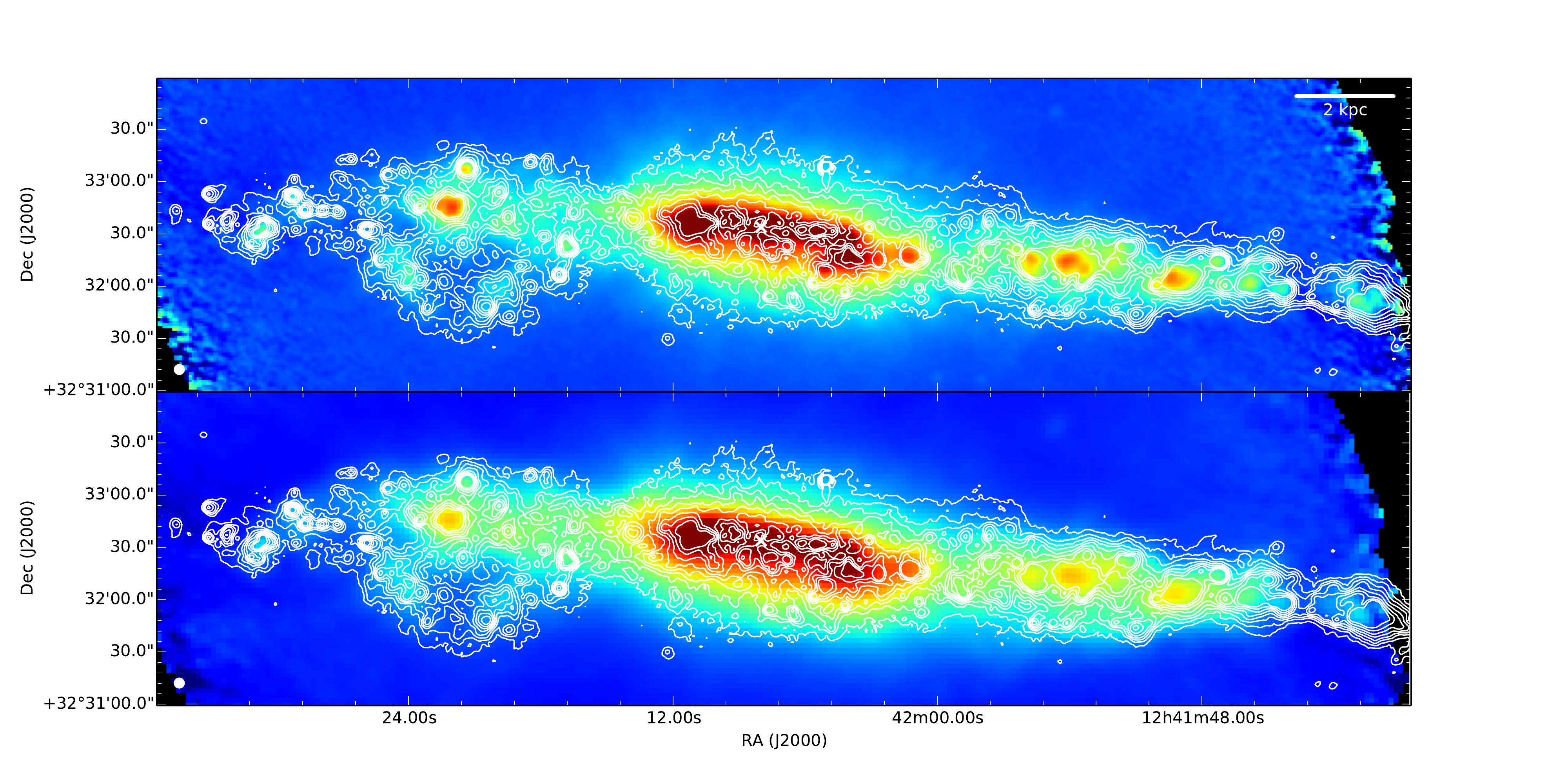} 
\caption{H$\alpha$ emission from the warm ionized gas overlaid as
  contours on the PACS images at 70 $\micron$ (top panel) and 160
  $\micron$ (bottom). The PACS maps are shown on a pixel scale of 1.4
  and 2.85\arcsec\ pixel$^{-1}$ at 70 and 160 \micron,
  respectively. The contour levels are 20, 50, 100, 250, 1000, and
  2000 times 4$\times 10^{34}$~ergs~s$^{-1}$.  North is up and east is
  to the left in both panels. The horizontal line in the upper right
  corner of the upper panel represents 2 kpc.  The beam size at each
  wavelength is indicated by a white filled circle in the bottom left
  corner of each panel. In all of the panels, the white ``X"
    indicates the position of the infrared nucleus
    \citep{1978PASP...90...28A,1981PASP...93..535A}. \label{NGC4631_PACS_Halpha}}
\end{figure*}

\begin{figure*}
\epsscale{1.0}
\plotone{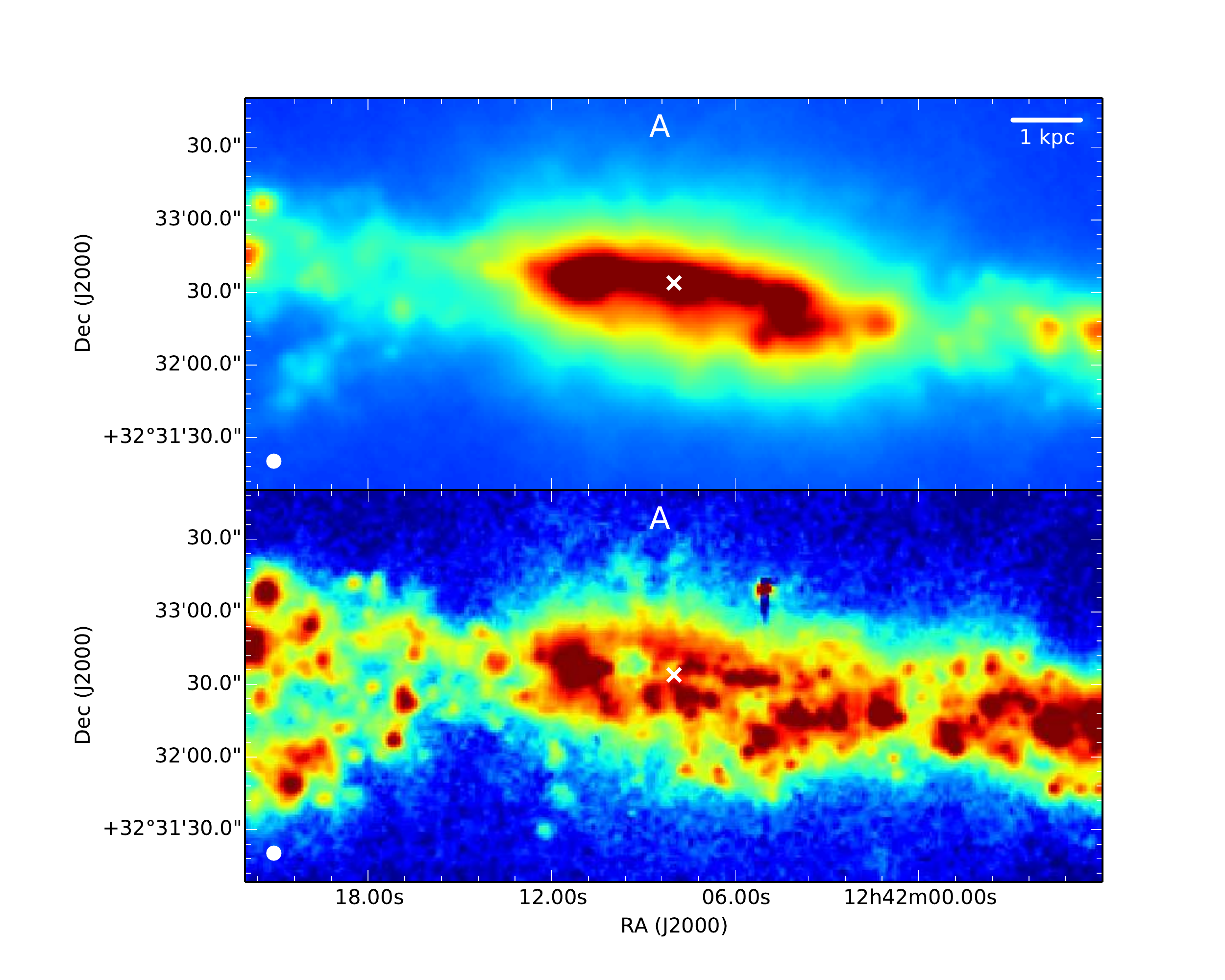} 
\caption{A close-up view of the central region of NGC~4631, comparing
  the PACS image at 70~\micron\ ({\it upper-panel}) and the H$\alpha$
  map ({\it lower-panel}). The label ``A'' indicates the position of
  the H$\alpha$ double-worm structure in both images. The PACS 70
  \micron\ map is presented on a pixel scale of
  1.4\arcsec\ pixel$^{-1}$, whereas the H$\alpha$ map is shown on a
  pixel scale of 1.2\arcsec\ pixel$^{-1}$.  North is up and east is to
  the left in all of the images. The horizontal line in the upper
  right corner of the upper panel represents 1 kpc. The beam size at
  each wavelength is indicated by a white filled circle in the bottom
  left corner of each panel.  In all of the panels, the white
    ``X" indicates the position of the infrared nucleus
    \citep{1978PASP...90...28A,1981PASP...93..535A}. \label{NGC4631_PACS_Halpha_zoom}}
\end{figure*}

\begin{figure*}
\epsscale{1.0}
\plotone{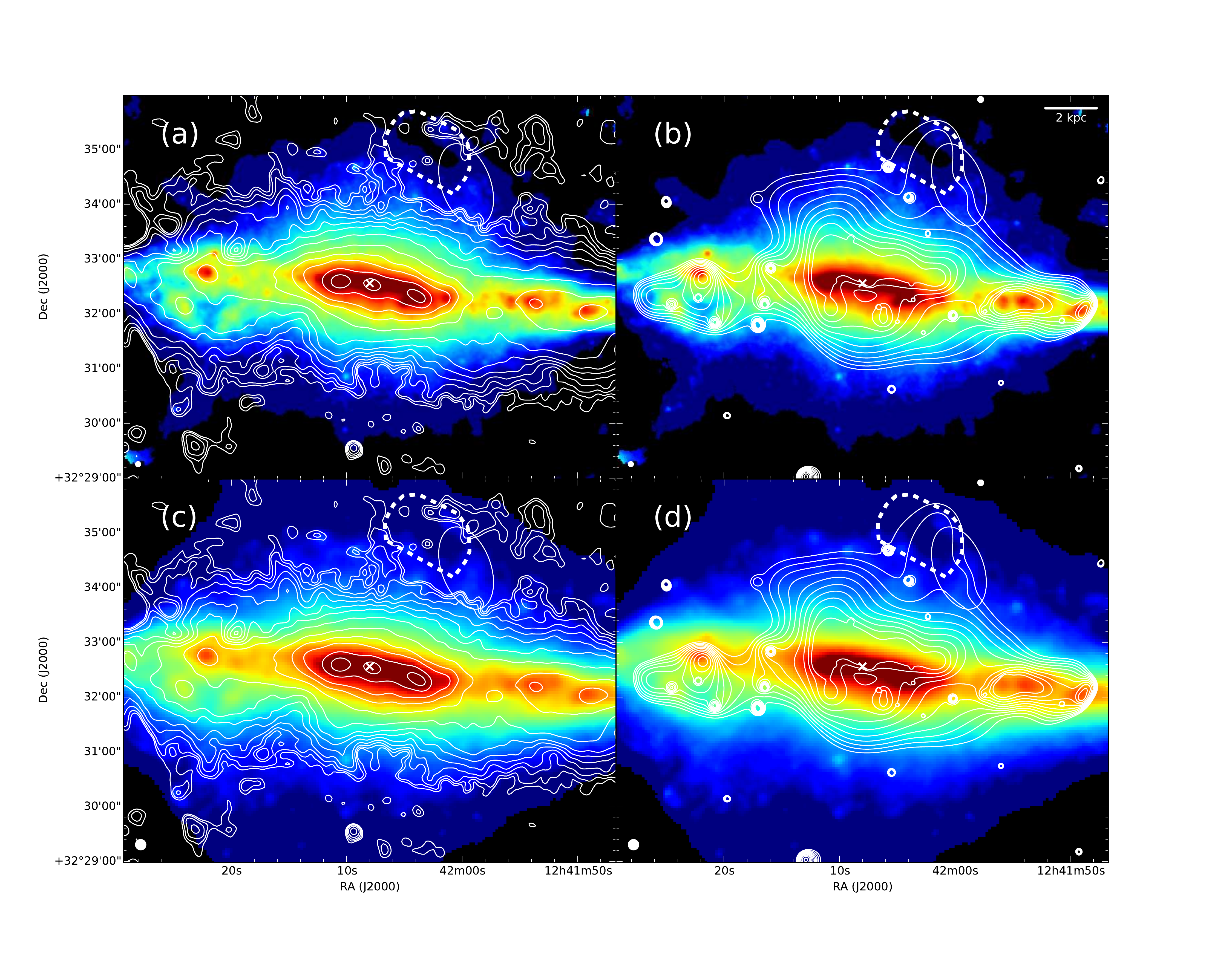} 
\caption{Adaptively smoothed PACS flux maps at 70 and 160 $\micron$
  compared with radio-continuum and soft X-ray images.  {\it (a)} 70
  $\micron$ image with a pixel scale of 1.4\arcsec\ pixel$^{-1}$ is
  compared with the Robust 2 weighted image at 1.5 GHz, shown as
  contours. {\it (b)} 70 $\micron$ smoothed map with a pixel scale of
  1.4\arcsec\ pixel$^{-1}$ is compared with the {\it Chandra} soft
  X-ray (0.3 -- 2.0~keV) emission, shown as contours. {\it (c)} and
  {\it (d)} Same as {\it (a)} and {\it (b)}, but with the 160
  $\micron$ smoothed map on a pixel scale of
  2.85\arcsec\ pixel$^{-1}$. The contours for the 1.5 GHz image are at
  9 (2$\sigma$), 13.5, 20, 30, 45, 65, 90, 150, 300, 500, 750, 1500,
  and 2500 $\times$ 10 $\mu$Jy beam$^{-1}$. The contours for the X-ray
  image are 0.1, 0.107, 0.114, 0.123, 0.135, 0.151, 0.172, 0.202,
  0.243 and 0.301 photon cm$^{-2}$ s$^{-1}$ arcsec$^{-2}$.  North is
  up and east to the left. The horizontal line in the upper right
  corner of panel (b) represents 2 kpc. The white ellipse indicates
  the position of the companion galaxy NGC~4627. The beam size at 70
  and 160 $\micron$ is indicated by a white filled circle in the
  bottom left corner of each panel. White dashed curves show the
    FIR loop discussed in Section~\ref{adata}. In all of the panels,
    the white ``X" indicates the position of the infrared nucleus
    \citep{1978PASP...90...28A,1981PASP...93..535A}.\label{comp_PACS_x-ray_radio_smooth}}
\end{figure*}

\begin{figure}
\epsscale{0.6}
\plotone{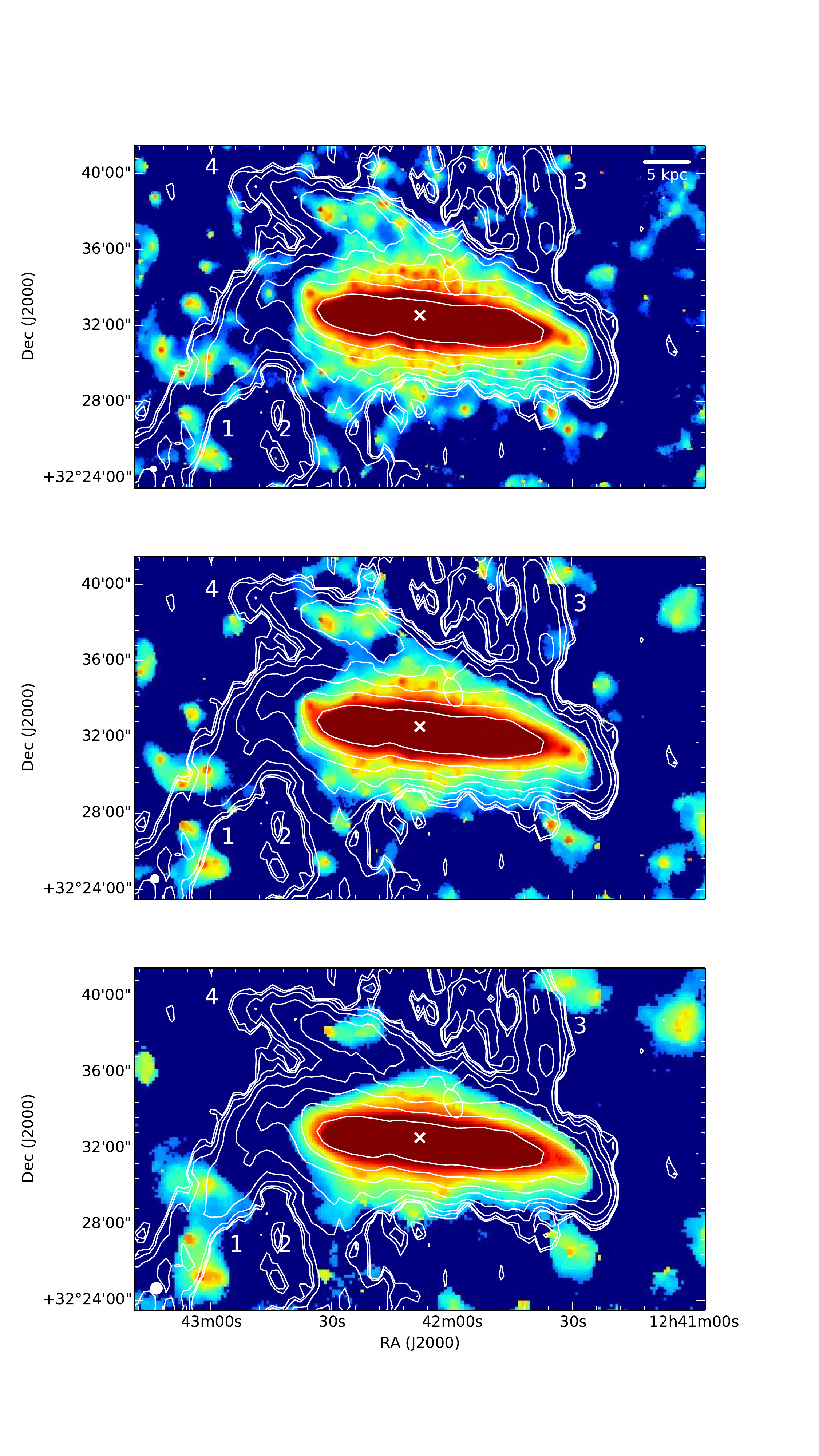} 
\caption{ The H~I 21-cm emission at 45\arcsec\ $\times$
  87\arcsec\ resolution (shown as contours) overlaid on the adaptively
  smoothed SPIRE images at 250 (upper panel), 350 (middle panel), and
  500 (lower panel) $\micron$ with a pixel scale of 4.5, 6.26, and
  9\arcsec\ pixel$^{-1}$ (FWHM/4), respectively. In column density
  units at the beam center, the contour levels are 0.5, 0.6, 0.8 ,
  1.3, 2.6, 5.7, 12.9, 30.1 and 70.7 $\times 10^{20} {\rm cm^{-2}}$.
  Only fluxes above 3 $\sigma$ are shown in these panels. North is up
  and east is to the left in all of the images. The horizontal line in
  the upper right corner of the upper panel represents 5 kpc. The
  white ellipse indicates the position of the companion galaxy
  NGC~4627. The beam size at each wavelength is indicated by a white
  filled circle in the bottom left corner of each panel. In all
    of the panels, the white ``X" indicates the position of the
    infrared nucleus
    \citep{1978PASP...90...28A,1981PASP...93..535A}.\label{NGC4631_SPIRE_HI_LR_smooth_units}}
\end{figure}

\begin{figure}
\epsscale{0.4}
\plotone{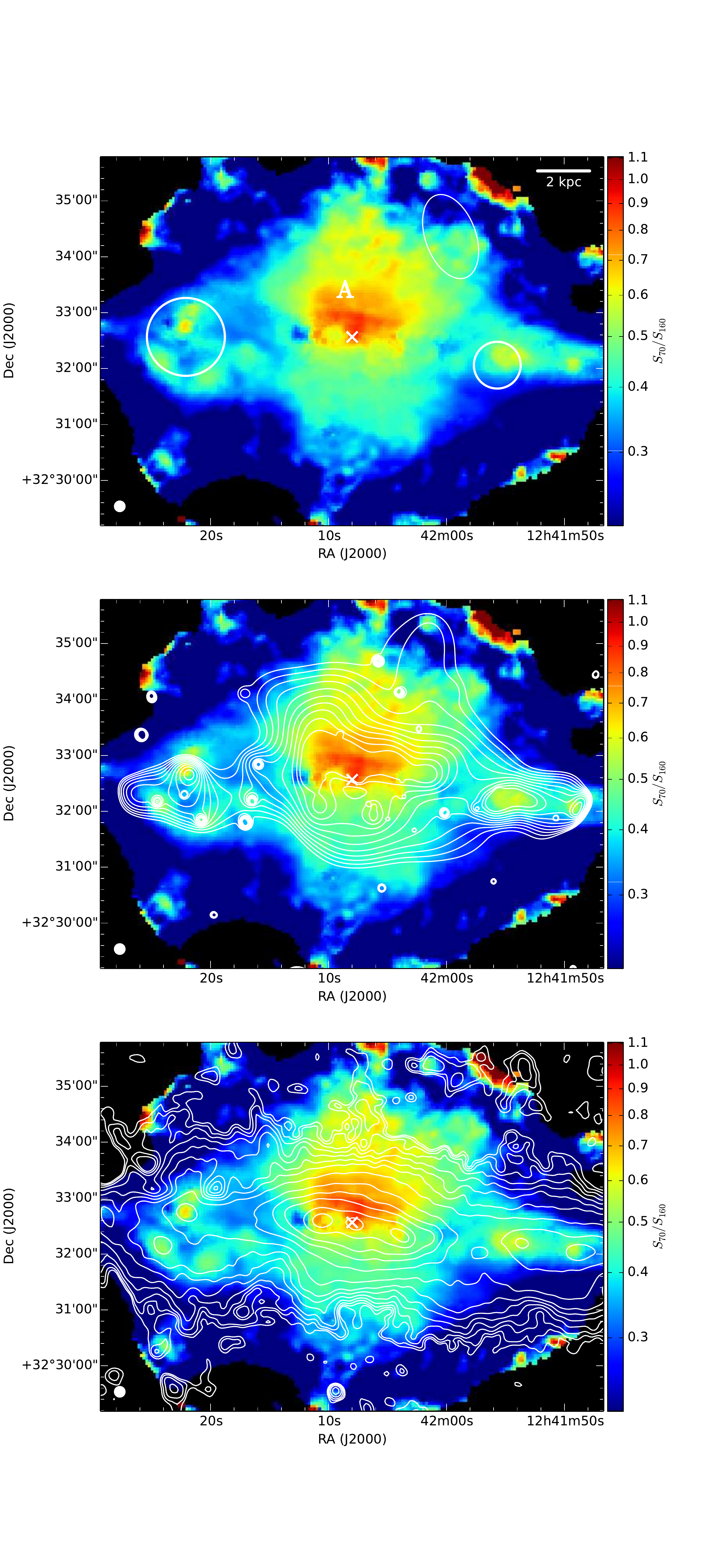} 
\caption{The $S_{70}/S_{160}$ ratio map of the adaptively smoothed
  PACS maps shown in Figure~\ref{comp_PACS_x-ray_radio_smooth}. The
  upper panel shows the PACS ratio, $S_{70}/S_{160}$, with a pixel
  scale of 2.85\arcsec\ pixel$^{-1}$. The middle panel shows the soft
  X-ray emission \citep{2004ApJS..151..193S}, shown as contours, over
  the $S_{70}/S_{160}$ ratio map. The contours are 0.1, 0.107, 0.114,
  0.123, 0.135, 0.151, 0.172, 0.202, 0.243 and 0.301 photon cm$^{-2}$
  s$^{-1}$ arcsec$^{-2}$.  The lower panel shows the Robust 2 weighted
  1.5 GHz image \citep{2012AJ....144...44I}, shown as contours, over
  the $S_{70}/S_{160}$ ratio map. The contours are at 9 (2$\sigma$),
  13.5, 20, 30, 45, 65, 90, 150, 300, 500, 750, 1500, and 2500
  $\times$ 10 $\mu$Jy beam$^{-1}$.  North is up and east is to the
  left. The horizontal line in the upper right corner of the upper
  panel represents 2 kpc. Note the X-ray and radio loop located
  $\sim$6 kpc north of the galaxy center. The white ellipse indicates
  the position of the companion galaxy NGC~4627. The white circles
  represent the approximate size of the H~I supershells
  \citep{1993AJ....105.2098R}. The label ``A'' indicates the
    position of the H$\alpha$ double-worm structure (see
    Figure~\ref{NGC4631_PACS_Halpha_zoom}) and the white ``X"
    indicates the position of the infrared nucleus
    \citep{1978PASP...90...28A,1981PASP...93..535A}. The beam size
  at 160 $\micron$ is indicated by a white filled circle in the bottom
  left corner of each
  panel. \label{ratio_radio_and_x-ray_smooth_flat_s3}}
\end{figure}

\begin{figure}
\epsscale{0.8}
\plotone{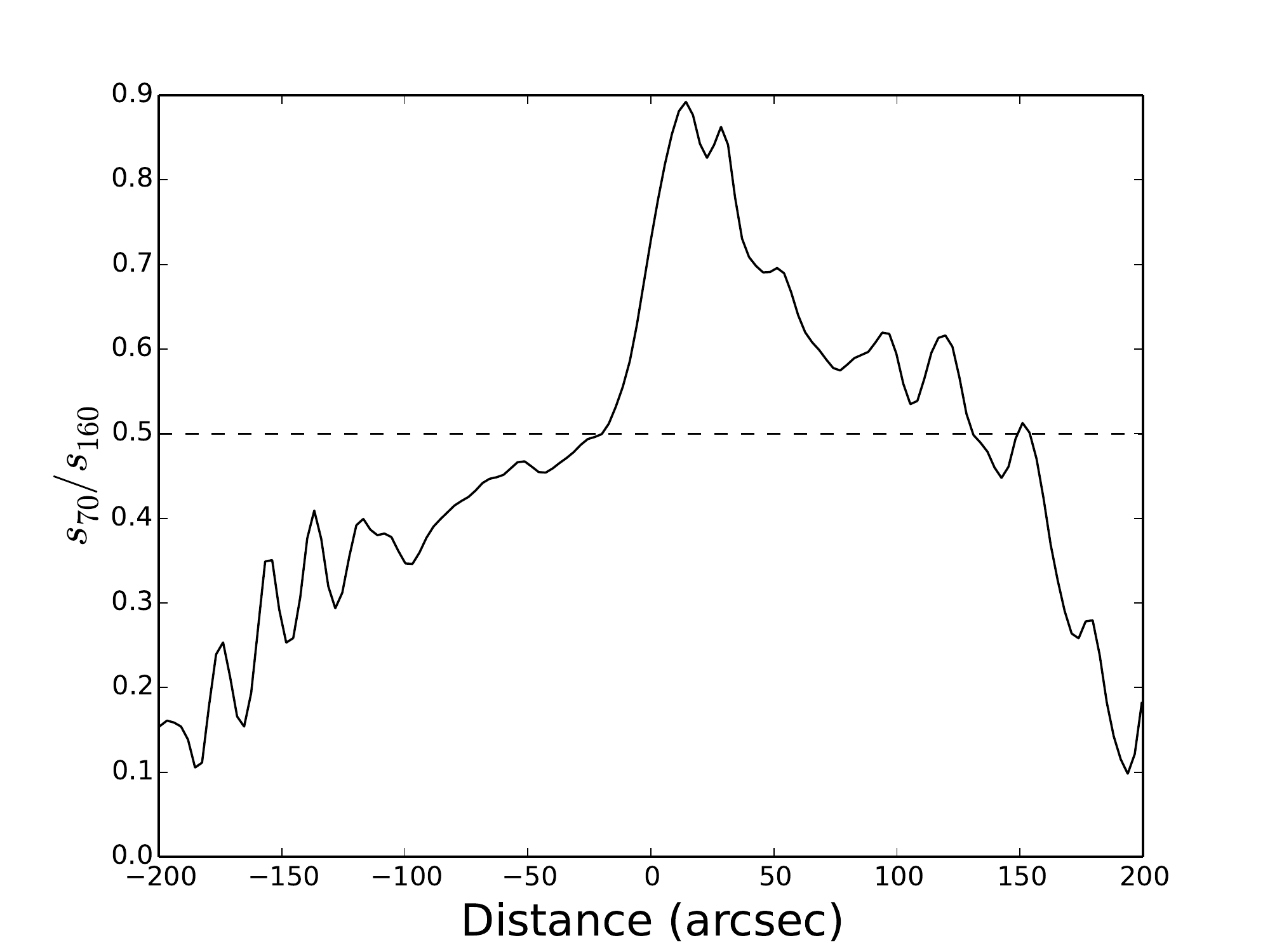} 
\caption{Vertical profile of the PACS $S_{70}/S_{160}$ ratio with a
  slice width of one pixel, 2.85\arcsec, perpendicular to the mid-plane
  passing through the galaxy center of NGC~4631. The origin of the
horizontal axis corresponds to the disk mid-plane. The horizontal
dotted line corresponds to $S_{70}/S_{160}$ = 0.5. Values above this
line mark the position of the superbubble region, shown in the next
figure. \label{radial_profile}}
\end{figure}

\begin{figure}
\epsscale{0.6}
\plotone{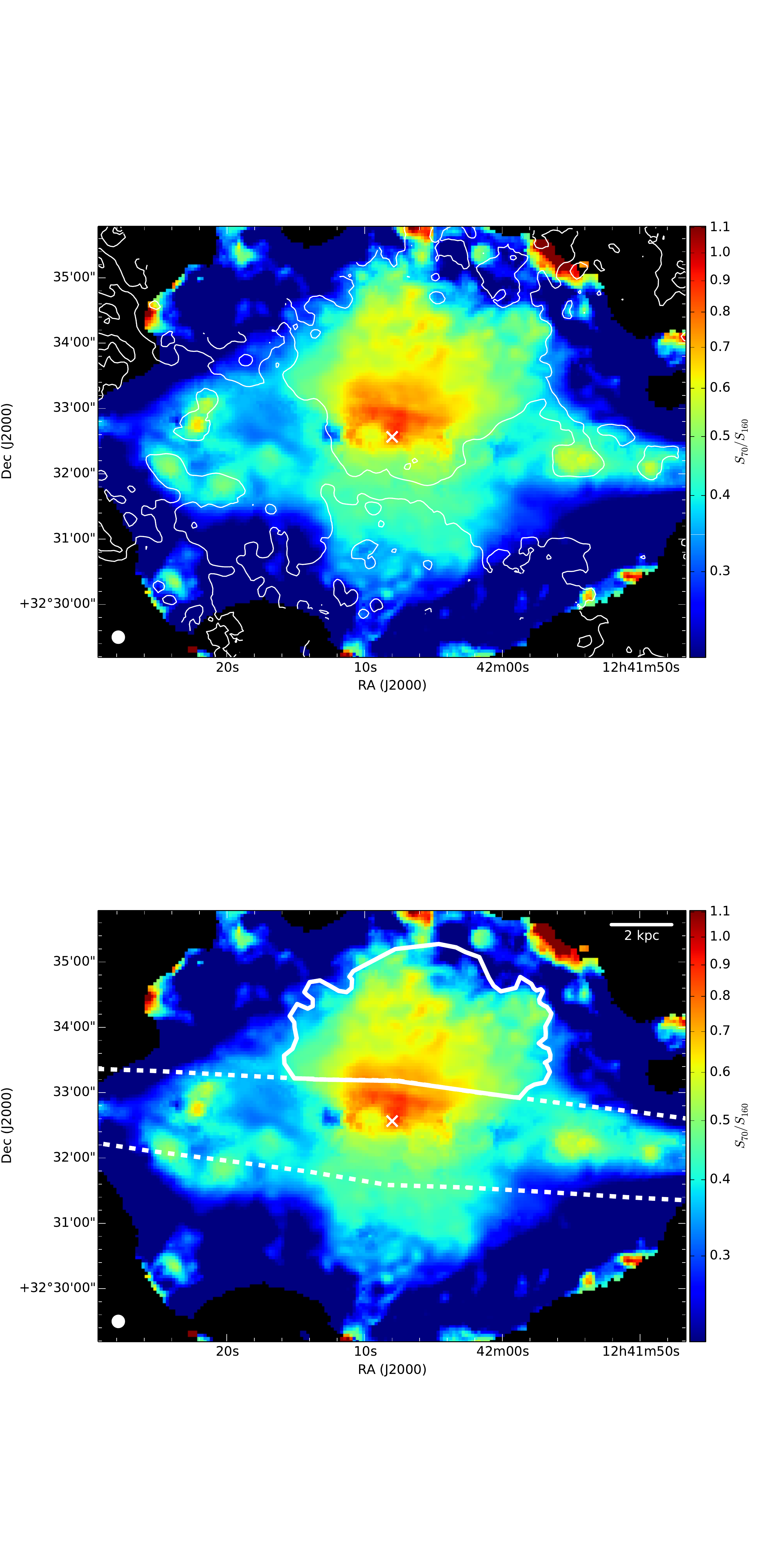} 
\caption{The $S_{70}/S_{160}$ ratio map derived from the adaptively
  smoothed PACS maps shown in
  Figure~\ref{comp_PACS_x-ray_radio_smooth}. The pixel scale is
  2.85\arcsec\ pixel$^{-1}$.  The contours in the upper panel
  represent a value of $S_{70}/S_{160}$ = 0.5, used to define the
  superbubble region.  The lower panel shows the region that is used
  to calculate the flux in the superbubble (solid line) and the disk
  region within one vertical scale height of the galaxy mid-plane,
  derived from IRAC 4.5 \micron\ image \citep[dashed line,
  ][]{2013ApJ...774..126M}.  North is up and east is to the left. The
  horizontal line in the upper right corner of the lower panel
  represents 2 kpc. The beam size at 160 $\micron$ is indicated by a
  white filled circle in the bottom left corner of each panel.  In
    all of the panels, the white ``X" indicates the position of the
    infrared nucleus
    \citep{1978PASP...90...28A,1981PASP...93..535A}.\label{ratio_smooth_bubble}}
\end{figure}

\begin{figure}
\epsscale{1.0}
\plotone{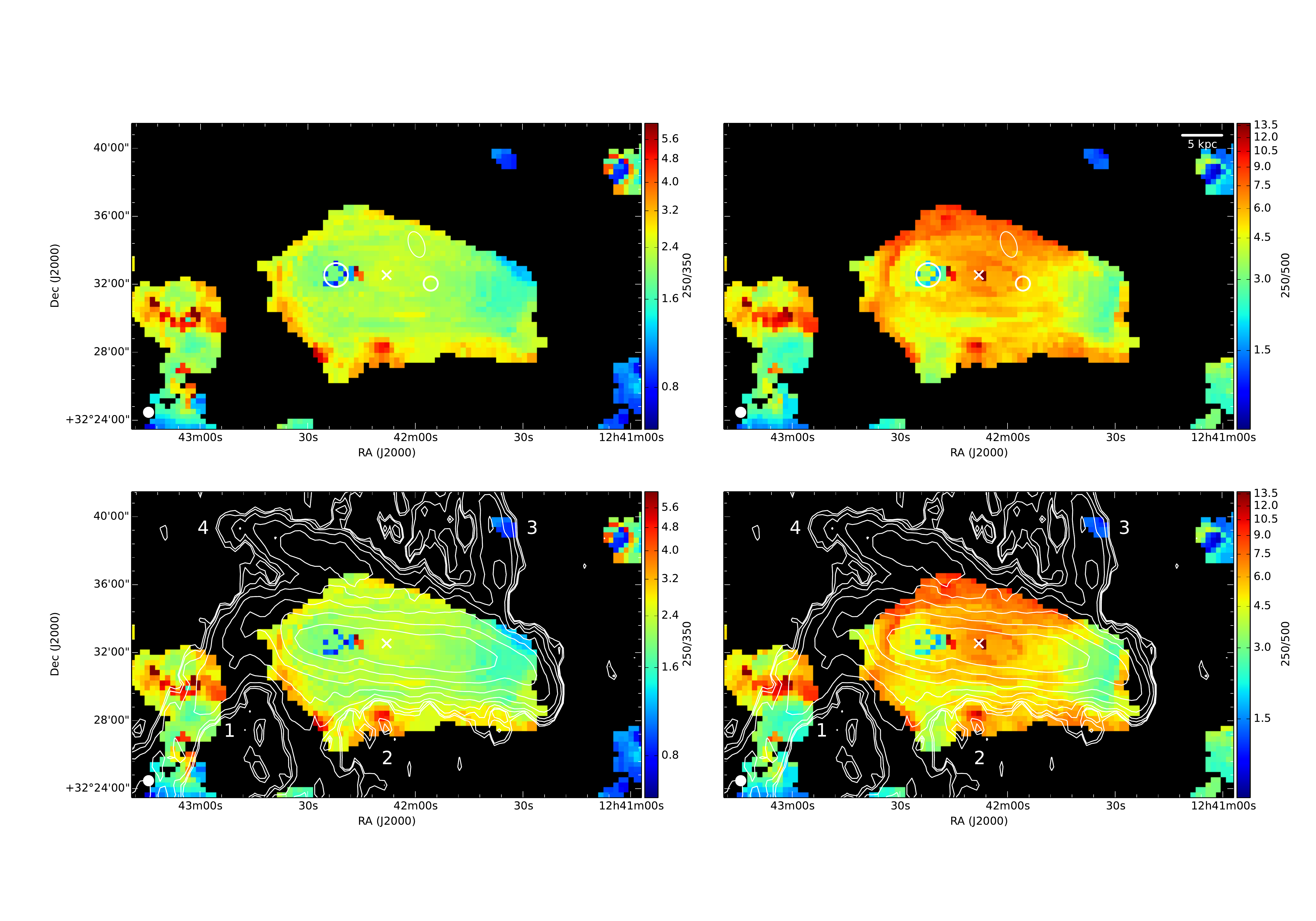}  
\caption{Left panels: the $S_{250}/S_{350}$ ratio map derived from the
  adaptively smoothed SPIRE maps shown in
  Figure~\ref{NGC4631_SPIRE_HI_LR_smooth_units}. Right panels: same as
  left panel but the $S_{250}/S_{500}$ ratio map. The upper row shows
  the ratio maps at a pixel scale of 18\arcsec\ pixel$^{-1}$. The
  lower rows compares these maps with the H~I 21-cm emission (as
  contours) at a resolution 45$\arcsec\ \times$ 87\arcsec.  In column
  density units at the beam center, the contour levels are 0.5, 0.6,
  0.8 , 1.3, 2.6, 5.7, 12.9, 30.1 and 70.7 $\times 10^{20} {\rm
    cm^{-2}}$. North is up and east to the left. The horizontal line
  in the upper right corner of the upper right panel represents 5
  kpc. The white circles represent the approximate size of the H~I
  supershells \citep{1993AJ....105.2098R}. The white ellipse indicates
  the position of the companion galaxy NGC~4627. The beam size at 500
  $\micron$ is indicated by a white filled circle in the bottom left
  corner of each panel. In all of the panels, the white ``X"
    indicates the position of the infrared nucleus
    \citep{1978PASP...90...28A,1981PASP...93..535A}.\label{NGC4631_SPIRE_ratios_250_500_250_350_smooth}}
\end{figure}

\begin{figure}
\epsscale{0.8}
\plotone{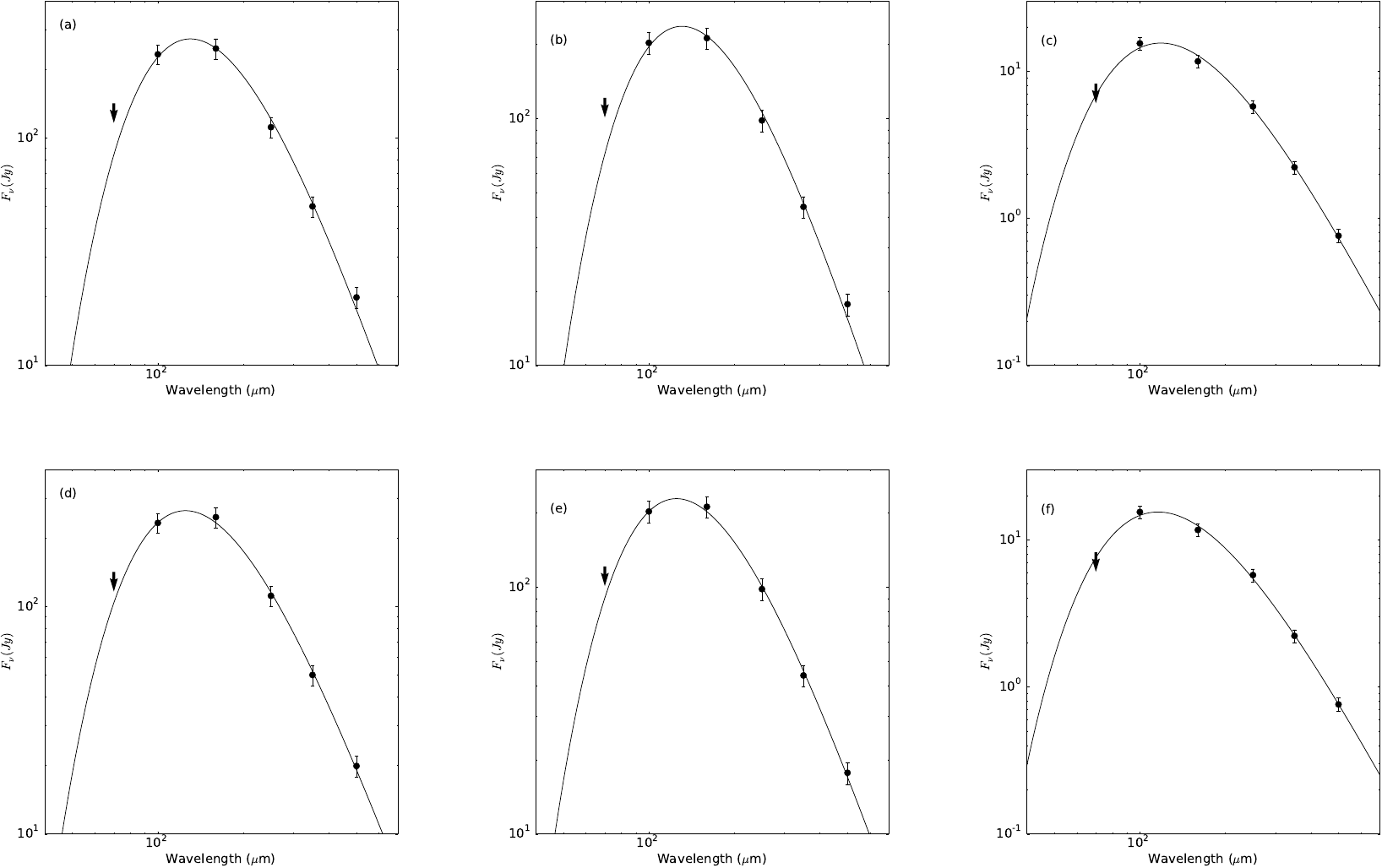} 
\caption{Single-temperature MBB fit to the global (panels a and d),
  disk (panels b and e), and superbubble (panels c and f) spectral
  energy distributions in the 70-500 \micron\ range. The upper panels
  show the fit with $\beta = 2.0$ (panels a,b and c) whereas the lower
  panels show the fit with the spectral index $\beta$ left as a free
  parameter (panels d, e and f). See Table~\ref{MBB} for details on
  the various fits.  \label{figure_1}}
\end{figure}

\begin{figure}
\epsscale{1.0}
\plottwo{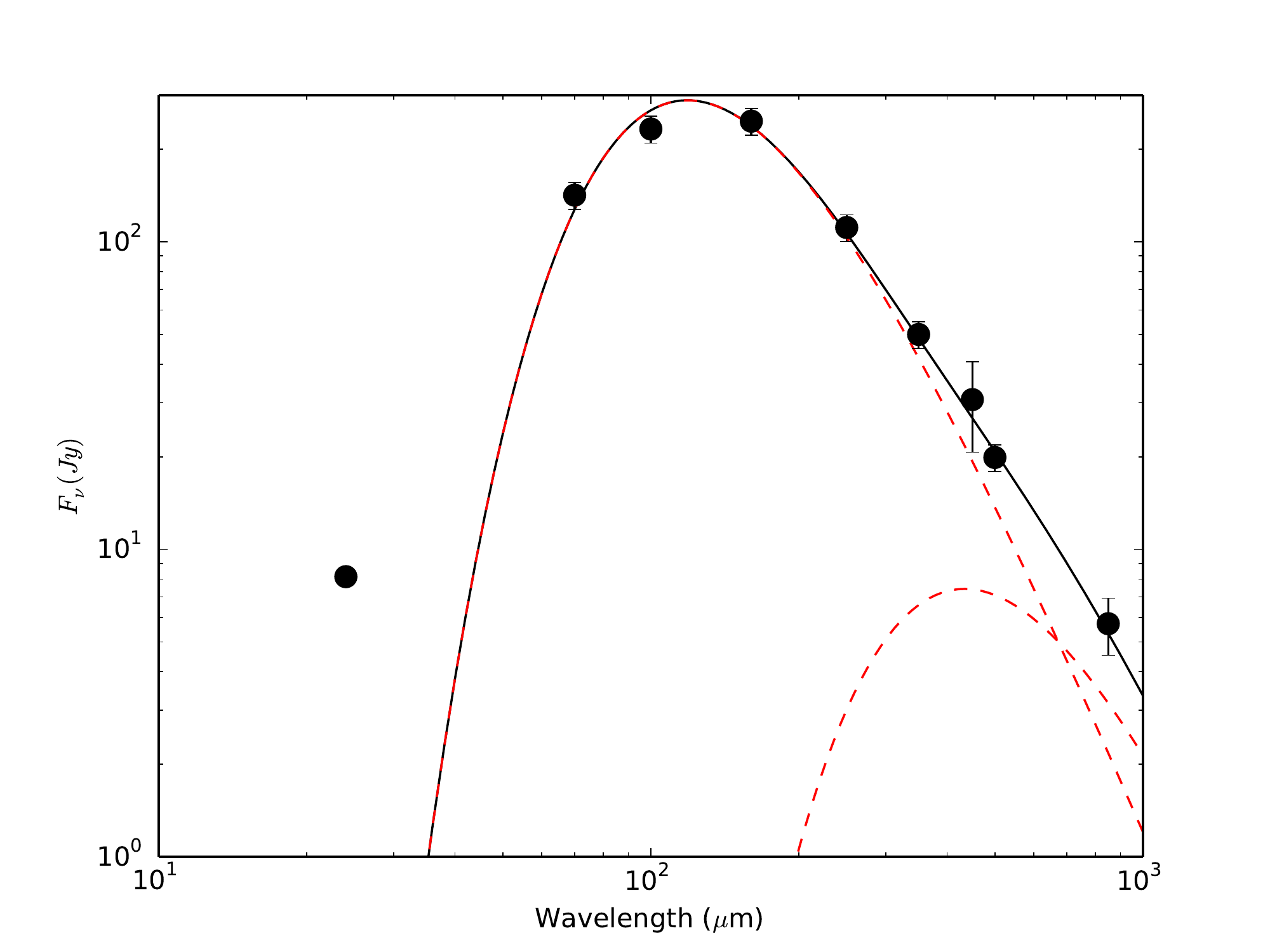}{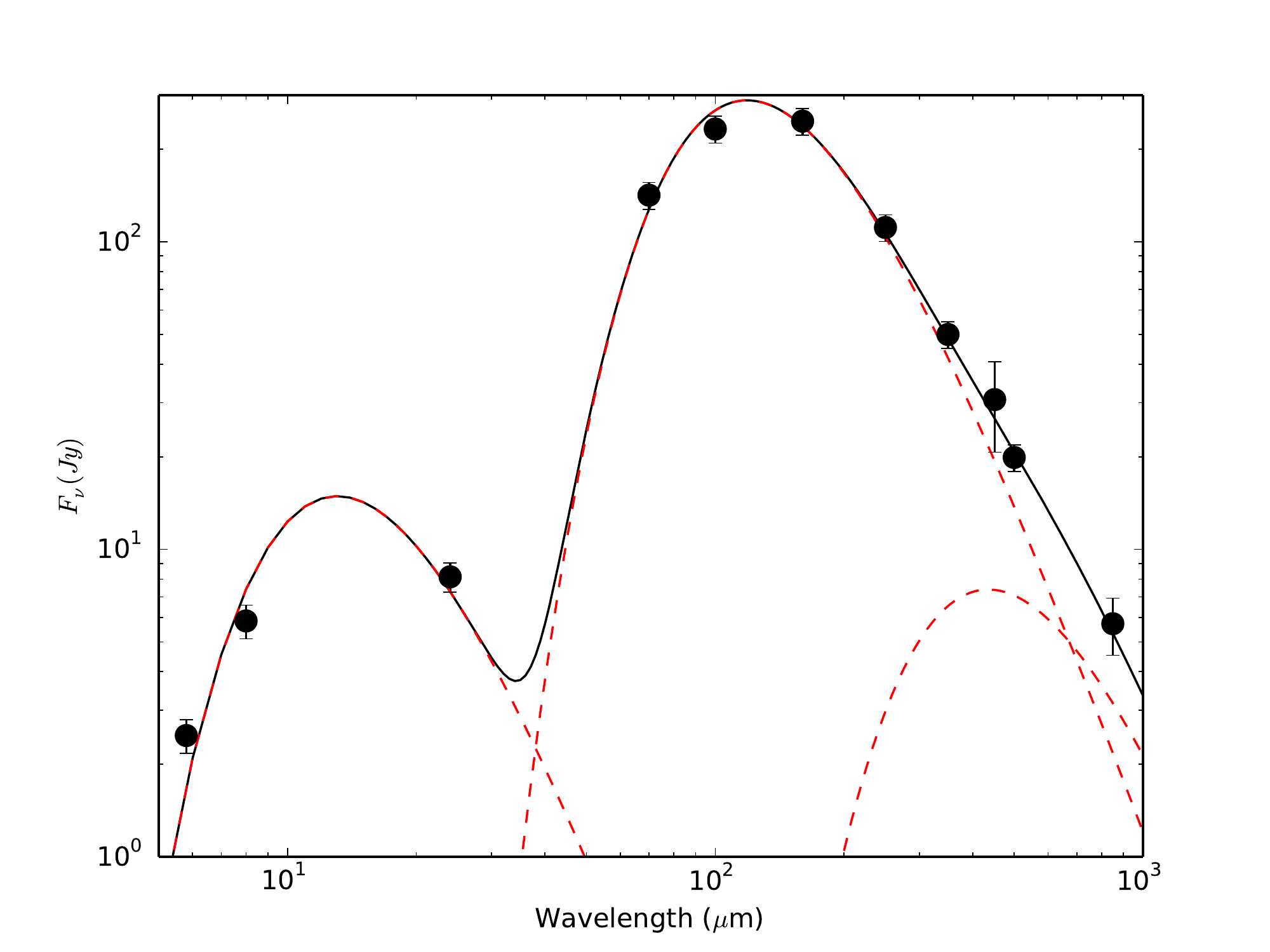}
\caption{Left panel: Two-temperature MBB fit with $\beta = 2.0$
  (fixed) to the global spectral energy distribution. For the sake of
  comparison, we also show the flux density at 24 $\micron$
  \citep[{\it Spitzer} MIPS, ][]{2009ApJ...703..517D}. However, this
  data point was not included in the fit. Right panel:
  three-temperature MBB fit with $\beta = 2.0$ (fixed) to the global
  spectral energy distribution. See Table 3 for the results from these
  fits. \label{two_comp_three_comp}}
\end{figure}

\begin{figure}
\epsscale{0.8} \plotone{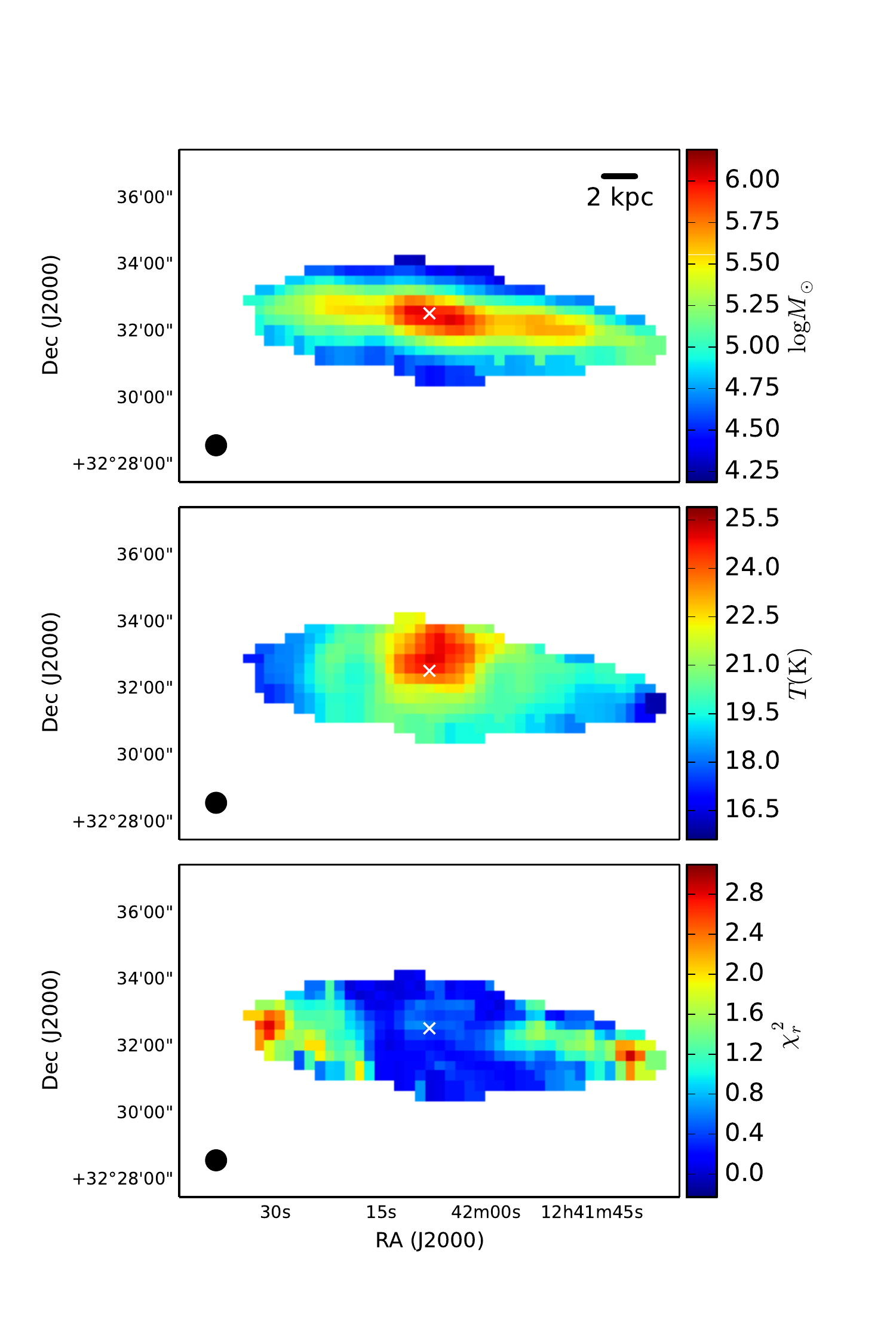}
\caption{Pixel-by-pixel SED fits where the fluxes at 70-500
  \micron\ from each pixel were fit with a single-temperature MBB with
  $\beta=2.0$ (fixed).  Note that the 70 $\micron$ flux is used as an
  upper limit, see text for detail. The pixel scale is
  18\arcsec\ pixel$^{-1}$. North is up and east is to the left.  The
  horizontal line in the upper right corner of the upper panel
  represents 2 kpc. The beam size at 500 $\micron$ is indicated by a
  black filled circle in the bottom left corner of each panel.  In
    all of the panels, the white ``X" indicates the position of the
    infrared nucleus
    \citep{1978PASP...90...28A,1981PASP...93..535A}.\label{mosaic_ngc4631_NO70_SED_fix_beta}}
\end{figure}

\begin{figure}
\epsscale{0.8}
\plotone{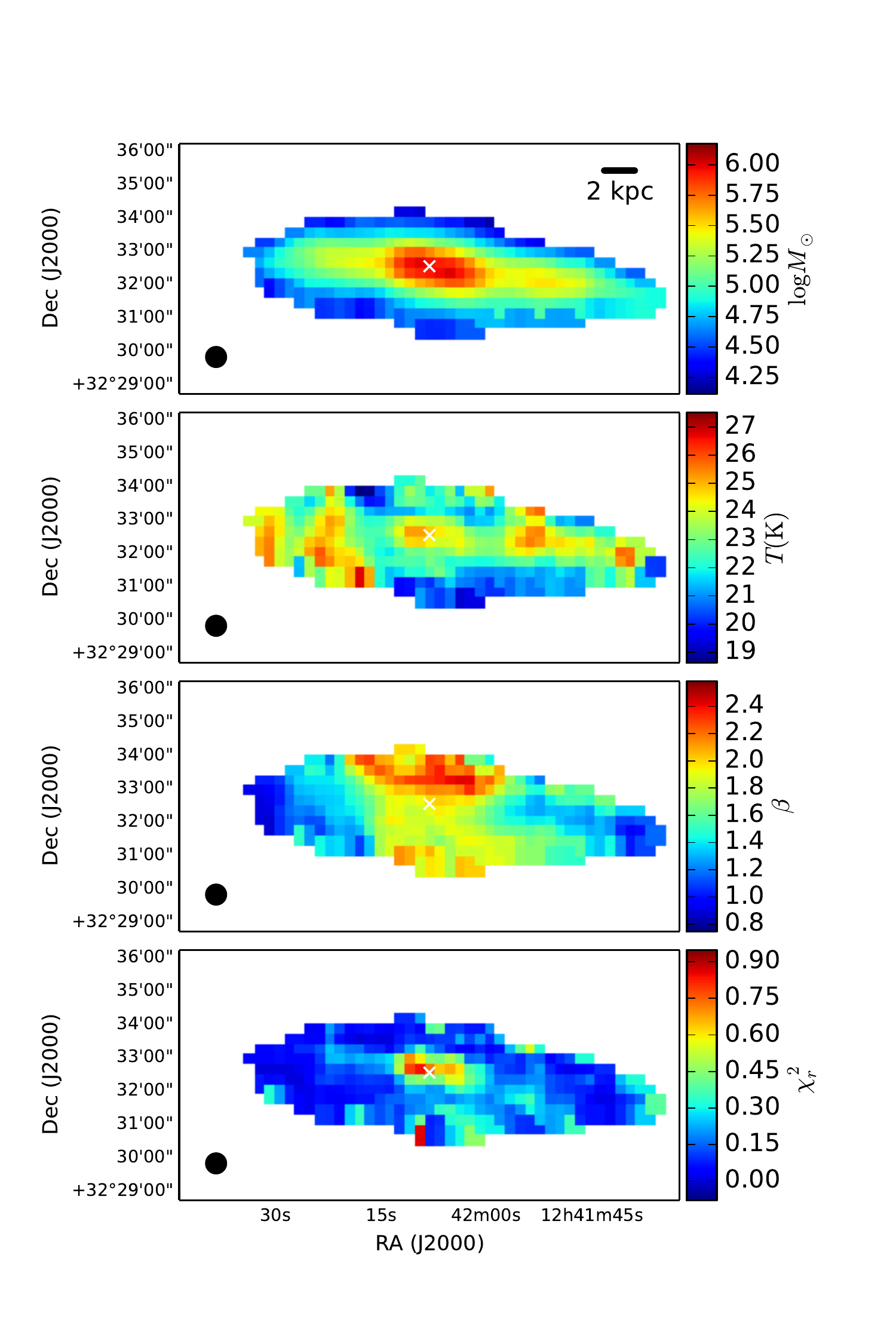} 
\caption{Same as Figure~\ref{mosaic_ngc4631_NO70_SED_fix_beta}, except
  that $\beta$ is now a free parameter. Note that the 70$\micron$ flux
  is used as an upper limit, see text for detail. The pixel scale is
  18\arcsec\ pixel$^{-1}$. North is up and east is to the left. The
  horizontal line in the upper right corner of the upper panel
  represents 2 kpc. The beam size at 500 $\micron$ is indicated by a
  black filled circle in the bottom left corner of each panel. In
    all of the panels, the white ``X" indicates the position of the
    infrared nucleus
    \citep{1978PASP...90...28A,1981PASP...93..535A}.\label{mosaic_ngc4631_NO70_SED}}
\end{figure}

\begin{figure}
\epsscale{1.2}
\plottwo{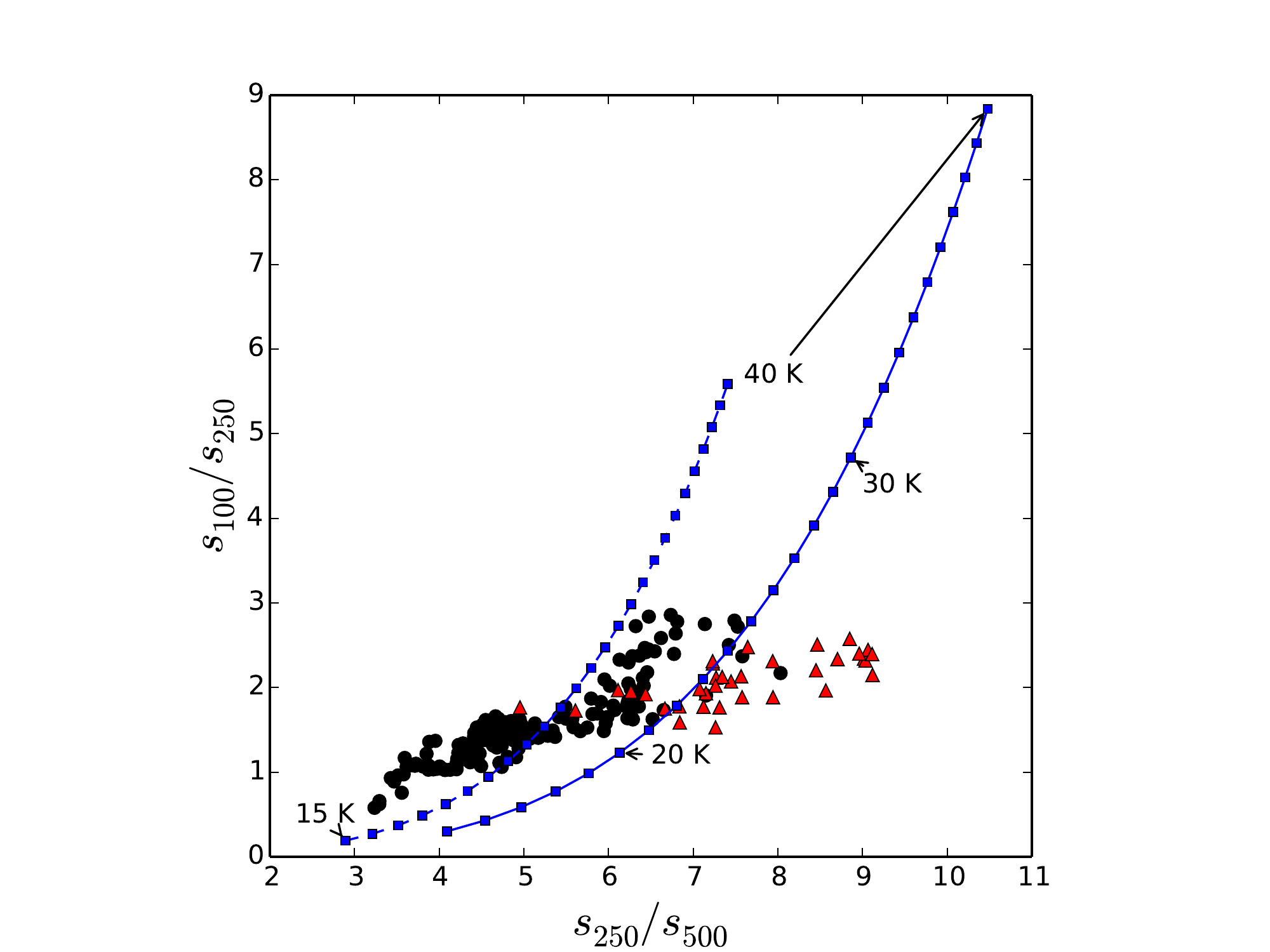}{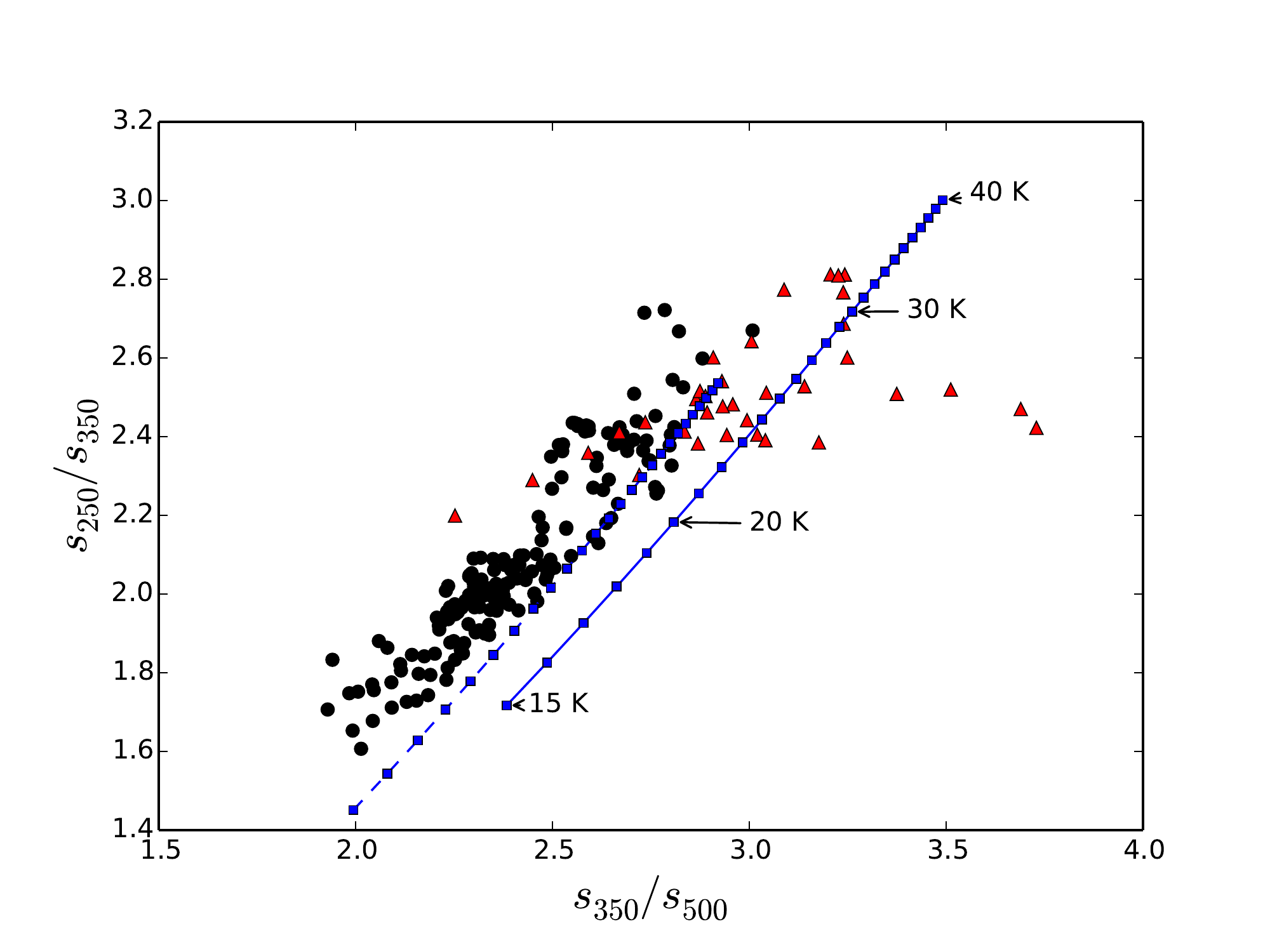}
\caption{Left panel: PACS-SPIRE color-color diagram,
  $S_{250}$/$S_{500}$ versus $S_{100}$/$S_{250}$, for each pixel in
  NGC~4631. Right panel: SPIRE color-color diagram,
  $S_{250}$/$S_{350}$ versus $S_{350}$/$S_{500}$ ratios, for each
  pixel in NGC~4631. The black circles represent the disk emission,
  whereas the red triangles represent emission in the superbubble
  region.  See Figure~\ref{ratio_smooth_bubble} for the limits of the
  disk and superbubble regions. All of the images were convolved to
  the lower resolution of the SPIRE 500 \micron\ data (FWHM/2,
  18\arcsec) before making this plot.  For comparison, the expected
  relationship from a single modified blackbody with a dust emissivity
  of $\beta = 2.0$ ({\it solid line}) and $\beta = 1.5$ ({\it dashed
    line}) is also shown. \label{ratio1_ratio2}}
\end{figure}

\begin{figure}
\epsscale{0.8}
\plotone{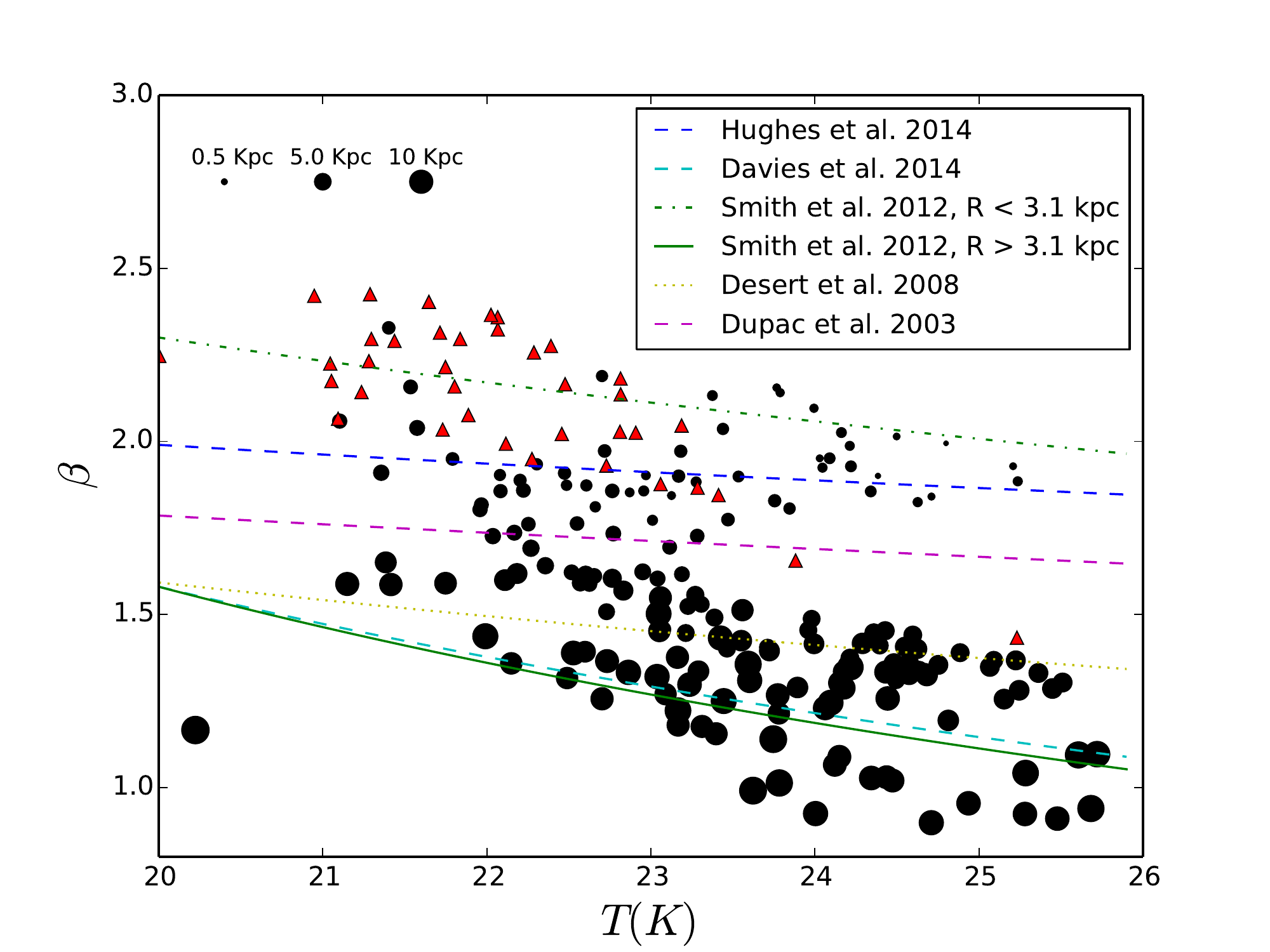} 
\caption{The observed anti-correlation between $\beta$ and the dust
  temperature from the single-temperature MBB for the pixel-by-pixel
  SED fits. Symbols are the same as in Figure~\ref{ratio1_ratio2},
  except that the symbol size (diameter) for the disk
    measurements is now proportional to the radial distance from the
    center of the galaxy.  The radial distance scale is shown in the
    upper left corner, where three different black circles are shown
    representing radial distances of 0.5, 5, and 10 kpc.  We compared
  our data with the best fit of some of the $\beta$-T relationships
  found in the literature: 1) resolved analysis of NGC~891
  \citep[dashed blue line, ][]{2014A&A...565A...4H}; 2) the {\it
    Herschel} Virgo Cluster Survey \citep[dashed cyan
    line,][]{2014MNRAS.438.1922D}; 3) the green solid and dot-dashed
  lines represent the best $\beta$-T fit from the resolved analysis of
  M31 \citep{2012ApJ...756...40S} for a radial distance of $3.1 < R <
  15 $ and $R < 3.1$~kpc, respectively; 4) ARCHEOPS sources
  \citep[yellow dotted line, ][]{2008A&A...481..411D} and, 5)
  PRONAOS-based data for different regions of the ISM
  \citep[][]{2003A&A...404L..11D}.  \label{beta-temp}}
\end{figure}
%The dashed blue line is the best fit for NGC~891 \citep{2014A&A...565A...4H} and the dashed cyan line represents the best fit from the {\it Herschel} Virgo Cluster Survey \citep{2014MNRAS.438.1922D}. The green solid and dot-dashed line represent the best $\beta$-T fit from the resolved analysis of M31 \citep{2012ApJ...756...40S} for a radial distance of  $3.1 < R < 15 $ and $R < 3.1$~kpc, respectively.  The best fit for ARCHEOPS sources is presented  as a yellow dotted line \citep{2008A&A...481..411D} and, finally,  the magenta dashed line represents the PRONAOS-based data for different regions of the ISM \citep[][]{2003A&A...404L..11D}

\begin{figure}
\epsscale{0.8}
\plotone{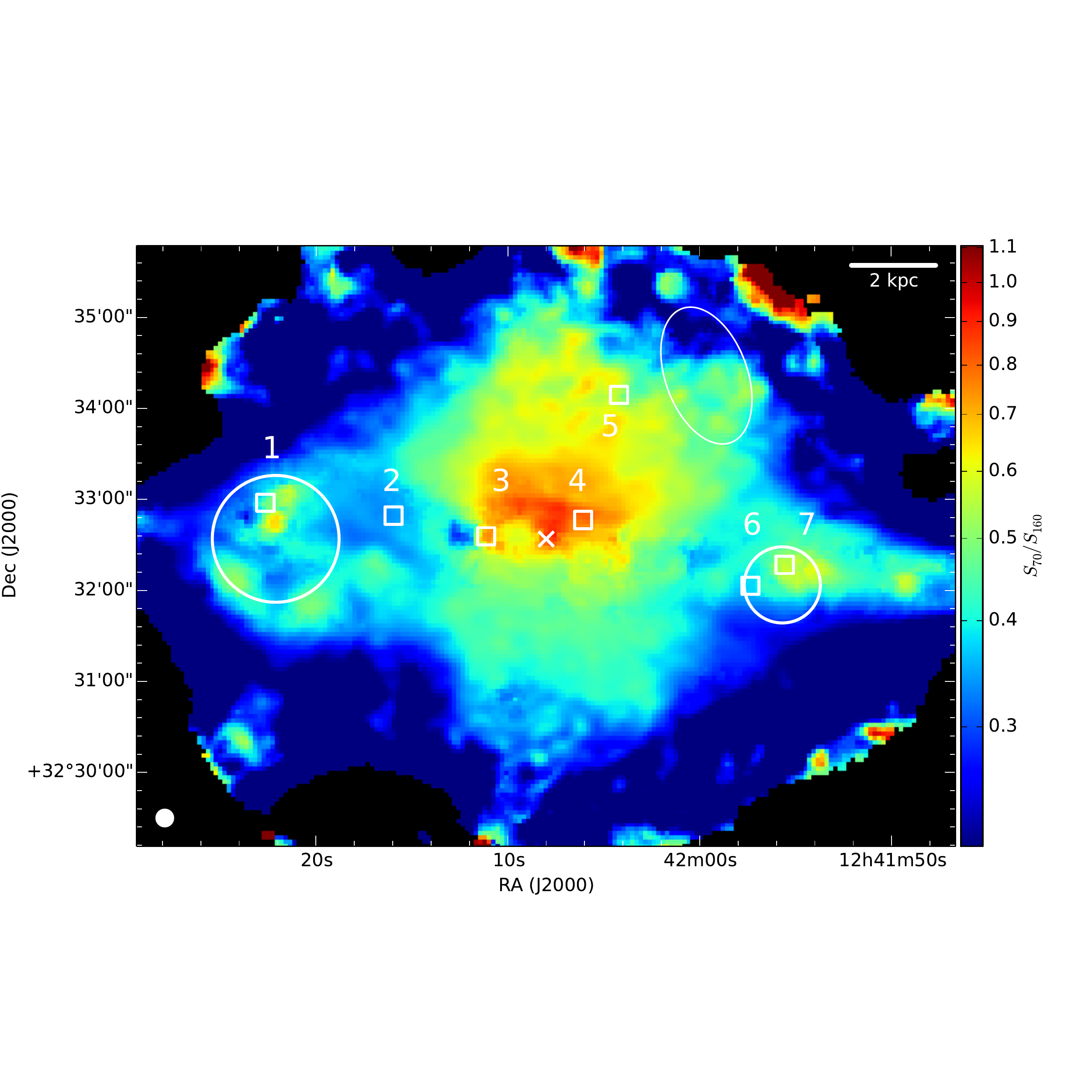} 
\caption{The $S_{70}/S_{160}$ ratio map of the adaptively smoothed
  PACS maps shown in Figure~\ref{comp_PACS_x-ray_radio_smooth}. The
  open white squares mark the positions of the brightest X-ray point
  sources. The white dashed circles represent the approximate size of
  the H~I supershells \citep{1993AJ....105.2098R}.  The white ellipse
  indicates the position of the companion galaxy NGC~4627. The beam
  size at 160 $\micron$ is indicated by a white filled circle in the
  bottom left corner. The white ``X" indicates the position of the
    infrared nucleus
    \citep{1978PASP...90...28A,1981PASP...93..535A}.\label{ratio_radio_and_x-ray_smooth_flat_s10_apendix}}
\end{figure}

\clearpage

\bibliographystyle{apj}

\bibliography{ms}

\begin{thebibliography}{104}
\expandafter\ifx\csname natexlab\endcsname\relax\def\natexlab#1{#1}\fi

\bibitem[{{Aaronson}(1978)}]{1978PASP...90...28A}
{Aaronson}, M. 1978, \pasp, 90, 28

\bibitem[{{Aaronson}(1981)}]{1981PASP...93..535A}
---. 1981, \pasp, 93, 535

\bibitem[{{Alton} {et~al.}(1999){Alton}, {Davies}, \&
  {Bianchi}}]{1999A&A...343...51A}
{Alton}, P.~B., {Davies}, J.~I., \& {Bianchi}, S. 1999, \aap, 343, 51

\bibitem[{{Aniano} {et~al.}(2011){Aniano}, {Draine}, {Gordon}, \&
  {Sandstrom}}]{2011PASP..123.1218A}
{Aniano}, G., {Draine}, B.~T., {Gordon}, K.~D., \& {Sandstrom}, K. 2011, \pasp,
  123, 1218

\bibitem[{{Bahcall}(1984)}]{1984ApJ...276..169B}
{Bahcall}, J.~N. 1984, \apj, 276, 169

\bibitem[{{Bendo} {et~al.}(2006){Bendo}, {Dale}, {Draine}, {Engelbracht},
  {Kennicutt}, {Calzetti}, {Gordon}, {Helou}, {Hollenbach}, {Li}, {Murphy},
  {Prescott}, \& {Smith}}]{2006ApJ...652..283B}
{Bendo}, G.~J., {Dale}, D.~A., {Draine}, B.~T., {Engelbracht}, C.~W.,
  {Kennicutt}, Jr., R.~C., {Calzetti}, D., {Gordon}, K.~D., {Helou}, G.,
  {Hollenbach}, D., {Li}, A., {Murphy}, E.~J., {Prescott}, M.~K.~M., \&
  {Smith}, J.-D.~T. 2006, \apj, 652, 283

\bibitem[{{Bendo} {et~al.}(2003){Bendo}, {Joseph}, {Wells}, {Gallais}, {Haas},
  {Heras}, {Klaas}, {Laureijs}, {Leech}, {Lemke}, {Metcalfe}, {Rowan-Robinson},
  {Schulz}, \& {Telesco}}]{2003AJ....125.2361B}
{Bendo}, G.~J., {Joseph}, R.~D., {Wells}, M., {Gallais}, P., {Haas}, M.,
  {Heras}, A.~M., {Klaas}, U., {Laureijs}, R.~J., {Leech}, K., {Lemke}, D.,
  {Metcalfe}, L., {Rowan-Robinson}, M., {Schulz}, B., \& {Telesco}, C. 2003,
  \aj, 125, 2361

\bibitem[{{Bianchi}(2013)}]{2013A&A...552A..89B}
{Bianchi}, S. 2013, \aap, 552, A89

\bibitem[{{Bolatto} {et~al.}(2013){Bolatto}, {Warren}, {Leroy}, {Walter},
  {Veilleux}, {Ostriker}, {Ott}, {Zwaan}, {Fisher}, {Weiss}, {Rosolowsky}, \&
  {Hodge}}]{2013Natur.499..450B}
{Bolatto}, A.~D., {Warren}, S.~R., {Leroy}, A.~K., {Walter}, F., {Veilleux},
  S., {Ostriker}, E.~C., {Ott}, J., {Zwaan}, M., {Fisher}, D.~B., {Weiss}, A.,
  {Rosolowsky}, E., \& {Hodge}, J. 2013, \nat, 499, 450

\bibitem[{{Bracco} {et~al.}(2011){Bracco}, {Cooray}, {Veneziani}, {Amblard},
  {Serra}, {Wardlow}, {Thompson}, {White}, {Auld}, {Baes}, {Bertoldi},
  {Buttiglione}, {Cava}, {Clements}, {Dariush}, {de Zotti}, {Dunne}, {Dye},
  {Eales}, {Fritz}, {Gomez}, {Hopwood}, {Ibar}, {Ivison}, {Jarvis}, {Lagache},
  {Lee}, {Leeuw}, {Maddox}, {Micha{\l}owski}, {Pearson}, {Pohlen}, {Rigby},
  {Rodighiero}, {Smith}, {Temi}, {Vaccari}, \& {van der
  Werf}}]{2011MNRAS.412.1151B}
{Bracco}, A., {Cooray}, A., {Veneziani}, M., {Amblard}, A., {Serra}, P.,
  {Wardlow}, J., {Thompson}, M.~A., {White}, G., {Auld}, R., {Baes}, M.,
  {Bertoldi}, F., {Buttiglione}, S., {Cava}, A., {Clements}, D.~L., {Dariush},
  A., {de Zotti}, G., {Dunne}, L., {Dye}, S., {Eales}, S., {Fritz}, J.,
  {Gomez}, H., {Hopwood}, R., {Ibar}, I., {Ivison}, R.~J., {Jarvis}, M.,
  {Lagache}, G., {Lee}, M.~G., {Leeuw}, L., {Maddox}, S., {Micha{\l}owski}, M.,
  {Pearson}, C., {Pohlen}, M., {Rigby}, E., {Rodighiero}, G., {Smith},
  D.~J.~B., {Temi}, P., {Vaccari}, M., \& {van der Werf}, P. 2011, \mnras, 412,
  1151

\bibitem[{{Braine} {et~al.}(1995){Braine}, {Kruegel}, {Sievers}, \&
  {Wielebinski}}]{1995A&A...295L..55B}
{Braine}, J., {Kruegel}, E., {Sievers}, A., \& {Wielebinski}, R. 1995, \aap,
  295, L55

\bibitem[{{Calzetti} {et~al.}(2010){Calzetti}, {Wu}, {Hong}, {Kennicutt},
  {Lee}, {Dale}, {Engelbracht}, {van Zee}, {Draine}, {Hao}, {Gordon},
  {Moustakas}, {Murphy}, {Regan}, {Begum}, {Block}, {Dalcanton}, {Funes}, {Gil
  de Paz}, {Johnson}, {Sakai}, {Skillman}, {Walter}, {Weisz}, {Williams}, \&
  {Wu}}]{2010ApJ...714.1256C}
{Calzetti}, D., {Wu}, S.-Y., {Hong}, S., {Kennicutt}, R.~C., {Lee}, J.~C.,
  {Dale}, D.~A., {Engelbracht}, C.~W., {van Zee}, L., {Draine}, B.~T., {Hao},
  C.-N., {Gordon}, K.~D., {Moustakas}, J., {Murphy}, E.~J., {Regan}, M.,
  {Begum}, A., {Block}, M., {Dalcanton}, J., {Funes}, J., {Gil de Paz}, A.,
  {Johnson}, B., {Sakai}, S., {Skillman}, E., {Walter}, F., {Weisz}, D.,
  {Williams}, B., \& {Wu}, Y. 2010, \apj, 714, 1256

\bibitem[{{Carpano} {et~al.}(2007){Carpano}, {Pollock}, {King}, {Wilms}, \&
  {Ehle}}]{2007A&A...471L..55C}
{Carpano}, S., {Pollock}, A.~M.~T., {King}, A.~R., {Wilms}, J., \& {Ehle}, M.
  2007, \aap, 471, L55

\bibitem[{{Cicone} {et~al.}(2014){Cicone}, {Maiolino}, {Sturm},
  {Graci{\'a}-Carpio}, {Feruglio}, {Neri}, {Aalto}, {Davies}, {Fiore},
  {Fischer}, {Garc{\'{\i}}a-Burillo}, {Gonz{\'a}lez-Alfonso},
  {Hailey-Dunsheath}, {Piconcelli}, \& {Veilleux}}]{2014A&A...562A..21C}
{Cicone}, C., {Maiolino}, R., {Sturm}, E., {Graci{\'a}-Carpio}, J., {Feruglio},
  C., {Neri}, R., {Aalto}, S., {Davies}, R., {Fiore}, F., {Fischer}, J.,
  {Garc{\'{\i}}a-Burillo}, S., {Gonz{\'a}lez-Alfonso}, E., {Hailey-Dunsheath},
  S., {Piconcelli}, E., \& {Veilleux}, S. 2014, \aap, 562, A21

\bibitem[{{Combes}(1978)}]{1978A&A....65...47C}
{Combes}, F. 1978, \aap, 65, 47

\bibitem[{{Cooper} {et~al.}(2008){Cooper}, {Bicknell}, {Sutherland}, \&
  {Bland-Hawthorn}}]{2008ApJ...674..157C}
{Cooper}, J.~L., {Bicknell}, G.~V., {Sutherland}, R.~S., \& {Bland-Hawthorn},
  J. 2008, \apj, 674, 157

\bibitem[{{Coupeaud} {et~al.}(2011){Coupeaud}, {Demyk}, {Meny}, {Nayral},
  {Delpech}, {Leroux}, {Depecker}, {Creff}, {Brubach}, \&
  {Roy}}]{2011A&A...535A.124C}
{Coupeaud}, A., {Demyk}, K., {Meny}, C., {Nayral}, C., {Delpech}, F., {Leroux},
  H., {Depecker}, C., {Creff}, G., {Brubach}, J.-B., \& {Roy}, P. 2011, \aap,
  535, A124

\bibitem[{{Crillon} \& {Monnet}(1969)}]{1969A&A.....2....1C}
{Crillon}, R., \& {Monnet}, G. 1969, \aap, 2, 1

\bibitem[{{Dale} {et~al.}(2012){Dale}, {Aniano}, {Engelbracht}, {Hinz},
  {Krause}, {Montiel}, {Roussel}, {Appleton}, {Armus}, {Beir{\~a}o}, {Bolatto},
  {Brandl}, {Calzetti}, {Crocker}, {Croxall}, {Draine}, {Galametz}, {Gordon},
  {Groves}, {Hao}, {Helou}, {Hunt}, {Johnson}, {Kennicutt}, {Koda}, {Leroy},
  {Li}, {Meidt}, {Miller}, {Murphy}, {Rahman}, {Rix}, {Sandstrom}, {Sauvage},
  {Schinnerer}, {Skibba}, {Smith}, {Tabatabaei}, {Walter}, {Wilson}, {Wolfire},
  \& {Zibetti}}]{2012ApJ...745...95D}
{Dale}, D.~A., {Aniano}, G., {Engelbracht}, C.~W., {Hinz}, J.~L., {Krause}, O.,
  {Montiel}, E.~J., {Roussel}, H., {Appleton}, P.~N., {Armus}, L.,
  {Beir{\~a}o}, P., {Bolatto}, A.~D., {Brandl}, B.~R., {Calzetti}, D.,
  {Crocker}, A.~F., {Croxall}, K.~V., {Draine}, B.~T., {Galametz}, M.,
  {Gordon}, K.~D., {Groves}, B.~A., {Hao}, C.-N., {Helou}, G., {Hunt}, L.~K.,
  {Johnson}, B.~D., {Kennicutt}, R.~C., {Koda}, J., {Leroy}, A.~K., {Li}, Y.,
  {Meidt}, S.~E., {Miller}, A.~E., {Murphy}, E.~J., {Rahman}, N., {Rix}, H.-W.,
  {Sandstrom}, K.~M., {Sauvage}, M., {Schinnerer}, E., {Skibba}, R.~A.,
  {Smith}, J.-D.~T., {Tabatabaei}, F.~S., {Walter}, F., {Wilson}, C.~D.,
  {Wolfire}, M.~G., \& {Zibetti}, S. 2012, \apj, 745, 95

\bibitem[{{Dale} {et~al.}(2005){Dale}, {Bendo}, {Engelbracht}, {Gordon},
  {Regan}, {Armus}, {Cannon}, {Calzetti}, {Draine}, {Helou}, {Joseph},
  {Kennicutt}, {Li}, {Murphy}, {Roussel}, {Walter}, {Hanson}, {Hollenbach},
  {Jarrett}, {Kewley}, {Lamanna}, {Leitherer}, {Meyer}, {Rieke}, {Rieke},
  {Sheth}, {Smith}, \& {Thornley}}]{2005ApJ...633..857D}
{Dale}, D.~A., {Bendo}, G.~J., {Engelbracht}, C.~W., {Gordon}, K.~D., {Regan},
  M.~W., {Armus}, L., {Cannon}, J.~M., {Calzetti}, D., {Draine}, B.~T.,
  {Helou}, G., {Joseph}, R.~D., {Kennicutt}, R.~C., {Li}, A., {Murphy}, E.~J.,
  {Roussel}, H., {Walter}, F., {Hanson}, H.~M., {Hollenbach}, D.~J., {Jarrett},
  T.~H., {Kewley}, L.~J., {Lamanna}, C.~A., {Leitherer}, C., {Meyer}, M.~J.,
  {Rieke}, G.~H., {Rieke}, M.~J., {Sheth}, K., {Smith}, J.~D.~T., \&
  {Thornley}, M.~D. 2005, \apj, 633, 857

\bibitem[{{Dale} {et~al.}(2009){Dale}, {Cohen}, {Johnson}, {Schuster},
  {Calzetti}, {Engelbracht}, {Gil de Paz}, {Kennicutt}, {Lee}, {Begum},
  {Block}, {Dalcanton}, {Funes}, {Gordon}, {Johnson}, {Marble}, {Sakai},
  {Skillman}, {van Zee}, {Walter}, {Weisz}, {Williams}, {Wu}, \&
  {Wu}}]{2009ApJ...703..517D}
{Dale}, D.~A., {Cohen}, S.~A., {Johnson}, L.~C., {Schuster}, M.~D., {Calzetti},
  D., {Engelbracht}, C.~W., {Gil de Paz}, A., {Kennicutt}, R.~C., {Lee}, J.~C.,
  {Begum}, A., {Block}, M., {Dalcanton}, J.~J., {Funes}, J.~G., {Gordon},
  K.~D., {Johnson}, B.~D., {Marble}, A.~R., {Sakai}, S., {Skillman}, E.~D.,
  {van Zee}, L., {Walter}, F., {Weisz}, D.~R., {Williams}, B., {Wu}, S.-Y., \&
  {Wu}, Y. 2009, \apj, 703, 517

\bibitem[{{Dale} {et~al.}(2007){Dale}, {Gil de Paz}, {Gordon}, {Hanson},
  {Armus}, {Bendo}, {Bianchi}, {Block}, {Boissier}, {Boselli}, {Buckalew},
  {Buat}, {Burgarella}, {Calzetti}, {Cannon}, {Engelbracht}, {Helou},
  {Hollenbach}, {Jarrett}, {Kennicutt}, {Leitherer}, {Li}, {Madore}, {Martin},
  {Meyer}, {Murphy}, {Regan}, {Roussel}, {Smith}, {Sosey}, {Thilker}, \&
  {Walter}}]{2007ApJ...655..863D}
{Dale}, D.~A., {Gil de Paz}, A., {Gordon}, K.~D., {Hanson}, H.~M., {Armus}, L.,
  {Bendo}, G.~J., {Bianchi}, L., {Block}, M., {Boissier}, S., {Boselli}, A.,
  {Buckalew}, B.~A., {Buat}, V., {Burgarella}, D., {Calzetti}, D., {Cannon},
  J.~M., {Engelbracht}, C.~W., {Helou}, G., {Hollenbach}, D.~J., {Jarrett},
  T.~H., {Kennicutt}, R.~C., {Leitherer}, C., {Li}, A., {Madore}, B.~F.,
  {Martin}, D.~C., {Meyer}, M.~J., {Murphy}, E.~J., {Regan}, M.~W., {Roussel},
  H., {Smith}, J.~D.~T., {Sosey}, M.~L., {Thilker}, D.~A., \& {Walter}, F.
  2007, \apj, 655, 863

\bibitem[{{Davies} {et~al.}(2014){Davies}, {Bianchi}, {Baes}, {Bendo},
  {Clemens}, {De Looze}, {Alighieri}, {Fritz}, {Fuller}, {Pappalardo},
  {Hughes}, {Madden}, {Smith}, {Verstappen}, \&
  {Vlahakis}}]{2014MNRAS.438.1922D}
{Davies}, J.~I., {Bianchi}, S., {Baes}, M., {Bendo}, G.~J., {Clemens}, M., {De
  Looze}, I., {Alighieri}, S.~d.~S., {Fritz}, J., {Fuller}, C., {Pappalardo},
  C., {Hughes}, T.~M., {Madden}, S., {Smith}, M.~W.~L., {Verstappen}, J., \&
  {Vlahakis}, C. 2014, \mnras, 438, 1922

\bibitem[{{D{\'e}sert} {et~al.}(2008){D{\'e}sert}, {Mac{\'{\i}}as-P{\'e}rez},
  {Mayet}, {Giardino}, {Renault}, {Aumont}, {Beno{\^i}t}, {Bernard},
  {Ponthieu}, \& {Tristram}}]{2008A&A...481..411D}
{D{\'e}sert}, F.-X., {Mac{\'{\i}}as-P{\'e}rez}, J.~F., {Mayet}, F., {Giardino},
  G., {Renault}, C., {Aumont}, J., {Beno{\^i}t}, A., {Bernard}, J.-P.,
  {Ponthieu}, N., \& {Tristram}, M. 2008, \aap, 481, 411

\bibitem[{{Donahue} {et~al.}(1995){Donahue}, {Aldering}, \&
  {Stocke}}]{1995ApJ...450L..45D}
{Donahue}, M., {Aldering}, G., \& {Stocke}, J.~T. 1995, \apjl, 450, L45

\bibitem[{{Draine}(2003)}]{2003ARA&A..41..241D}
{Draine}, B.~T. 2003, \araa, 41, 241

\bibitem[{{Draine} {et~al.}(2007){Draine}, {Dale}, {Bendo}, {Gordon}, {Smith},
  {Armus}, {Engelbracht}, {Helou}, {Kennicutt}, {Li}, {Roussel}, {Walter},
  {Calzetti}, {Moustakas}, {Murphy}, {Rieke}, {Bot}, {Hollenbach}, {Sheth}, \&
  {Teplitz}}]{2007ApJ...663..866D}
{Draine}, B.~T., {Dale}, D.~A., {Bendo}, G., {Gordon}, K.~D., {Smith},
  J.~D.~T., {Armus}, L., {Engelbracht}, C.~W., {Helou}, G., {Kennicutt}, Jr.,
  R.~C., {Li}, A., {Roussel}, H., {Walter}, F., {Calzetti}, D., {Moustakas},
  J., {Murphy}, E.~J., {Rieke}, G.~H., {Bot}, C., {Hollenbach}, D.~J., {Sheth},
  K., \& {Teplitz}, H.~I. 2007, \apj, 663, 866

\bibitem[{{Draine} \& {Li}(2007)}]{2007ApJ...657..810D}
{Draine}, B.~T., \& {Li}, A. 2007, \apj, 657, 810

\bibitem[{{Dumke} {et~al.}(2004){Dumke}, {Krause}, \&
  {Wielebinski}}]{2004A&A...414..475D}
{Dumke}, M., {Krause}, M., \& {Wielebinski}, R. 2004, \aap, 414, 475

\bibitem[{{Dumke} {et~al.}(2001){Dumke}, {Nieten}, {Thuma}, {Wielebinski}, \&
  {Walsh}}]{2001A&A...373..853D}
{Dumke}, M., {Nieten}, C., {Thuma}, G., {Wielebinski}, R., \& {Walsh}, W. 2001,
  \aap, 373, 853

\bibitem[{{Dupac} {et~al.}(2003){Dupac}, {Bernard}, {Boudet}, {Giard},
  {Lamarre}, {M{\'e}ny}, {Pajot}, {Ristorcelli}, {Serra}, {Stepnik}, \&
  {Torre}}]{2003A&A...404L..11D}
{Dupac}, X., {Bernard}, J.-P., {Boudet}, N., {Giard}, M., {Lamarre}, J.-M.,
  {M{\'e}ny}, C., {Pajot}, F., {Ristorcelli}, I., {Serra}, G., {Stepnik}, B.,
  \& {Torre}, J.-P. 2003, \aap, 404, L11

\bibitem[{{Ekers} \& {Sancisi}(1977)}]{1977A&A....54..973E}
{Ekers}, R.~D., \& {Sancisi}, R. 1977, \aap, 54, 973

\bibitem[{{Fabbiano} \& {Trinchieri}(1987)}]{1987ApJ...315...46F}
{Fabbiano}, G., \& {Trinchieri}, G. 1987, \apj, 315, 46

\bibitem[{{Golla} {et~al.}(1996){Golla}, {Dettmar}, \&
  {Domgoergen}}]{1996A&A...313..439G}
{Golla}, G., {Dettmar}, R.-J., \& {Domgoergen}, H. 1996, \aap, 313, 439

\bibitem[{{Golla} \& {Hummel}(1994)}]{1994A&A...284..777G}
{Golla}, G., \& {Hummel}, E. 1994, \aap, 284, 777

\bibitem[{{Golla} \& {Wielebinski}(1994)}]{1994A&A...286..733G}
{Golla}, G., \& {Wielebinski}, R. 1994, \aap, 286, 733

\bibitem[{{Griffin} {et~al.}(2010){Griffin}, {Abergel}, {Abreu}, {Ade},
  {Andr{\'e}}, {Augueres}, {Babbedge}, {Bae}, {Baillie}, {Baluteau}, {Barlow},
  {Bendo}, {Benielli}, {Bock}, {Bonhomme}, {Brisbin}, {Brockley-Blatt},
  {Caldwell}, {Cara}, {Castro-Rodriguez}, {Cerulli}, {Chanial}, {Chen},
  {Clark}, {Clements}, {Clerc}, {Coker}, {Communal}, {Conversi}, {Cox},
  {Crumb}, {Cunningham}, {Daly}, {Davis}, {de Antoni}, {Delderfield}, {Devin},
  {di Giorgio}, {Didschuns}, {Dohlen}, {Donati}, {Dowell}, {Dowell}, {Duband},
  {Dumaye}, {Emery}, {Ferlet}, {Ferrand}, {Fontignie}, {Fox}, {Franceschini},
  {Frerking}, {Fulton}, {Garcia}, {Gastaud}, {Gear}, {Glenn}, {Goizel},
  {Griffin}, {Grundy}, {Guest}, {Guillemet}, {Hargrave}, {Harwit}, {Hastings},
  {Hatziminaoglou}, {Herman}, {Hinde}, {Hristov}, {Huang}, {Imhof}, {Isaak},
  {Israelsson}, {Ivison}, {Jennings}, {Kiernan}, {King}, {Lange}, {Latter},
  {Laurent}, {Laurent}, {Leeks}, {Lellouch}, {Levenson}, {Li}, {Li},
  {Lilienthal}, {Lim}, {Liu}, {Lu}, {Madden}, {Mainetti}, {Marliani}, {McKay},
  {Mercier}, {Molinari}, {Morris}, {Moseley}, {Mulder}, {Mur}, {Naylor},
  {Nguyen}, {O'Halloran}, {Oliver}, {Olofsson}, {Olofsson}, {Orfei}, {Page},
  {Pain}, {Panuzzo}, {Papageorgiou}, {Parks}, {Parr-Burman}, {Pearce},
  {Pearson}, {P{\'e}rez-Fournon}, {Pinsard}, {Pisano}, {Podosek}, {Pohlen},
  {Polehampton}, {Pouliquen}, {Rigopoulou}, {Rizzo}, {Roseboom}, {Roussel},
  {Rowan-Robinson}, {Rownd}, {Saraceno}, {Sauvage}, {Savage}, {Savini},
  {Sawyer}, {Scharmberg}, {Schmitt}, {Schneider}, {Schulz}, {Schwartz},
  {Shafer}, {Shupe}, {Sibthorpe}, {Sidher}, {Smith}, {Smith}, {Smith},
  {Spencer}, {Stobie}, {Sudiwala}, {Sukhatme}, {Surace}, {Stevens}, {Swinyard},
  {Trichas}, {Tourette}, {Triou}, {Tseng}, {Tucker}, {Turner}, {Vaccari},
  {Valtchanov}, {Vigroux}, {Virique}, {Voellmer}, {Walker}, {Ward}, {Waskett},
  {Weilert}, {Wesson}, {White}, {Whitehouse}, {Wilson}, {Winter}, {Woodcraft},
  {Wright}, {Xu}, {Zavagno}, {Zemcov}, {Zhang}, \&
  {Zonca}}]{2010A&A...518L...3G}
{Griffin}, M.~J., {Abergel}, A., {Abreu}, A., {Ade}, P.~A.~R., {Andr{\'e}}, P.,
  {Augueres}, J.-L., {Babbedge}, T., {Bae}, Y., {Baillie}, T., {Baluteau},
  J.-P., {Barlow}, M.~J., {Bendo}, G., {Benielli}, D., {Bock}, J.~J.,
  {Bonhomme}, P., {Brisbin}, D., {Brockley-Blatt}, C., {Caldwell}, M., {Cara},
  C., {Castro-Rodriguez}, N., {Cerulli}, R., {Chanial}, P., {Chen}, S.,
  {Clark}, E., {Clements}, D.~L., {Clerc}, L., {Coker}, J., {Communal}, D.,
  {Conversi}, L., {Cox}, P., {Crumb}, D., {Cunningham}, C., {Daly}, F.,
  {Davis}, G.~R., {de Antoni}, P., {Delderfield}, J., {Devin}, N., {di
  Giorgio}, A., {Didschuns}, I., {Dohlen}, K., {Donati}, M., {Dowell}, A.,
  {Dowell}, C.~D., {Duband}, L., {Dumaye}, L., {Emery}, R.~J., {Ferlet}, M.,
  {Ferrand}, D., {Fontignie}, J., {Fox}, M., {Franceschini}, A., {Frerking},
  M., {Fulton}, T., {Garcia}, J., {Gastaud}, R., {Gear}, W.~K., {Glenn}, J.,
  {Goizel}, A., {Griffin}, D.~K., {Grundy}, T., {Guest}, S., {Guillemet}, L.,
  {Hargrave}, P.~C., {Harwit}, M., {Hastings}, P., {Hatziminaoglou}, E.,
  {Herman}, M., {Hinde}, B., {Hristov}, V., {Huang}, M., {Imhof}, P., {Isaak},
  K.~J., {Israelsson}, U., {Ivison}, R.~J., {Jennings}, D., {Kiernan}, B.,
  {King}, K.~J., {Lange}, A.~E., {Latter}, W., {Laurent}, G., {Laurent}, P.,
  {Leeks}, S.~J., {Lellouch}, E., {Levenson}, L., {Li}, B., {Li}, J.,
  {Lilienthal}, J., {Lim}, T., {Liu}, S.~J., {Lu}, N., {Madden}, S.,
  {Mainetti}, G., {Marliani}, P., {McKay}, D., {Mercier}, K., {Molinari}, S.,
  {Morris}, H., {Moseley}, H., {Mulder}, J., {Mur}, M., {Naylor}, D.~A.,
  {Nguyen}, H., {O'Halloran}, B., {Oliver}, S., {Olofsson}, G., {Olofsson},
  H.-G., {Orfei}, R., {Page}, M.~J., {Pain}, I., {Panuzzo}, P., {Papageorgiou},
  A., {Parks}, G., {Parr-Burman}, P., {Pearce}, A., {Pearson}, C.,
  {P{\'e}rez-Fournon}, I., {Pinsard}, F., {Pisano}, G., {Podosek}, J.,
  {Pohlen}, M., {Polehampton}, E.~T., {Pouliquen}, D., {Rigopoulou}, D.,
  {Rizzo}, D., {Roseboom}, I.~G., {Roussel}, H., {Rowan-Robinson}, M., {Rownd},
  B., {Saraceno}, P., {Sauvage}, M., {Savage}, R., {Savini}, G., {Sawyer}, E.,
  {Scharmberg}, C., {Schmitt}, D., {Schneider}, N., {Schulz}, B., {Schwartz},
  A., {Shafer}, R., {Shupe}, D.~L., {Sibthorpe}, B., {Sidher}, S., {Smith}, A.,
  {Smith}, A.~J., {Smith}, D., {Spencer}, L., {Stobie}, B., {Sudiwala}, R.,
  {Sukhatme}, K., {Surace}, C., {Stevens}, J.~A., {Swinyard}, B.~M., {Trichas},
  M., {Tourette}, T., {Triou}, H., {Tseng}, S., {Tucker}, C., {Turner}, A.,
  {Vaccari}, M., {Valtchanov}, I., {Vigroux}, L., {Virique}, E., {Voellmer},
  G., {Walker}, H., {Ward}, R., {Waskett}, T., {Weilert}, M., {Wesson}, R.,
  {White}, G.~J., {Whitehouse}, N., {Wilson}, C.~D., {Winter}, B., {Woodcraft},
  A.~L., {Wright}, G.~S., {Xu}, C.~K., {Zavagno}, A., {Zemcov}, M., {Zhang},
  L., \& {Zonca}, E. 2010, \aap, 518, L3

\bibitem[{{Hoopes} {et~al.}(1999){Hoopes}, {Walterbos}, \&
  {Rand}}]{1999ApJ...522..669H}
{Hoopes}, C.~G., {Walterbos}, R.~A.~M., \& {Rand}, R.~J. 1999, \apj, 522, 669

\bibitem[{{Howk} \& {Savage}(1997)}]{1997AJ....114.2463H}
{Howk}, J.~C., \& {Savage}, B.~D. 1997, \aj, 114, 2463

\bibitem[{{Hughes} {et~al.}(2014){Hughes}, {Baes}, {Fritz}, {Smith}, {Parkin},
  {Gentile}, {Bendo}, {Wilson}, {Allaert}, {Bianchi}, {De Looze}, {Verstappen},
  {Viaene}, {Boquien}, {Boselli}, {Clements}, {Davies}, {Galametz}, {Madden},
  {R{\'e}my-Ruyer}, \& {Spinoglio}}]{2014A&A...565A...4H}
{Hughes}, T.~M., {Baes}, M., {Fritz}, J., {Smith}, M.~W.~L., {Parkin}, T.~J.,
  {Gentile}, G., {Bendo}, G.~J., {Wilson}, C.~D., {Allaert}, F., {Bianchi}, S.,
  {De Looze}, I., {Verstappen}, J., {Viaene}, S., {Boquien}, M., {Boselli}, A.,
  {Clements}, D.~L., {Davies}, J.~I., {Galametz}, M., {Madden}, S.~C.,
  {R{\'e}my-Ruyer}, A., \& {Spinoglio}, L. 2014, \aap, 565, A4

\bibitem[{{Hummel} {et~al.}(1991){Hummel}, {Beck}, \&
  {Dahlem}}]{1991A&A...248...23H}
{Hummel}, E., {Beck}, R., \& {Dahlem}, M. 1991, \aap, 248, 23

\bibitem[{{Hummel} \& {Dettmar}(1990)}]{1990A&A...236...33H}
{Hummel}, E., \& {Dettmar}, R.-J. 1990, \aap, 236, 33

\bibitem[{{Hummel} {et~al.}(1988){Hummel}, {Lesch}, {Wielebinski}, \&
  {Schlickeiser}}]{1988A&A...197L..29H}
{Hummel}, E., {Lesch}, H., {Wielebinski}, R., \& {Schlickeiser}, R. 1988, \aap,
  197, L29

\bibitem[{{Irwin} {et~al.}(2012){Irwin}, {Beck}, {Benjamin}, {Dettmar},
  {English}, {Heald}, {Henriksen}, {Johnson}, {Krause}, {Li}, {Miskolczi},
  {Mora}, {Murphy}, {Oosterloo}, {Porter}, {Rand}, {Saikia}, {Schmidt},
  {Strong}, {Walterbos}, {Wang}, \& {Wiegert}}]{2012AJ....144...44I}
{Irwin}, J., {Beck}, R., {Benjamin}, R.~A., {Dettmar}, R.-J., {English}, J.,
  {Heald}, G., {Henriksen}, R.~N., {Johnson}, M., {Krause}, M., {Li}, J.-T.,
  {Miskolczi}, A., {Mora}, S.~C., {Murphy}, E.~J., {Oosterloo}, T., {Porter},
  T.~A., {Rand}, R.~J., {Saikia}, D.~J., {Schmidt}, P., {Strong}, A.~W.,
  {Walterbos}, R., {Wang}, Q.~D., \& {Wiegert}, T. 2012, \aj, 144, 44

\bibitem[{{Irwin} {et~al.}(2011){Irwin}, {Wilson}, {Wiegert}, {Bendo},
  {Warren}, {Wang}, {Israel}, {Serjeant}, {Knapen}, {Brinks}, {Tilanus}, {van
  der Werf}, \& {M{\"u}hle}}]{2011MNRAS.410.1423I}
{Irwin}, J.~A., {Wilson}, C.~D., {Wiegert}, T., {Bendo}, G.~J., {Warren},
  B.~E., {Wang}, Q.~D., {Israel}, F.~P., {Serjeant}, S., {Knapen}, J.~H.,
  {Brinks}, E., {Tilanus}, R.~P.~J., {van der Werf}, P., \& {M{\"u}hle}, S.
  2011, \mnras, 410, 1423

\bibitem[{{Israel}(2009)}]{2009A&A...506..689I}
{Israel}, F.~P. 2009, \aap, 506, 689

\bibitem[{{Jurac} {et~al.}(1998){Jurac}, {Johnson}, \&
  {Donn}}]{1998ApJ...503..247J}
{Jurac}, S., {Johnson}, R.~E., \& {Donn}, B. 1998, \apj, 503, 247

\bibitem[{{Kelly} {et~al.}(2012){Kelly}, {Shetty}, {Stutz}, {Kauffmann},
  {Goodman}, \& {Launhardt}}]{2012ApJ...752...55K}
{Kelly}, B.~C., {Shetty}, R., {Stutz}, A.~M., {Kauffmann}, J., {Goodman},
  A.~A., \& {Launhardt}, R. 2012, \apj, 752, 55

\bibitem[{{Kennicutt} \& {Evans}(2012)}]{2012ARA&A..50..531K}
{Kennicutt}, R.~C., \& {Evans}, N.~J. 2012, \araa, 50, 531

\bibitem[{{Kennicutt}(1998)}]{1998ARA&A..36..189K}
{Kennicutt}, Jr., R.~C. 1998, \araa, 36, 189

\bibitem[{{Kennicutt} {et~al.}(2009){Kennicutt}, {Hao}, {Calzetti},
  {Moustakas}, {Dale}, {Bendo}, {Engelbracht}, {Johnson}, \&
  {Lee}}]{2009ApJ...703.1672K}
{Kennicutt}, Jr., R.~C., {Hao}, C.-N., {Calzetti}, D., {Moustakas}, J., {Dale},
  D.~A., {Bendo}, G., {Engelbracht}, C.~W., {Johnson}, B.~D., \& {Lee}, J.~C.
  2009, \apj, 703, 1672

\bibitem[{{Lynds} \& {Sandage}(1963)}]{1963ApJ...137.1005L}
{Lynds}, C.~R., \& {Sandage}, A.~R. 1963, \apj, 137, 1005

\bibitem[{{Martin} \& {Kern}(2001)}]{2001ApJ...555..258M}
{Martin}, C., \& {Kern}, B. 2001, \apj, 555, 258

\bibitem[{{McCormick} {et~al.}(2013){McCormick}, {Veilleux}, \&
  {Rupke}}]{2013ApJ...774..126M}
{McCormick}, A., {Veilleux}, S., \& {Rupke}, D.~S.~N. 2013, \apj, 774, 126

\bibitem[{{Mel{\'e}ndez} {et~al.}(2014){Mel{\'e}ndez}, {Mushotzky}, {Shimizu},
  {Barger}, \& {Cowie}}]{2014ApJ...794..152M}
{Mel{\'e}ndez}, M., {Mushotzky}, R.~F., {Shimizu}, T.~T., {Barger}, A.~J., \&
  {Cowie}, L.~L. 2014, \apj, 794, 152

\bibitem[{{Meny} {et~al.}(2007){Meny}, {Gromov}, {Boudet}, {Bernard},
  {Paradis}, \& {Nayral}}]{2007A&A...468..171M}
{Meny}, C., {Gromov}, V., {Boudet}, N., {Bernard}, J.-P., {Paradis}, D., \&
  {Nayral}, C. 2007, \aap, 468, 171

\bibitem[{{Mora} \& {Krause}(2013)}]{2013A&A...560A..42M}
{Mora}, S.~C., \& {Krause}, M. 2013, \aap, 560, A42

\bibitem[{{Neininger} \& {Dumke}(1999)}]{1999PNAS...96.5360N}
{Neininger}, N., \& {Dumke}, M. 1999, Proceedings of the National Academy of
  Science, 96, 5360

\bibitem[{{Ott}(2010)}]{2010ASPC..434..139O}
{Ott}, S. 2010, in Astronomical Society of the Pacific Conference Series, Vol.
  434, Astronomical Data Analysis Software and Systems XIX, ed. Y.~{Mizumoto},
  K.-I. {Morita}, \& M.~{Ohishi}, 139

\bibitem[{{Paradis} {et~al.}(2010){Paradis}, {Veneziani}, {Noriega-Crespo},
  {Paladini}, {Piacentini}, {Bernard}, {de Bernardis}, {Calzoletti},
  {Faustini}, {Martin}, {Masi}, {Montier}, {Natoli}, {Ristorcelli}, {Thompson},
  {Traficante}, \& {Molinari}}]{2010A&A...520L...8P}
{Paradis}, D., {Veneziani}, M., {Noriega-Crespo}, A., {Paladini}, R.,
  {Piacentini}, F., {Bernard}, J.~P., {de Bernardis}, P., {Calzoletti}, L.,
  {Faustini}, F., {Martin}, P., {Masi}, S., {Montier}, L., {Natoli}, P.,
  {Ristorcelli}, I., {Thompson}, M.~A., {Traficante}, A., \& {Molinari}, S.
  2010, \aap, 520, L8

\bibitem[{{Pilbratt} {et~al.}(2010){Pilbratt}, {Riedinger}, {Passvogel},
  {Crone}, {Doyle}, {Gageur}, {Heras}, {Jewell}, {Metcalfe}, {Ott}, \&
  {Schmidt}}]{2010A&A...518L...1P}
{Pilbratt}, G.~L., {Riedinger}, J.~R., {Passvogel}, T., {Crone}, G., {Doyle},
  D., {Gageur}, U., {Heras}, A.~M., {Jewell}, C., {Metcalfe}, L., {Ott}, S., \&
  {Schmidt}, M. 2010, \aap, 518, L1

\bibitem[{{Poglitsch} {et~al.}(2010){Poglitsch}, {Waelkens}, {Geis},
  {Feuchtgruber}, {Vandenbussche}, {Rodriguez}, {Krause}, {Renotte}, {van
  Hoof}, {Saraceno}, {Cepa}, {Kerschbaum}, {Agn{\`e}se}, {Ali}, {Altieri},
  {Andreani}, {Augueres}, {Balog}, {Barl}, {Bauer}, {Belbachir}, {Benedettini},
  {Billot}, {Boulade}, {Bischof}, {Blommaert}, {Callut}, {Cara}, {Cerulli},
  {Cesarsky}, {Contursi}, {Creten}, {De Meester}, {Doublier}, {Doumayrou},
  {Duband}, {Exter}, {Genzel}, {Gillis}, {Gr{\"o}zinger}, {Henning},
  {Herreros}, {Huygen}, {Inguscio}, {Jakob}, {Jamar}, {Jean}, {de Jong},
  {Katterloher}, {Kiss}, {Klaas}, {Lemke}, {Lutz}, {Madden}, {Marquet},
  {Martignac}, {Mazy}, {Merken}, {Montfort}, {Morbidelli}, {M{\"u}ller},
  {Nielbock}, {Okumura}, {Orfei}, {Ottensamer}, {Pezzuto}, {Popesso},
  {Putzeys}, {Regibo}, {Reveret}, {Royer}, {Sauvage}, {Schreiber}, {Stegmaier},
  {Schmitt}, {Schubert}, {Sturm}, {Thiel}, {Tofani}, {Vavrek}, {Wetzstein},
  {Wieprecht}, \& {Wiezorrek}}]{2010A&A...518L...2P}
{Poglitsch}, A., {Waelkens}, C., {Geis}, N., {Feuchtgruber}, H.,
  {Vandenbussche}, B., {Rodriguez}, L., {Krause}, O., {Renotte}, E., {van
  Hoof}, C., {Saraceno}, P., {Cepa}, J., {Kerschbaum}, F., {Agn{\`e}se}, P.,
  {Ali}, B., {Altieri}, B., {Andreani}, P., {Augueres}, J.-L., {Balog}, Z.,
  {Barl}, L., {Bauer}, O.~H., {Belbachir}, N., {Benedettini}, M., {Billot}, N.,
  {Boulade}, O., {Bischof}, H., {Blommaert}, J., {Callut}, E., {Cara}, C.,
  {Cerulli}, R., {Cesarsky}, D., {Contursi}, A., {Creten}, Y., {De Meester},
  W., {Doublier}, V., {Doumayrou}, E., {Duband}, L., {Exter}, K., {Genzel}, R.,
  {Gillis}, J.-M., {Gr{\"o}zinger}, U., {Henning}, T., {Herreros}, J.,
  {Huygen}, R., {Inguscio}, M., {Jakob}, G., {Jamar}, C., {Jean}, C., {de
  Jong}, J., {Katterloher}, R., {Kiss}, C., {Klaas}, U., {Lemke}, D., {Lutz},
  D., {Madden}, S., {Marquet}, B., {Martignac}, J., {Mazy}, A., {Merken}, P.,
  {Montfort}, F., {Morbidelli}, L., {M{\"u}ller}, T., {Nielbock}, M.,
  {Okumura}, K., {Orfei}, R., {Ottensamer}, R., {Pezzuto}, S., {Popesso}, P.,
  {Putzeys}, J., {Regibo}, S., {Reveret}, V., {Royer}, P., {Sauvage}, M.,
  {Schreiber}, J., {Stegmaier}, J., {Schmitt}, D., {Schubert}, J., {Sturm}, E.,
  {Thiel}, M., {Tofani}, G., {Vavrek}, R., {Wetzstein}, M., {Wieprecht}, E., \&
  {Wiezorrek}, E. 2010, \aap, 518, L2

\bibitem[{{Radburn-Smith} {et~al.}(2011){Radburn-Smith}, {de Jong}, {Seth},
  {Bailin}, {Bell}, {Brown}, {Bullock}, {Courteau}, {Dalcanton}, {Ferguson},
  {Goudfrooij}, {Holfeltz}, {Holwerda}, {Purcell}, {Sick}, {Streich}, {Vlajic},
  \& {Zucker}}]{2011ApJS..195...18R}
{Radburn-Smith}, D.~J., {de Jong}, R.~S., {Seth}, A.~C., {Bailin}, J., {Bell},
  E.~F., {Brown}, T.~M., {Bullock}, J.~S., {Courteau}, S., {Dalcanton}, J.~J.,
  {Ferguson}, H.~C., {Goudfrooij}, P., {Holfeltz}, S., {Holwerda}, B.~W.,
  {Purcell}, C., {Sick}, J., {Streich}, D., {Vlajic}, M., \& {Zucker}, D.~B.
  2011, \apjs, 195, 18

\bibitem[{{Rand}(1994)}]{1994A&A...285..833R}
{Rand}, R.~J. 1994, \aap, 285, 833

\bibitem[{{Rand}(2000)}]{2000ApJ...535..663R}
---. 2000, \apj, 535, 663

\bibitem[{{Rand} {et~al.}(1992){Rand}, {Kulkarni}, \&
  {Hester}}]{1992ApJ...396...97R}
{Rand}, R.~J., {Kulkarni}, S.~R., \& {Hester}, J.~J. 1992, \apj, 396, 97

\bibitem[{{Rand} \& {Stone}(1996)}]{1996AJ....111..190R}
{Rand}, R.~J., \& {Stone}, J.~M. 1996, \aj, 111, 190

\bibitem[{{Rand} \& {van der Hulst}(1993)}]{1993AJ....105.2098R}
{Rand}, R.~J., \& {van der Hulst}, J.~M. 1993, \aj, 105, 2098

\bibitem[{{Raymond} {et~al.}(2013){Raymond}, {Ghavamian}, {Williams}, {Blair},
  {Borkowski}, {Gaetz}, \& {Sankrit}}]{2013ApJ...778..161R}
{Raymond}, J.~C., {Ghavamian}, P., {Williams}, B.~J., {Blair}, W.~P.,
  {Borkowski}, K.~J., {Gaetz}, T.~J., \& {Sankrit}, R. 2013, \apj, 778, 161

\bibitem[{{Read} {et~al.}(1997){Read}, {Ponman}, \&
  {Strickland}}]{1997MNRAS.286..626R}
{Read}, A.~M., {Ponman}, T.~J., \& {Strickland}, D.~K. 1997, \mnras, 286, 626

\bibitem[{{Rice} {et~al.}(1988){Rice}, {Lonsdale}, {Soifer}, {Neugebauer},
  {Kopan}, {Lloyd}, {de Jong}, \& {Habing}}]{1988ApJS...68...91R}
{Rice}, W., {Lonsdale}, C.~J., {Soifer}, B.~T., {Neugebauer}, G., {Kopan},
  E.~L., {Lloyd}, L.~A., {de Jong}, T., \& {Habing}, H.~J. 1988, \apjs, 68, 91

\bibitem[{{Roberts}(1968)}]{1968ApJ...151..117R}
{Roberts}, M.~S. 1968, \apj, 151, 117

\bibitem[{{Roussel}(2013)}]{2013PASP..125.1126R}
{Roussel}, H. 2013, \pasp, 125, 1126

\bibitem[{{Roussel} {et~al.}(2010){Roussel}, {Wilson}, {Vigroux}, {Isaak},
  {Sauvage}, {Madden}, {Auld}, {Baes}, {Barlow}, {Bendo}, {Bock}, {Boselli},
  {Bradford}, {Buat}, {Castro-Rodriguez}, {Chanial}, {Charlot}, {Ciesla},
  {Clements}, {Cooray}, {Cormier}, {Cortese}, {Davies}, {Dwek}, {Eales},
  {Elbaz}, {Galametz}, {Galliano}, {Gear}, {Glenn}, {Gomez}, {Griffin}, {Hony},
  {Levenson}, {Lu}, {O'Halloran}, {Okumura}, {Oliver}, {Page}, {Panuzzo},
  {Papageorgiou}, {Parkin}, {Perez-Fournon}, {Pohlen}, {Rangwala}, {Rigby},
  {Rykala}, {Sacchi}, {Schulz}, {Schirm}, {Smith}, {Spinoglio}, {Stevens},
  {Srinivasan}, {Symeonidis}, {Trichas}, {Vaccari}, {Wozniak}, {Wright}, \&
  {Zeilinger}}]{2010A&A...518L..66R}
{Roussel}, H., {Wilson}, C.~D., {Vigroux}, L., {Isaak}, K.~G., {Sauvage}, M.,
  {Madden}, S.~C., {Auld}, R., {Baes}, M., {Barlow}, M.~J., {Bendo}, G.~J.,
  {Bock}, J.~J., {Boselli}, A., {Bradford}, M., {Buat}, V., {Castro-Rodriguez},
  N., {Chanial}, P., {Charlot}, S., {Ciesla}, L., {Clements}, D.~L., {Cooray},
  A., {Cormier}, D., {Cortese}, L., {Davies}, J.~I., {Dwek}, E., {Eales},
  S.~A., {Elbaz}, D., {Galametz}, M., {Galliano}, F., {Gear}, W.~K., {Glenn},
  J., {Gomez}, H.~L., {Griffin}, M., {Hony}, S., {Levenson}, L.~R., {Lu}, N.,
  {O'Halloran}, B., {Okumura}, K., {Oliver}, S., {Page}, M.~J., {Panuzzo}, P.,
  {Papageorgiou}, A., {Parkin}, T.~J., {Perez-Fournon}, I., {Pohlen}, M.,
  {Rangwala}, N., {Rigby}, E.~E., {Rykala}, A., {Sacchi}, N., {Schulz}, B.,
  {Schirm}, M.~R.~P., {Smith}, M.~W.~L., {Spinoglio}, L., {Stevens}, J.~A.,
  {Srinivasan}, S., {Symeonidis}, M., {Trichas}, M., {Vaccari}, M., {Wozniak},
  H., {Wright}, G.~S., \& {Zeilinger}, W.~W. 2010, \aap, 518, L66

\bibitem[{{Schechtman-Rook} \& {Hess}(2012)}]{2012ApJ...750..171S}
{Schechtman-Rook}, A., \& {Hess}, K.~M. 2012, \apj, 750, 171

\bibitem[{{Seth} {et~al.}(2005{\natexlab{a}}){Seth}, {Dalcanton}, \& {de
  Jong}}]{2005AJ....129.1331S}
{Seth}, A.~C., {Dalcanton}, J.~J., \& {de Jong}, R.~S. 2005{\natexlab{a}}, \aj,
  129, 1331

\bibitem[{{Seth} {et~al.}(2005{\natexlab{b}}){Seth}, {Dalcanton}, \& {de
  Jong}}]{2005AJ....130.1574S}
---. 2005{\natexlab{b}}, \aj, 130, 1574

\bibitem[{{Shetty} {et~al.}(2009{\natexlab{a}}){Shetty}, {Kauffmann}, {Schnee},
  \& {Goodman}}]{2009ApJ...696..676S}
{Shetty}, R., {Kauffmann}, J., {Schnee}, S., \& {Goodman}, A.~A.
  2009{\natexlab{a}}, \apj, 696, 676

\bibitem[{{Shetty} {et~al.}(2009{\natexlab{b}}){Shetty}, {Kauffmann}, {Schnee},
  {Goodman}, \& {Ercolano}}]{2009ApJ...696.2234S}
{Shetty}, R., {Kauffmann}, J., {Schnee}, S., {Goodman}, A.~A., \& {Ercolano},
  B. 2009{\natexlab{b}}, \apj, 696, 2234

\bibitem[{{Skibba} {et~al.}(2011){Skibba}, {Engelbracht}, {Dale}, {Hinz},
  {Zibetti}, {Crocker}, {Groves}, {Hunt}, {Johnson}, {Meidt}, {Murphy},
  {Appleton}, {Armus}, {Bolatto}, {Brandl}, {Calzetti}, {Croxall}, {Galametz},
  {Gordon}, {Kennicutt}, {Koda}, {Krause}, {Montiel}, {Rix}, {Roussel},
  {Sandstrom}, {Sauvage}, {Schinnerer}, {Smith}, {Walter}, {Wilson}, \&
  {Wolfire}}]{2011ApJ...738...89S}
{Skibba}, R.~A., {Engelbracht}, C.~W., {Dale}, D., {Hinz}, J., {Zibetti}, S.,
  {Crocker}, A., {Groves}, B., {Hunt}, L., {Johnson}, B.~D., {Meidt}, S.,
  {Murphy}, E., {Appleton}, P., {Armus}, L., {Bolatto}, A., {Brandl}, B.,
  {Calzetti}, D., {Croxall}, K., {Galametz}, M., {Gordon}, K.~D., {Kennicutt},
  R.~C., {Koda}, J., {Krause}, O., {Montiel}, E., {Rix}, H.-W., {Roussel}, H.,
  {Sandstrom}, K., {Sauvage}, M., {Schinnerer}, E., {Smith}, J.~D., {Walter},
  F., {Wilson}, C.~D., \& {Wolfire}, M. 2011, \apj, 738, 89

\bibitem[{{Smith} {et~al.}(2001){Smith}, {Collins}, {Waller}, {Roberts},
  {Smith}, {Bohlin}, {Cheng}, {Fanelli}, {Neff}, {O'Connell}, {Parise},
  {Smith}, \& {Stecher}}]{2001ApJ...546..829S}
{Smith}, A.~M., {Collins}, N.~R., {Waller}, W.~H., {Roberts}, M.~S., {Smith},
  D.~A., {Bohlin}, R.~C., {Cheng}, K.-P., {Fanelli}, M.~N., {Neff}, S.~G.,
  {O'Connell}, R.~W., {Parise}, R.~A., {Smith}, E.~P., \& {Stecher}, T.~P.
  2001, \apj, 546, 829

\bibitem[{{Smith} {et~al.}(2012){Smith}, {Eales}, {Gomez}, {Roman-Duval},
  {Fritz}, {Braun}, {Baes}, {Bendo}, {Blommaert}, {Boquien}, {Boselli},
  {Clements}, {Cooray}, {Cortese}, {de Looze}, {Ford}, {Gear}, {Gentile},
  {Gordon}, {Kirk}, {Lebouteiller}, {Madden}, {Mentuch}, {O'Halloran}, {Page},
  {Schulz}, {Spinoglio}, {Verstappen}, {Wilson}, \&
  {Thilker}}]{2012ApJ...756...40S}
{Smith}, M.~W.~L., {Eales}, S.~A., {Gomez}, H.~L., {Roman-Duval}, J., {Fritz},
  J., {Braun}, R., {Baes}, M., {Bendo}, G.~J., {Blommaert}, J.~A.~D.~L.,
  {Boquien}, M., {Boselli}, A., {Clements}, D.~L., {Cooray}, A.~R., {Cortese},
  L., {de Looze}, I., {Ford}, G.~P., {Gear}, W.~K., {Gentile}, G., {Gordon},
  K.~D., {Kirk}, J., {Lebouteiller}, V., {Madden}, S., {Mentuch}, E.,
  {O'Halloran}, B., {Page}, M.~J., {Schulz}, B., {Spinoglio}, L., {Verstappen},
  J., {Wilson}, C.~D., \& {Thilker}, D.~A. 2012, \apj, 756, 40

\bibitem[{{Soria} \& {Ghosh}(2009)}]{2009ApJ...696..287S}
{Soria}, R., \& {Ghosh}, K.~K. 2009, \apj, 696, 287

\bibitem[{{Stevens} {et~al.}(2005){Stevens}, {Amure}, \&
  {Gear}}]{2005MNRAS.357..361S}
{Stevens}, J.~A., {Amure}, M., \& {Gear}, W.~K. 2005, \mnras, 357, 361

\bibitem[{{Strickland} {et~al.}(2004){Strickland}, {Heckman}, {Colbert},
  {Hoopes}, \& {Weaver}}]{2004ApJS..151..193S}
{Strickland}, D.~K., {Heckman}, T.~M., {Colbert}, E.~J.~M., {Hoopes}, C.~G., \&
  {Weaver}, K.~A. 2004, \apjs, 151, 193

\bibitem[{{Sturm} {et~al.}(2011){Sturm}, {Gonz{\'a}lez-Alfonso}, {Veilleux},
  {Fischer}, {Graci{\'a}-Carpio}, {Hailey-Dunsheath}, {Contursi}, {Poglitsch},
  {Sternberg}, {Davies}, {Genzel}, {Lutz}, {Tacconi}, {Verma}, {Maiolino}, \&
  {de Jong}}]{2011ApJ...733L..16S}
{Sturm}, E., {Gonz{\'a}lez-Alfonso}, E., {Veilleux}, S., {Fischer}, J.,
  {Graci{\'a}-Carpio}, J., {Hailey-Dunsheath}, S., {Contursi}, A., {Poglitsch},
  A., {Sternberg}, A., {Davies}, R., {Genzel}, R., {Lutz}, D., {Tacconi}, L.,
  {Verma}, A., {Maiolino}, R., \& {de Jong}, J.~A. 2011, \apjl, 733, L16

\bibitem[{{Taylor} \& {Wang}(2003)}]{2003AJ....125.1204T}
{Taylor}, C.~L., \& {Wang}, Q.~D. 2003, \aj, 125, 1204

\bibitem[{{T{\"u}llmann} {et~al.}(2006{\natexlab{a}}){T{\"u}llmann},
  {Breitschwerdt}, {Rossa}, {Pietsch}, \& {Dettmar}}]{2006A&A...457..779T}
{T{\"u}llmann}, R., {Breitschwerdt}, D., {Rossa}, J., {Pietsch}, W., \&
  {Dettmar}, R.-J. 2006{\natexlab{a}}, \aap, 457, 779

\bibitem[{{T{\"u}llmann} {et~al.}(2006{\natexlab{b}}){T{\"u}llmann}, {Pietsch},
  {Rossa}, {Breitschwerdt}, \& {Dettmar}}]{2006A&A...448...43T}
{T{\"u}llmann}, R., {Pietsch}, W., {Rossa}, J., {Breitschwerdt}, D., \&
  {Dettmar}, R.-J. 2006{\natexlab{b}}, \aap, 448, 43

\bibitem[{{van der Kruit}(1988)}]{1988A&A...192..117V}
{van der Kruit}, P.~C. 1988, \aap, 192, 117

\bibitem[{{Veilleux} {et~al.}(2005){Veilleux}, {Cecil}, \&
  {Bland-Hawthorn}}]{2005ARA&A..43..769V}
{Veilleux}, S., {Cecil}, G., \& {Bland-Hawthorn}, J. 2005, \araa, 43, 769

\bibitem[{{Veilleux} {et~al.}(2013){Veilleux}, {Mel{\'e}ndez}, {Sturm},
  {Gracia-Carpio}, {Fischer}, {Gonz{\'a}lez-Alfonso}, {Contursi}, {Lutz},
  {Poglitsch}, {Davies}, {Genzel}, {Tacconi}, {de Jong}, {Sternberg}, {Netzer},
  {Hailey-Dunsheath}, {Verma}, {Rupke}, {Maiolino}, {Teng}, \&
  {Polisensky}}]{2013ApJ...776...27V}
{Veilleux}, S., {Mel{\'e}ndez}, M., {Sturm}, E., {Gracia-Carpio}, J.,
  {Fischer}, J., {Gonz{\'a}lez-Alfonso}, E., {Contursi}, A., {Lutz}, D.,
  {Poglitsch}, A., {Davies}, R., {Genzel}, R., {Tacconi}, L., {de Jong}, J.~A.,
  {Sternberg}, A., {Netzer}, H., {Hailey-Dunsheath}, S., {Verma}, A., {Rupke},
  D.~S.~N., {Maiolino}, R., {Teng}, S.~H., \& {Polisensky}, E. 2013, \apj, 776,
  27

\bibitem[{{Veneziani} {et~al.}(2010){Veneziani}, {Ade}, {Bock}, {Boscaleri},
  {Crill}, {de Bernardis}, {De Gasperis}, {de Oliveira-Costa}, {De Troia}, {Di
  Stefano}, {Ganga}, {Jones}, {Kisner}, {Lange}, {MacTavish}, {Masi},
  {Mauskopf}, {Montroy}, {Natoli}, {Netterfield}, {Pascale}, {Piacentini},
  {Pietrobon}, {Polenta}, {Ricciardi}, {Romeo}, \&
  {Ruhl}}]{2010ApJ...713..959V}
{Veneziani}, M., {Ade}, P.~A.~R., {Bock}, J.~J., {Boscaleri}, A., {Crill},
  B.~P., {de Bernardis}, P., {De Gasperis}, G., {de Oliveira-Costa}, A., {De
  Troia}, G., {Di Stefano}, G., {Ganga}, K.~M., {Jones}, W.~C., {Kisner},
  T.~S., {Lange}, A.~E., {MacTavish}, C.~J., {Masi}, S., {Mauskopf}, P.~D.,
  {Montroy}, T.~E., {Natoli}, P., {Netterfield}, C.~B., {Pascale}, E.,
  {Piacentini}, F., {Pietrobon}, D., {Polenta}, G., {Ricciardi}, S., {Romeo},
  G., \& {Ruhl}, J.~E. 2010, \apj, 713, 959

\bibitem[{{Vogler} \& {Pietsch}(1996)}]{1996A&A...311...35V}
{Vogler}, A., \& {Pietsch}, W. 1996, \aap, 311, 35

\bibitem[{{Wang} {et~al.}(2001){Wang}, {Immler}, {Walterbos}, {Lauroesch}, \&
  {Breitschwerdt}}]{2001ApJ...555L..99W}
{Wang}, Q.~D., {Immler}, S., {Walterbos}, R., {Lauroesch}, J.~T., \&
  {Breitschwerdt}, D. 2001, \apjl, 555, L99

\bibitem[{{Wang} {et~al.}(1995){Wang}, {Walterbos}, {Steakley}, {Norman}, \&
  {Braun}}]{1995ApJ...439..176W}
{Wang}, Q.~D., {Walterbos}, R.~A.~M., {Steakley}, M.~F., {Norman}, C.~A., \&
  {Braun}, R. 1995, \apj, 439, 176

\bibitem[{{Weliachew} {et~al.}(1978){Weliachew}, {Sancisi}, \&
  {Guelin}}]{1978A&A....65...37W}
{Weliachew}, L., {Sancisi}, R., \& {Guelin}, M. 1978, \aap, 65, 37

\bibitem[{{Winter} {et~al.}(2006){Winter}, {Mushotzky}, \&
  {Reynolds}}]{2006ApJ...649..730W}
{Winter}, L.~M., {Mushotzky}, R.~F., \& {Reynolds}, C.~S. 2006, \apj, 649, 730

\bibitem[{{Winter} {et~al.}(2007){Winter}, {Mushotzky}, \&
  {Reynolds}}]{2007ApJ...655..163W}
---. 2007, \apj, 655, 163

\bibitem[{{Yamasaki} {et~al.}(2009){Yamasaki}, {Sato}, {Mitsuishi}, \&
  {Ohashi}}]{2009PASJ...61S.291Y}
{Yamasaki}, N.~Y., {Sato}, K., {Mitsuishi}, I., \& {Ohashi}, T. 2009, \pasj,
  61, 291

\bibitem[{{Young} {et~al.}(1989){Young}, {Xie}, {Kenney}, \&
  {Rice}}]{1989ApJS...70..699Y}
{Young}, J.~S., {Xie}, S., {Kenney}, J.~D.~P., \& {Rice}, W.~L. 1989, \apjs,
  70, 699

\bibitem[{{Ysard} {et~al.}(2012){Ysard}, {Juvela}, {Demyk}, {Guillet},
  {Abergel}, {Bernard}, {Malinen}, {M{\'e}ny}, {Montier}, {Paradis},
  {Ristorcelli}, \& {Verstraete}}]{2012A&A...542A..21Y}
{Ysard}, N., {Juvela}, M., {Demyk}, K., {Guillet}, V., {Abergel}, A.,
  {Bernard}, J.-P., {Malinen}, J., {M{\'e}ny}, C., {Montier}, L., {Paradis},
  D., {Ristorcelli}, I., \& {Verstraete}, L. 2012, \aap, 542, A21

\bibitem[{{Zibetti}(2009)}]{2009arXiv0911.4956Z}
{Zibetti}, S. 2009, ArXiv e-prints

\bibitem[{{Zibetti} {et~al.}(2009){Zibetti}, {Charlot}, \&
  {Rix}}]{2009MNRAS.400.1181Z}
{Zibetti}, S., {Charlot}, S., \& {Rix}, H.-W. 2009, \mnras, 400, 1181

\end{thebibliography}

\end{document}